\documentclass[a4paper,10pt,twoside]{cpc-hepnp}
\usepackage{subfigure}
\usepackage{color}
\usepackage{epsfig}
\usepackage{indentfirst}
\usepackage{multirow}
\usepackage{bm}
\usepackage{fancyhdr}
\usepackage{enumerate}
\usepackage[perpage,symbol]{footmisc}
\usepackage[colorlinks,linkcolor=blue,anchorcolor=blue,citecolor=blue]{hyperref}
\usepackage{multicol}
\usepackage{booktabs}
\usepackage{amssymb,bm,mathrsfs,bbm,amscd}
\usepackage[tbtags]{amsmath}
\usepackage{lastpage}
\usepackage{epstopdf}
\usepackage{lineno}
\newcommand{\BR}{{\cal B}}

\newcommand{\pp}{\pi^+\pi^-}
\newcommand{\etap}{\eta^\prime}

\newcommand{\jpsi}{J/\psi}

\newcommand{\EE}{e^+e^-}
\newcommand{\MM}{\mu^+\mu^-}
\newcommand{\pip}{\pi^+}
\newcommand{\pim}{\pi^-}
\newcommand{\pin}{\pi^-}
\newcommand{\pio}{\pi^0}

\newcommand{\g}{\gamma}
\newcommand{\ar}{\rightarrow}
\newcommand{\ra}{\rightarrow}

\makeatletter
\newenvironment{tablehere}
  {\def\@captype{table}}
  {}
\newenvironment{figurehere}
  {\def\@captype{figure}}
  {}
\makeatother

\begin{document}

\fancyhead[c]{\small Chinese Physics C~~~Vol. X, No. X (X) X}
\fancyfoot[C]{\small X-\thepage}

\footnotetext[0]{Received X X X}

\title{An overview of $\eta$ and $\eta^{\prime}$ decays at BESIII\thanks{This article
    is supported in part by National Natural Science Foundation of
    China (NSFC) under Contract No. 11565006, 11675184, 11735014}}

\author{
Shuang-shi Fang $^{1,2,4}$\email{fangss@ihep.ac.cn}
\quad Andrzej Kupsc$^{3}$\email{Andrzej.Kupsc@physics.uu.se}%
\quad Dai-hui Wei$^{4}$\email{weidh@gxnu.edu.cn}
}
\maketitle

\address{

$^1$ Institute of High Energy Physics, Chinese Academy of Science, 100049 Beijing,
People's Republic of China \\
$^2$ University of Chinese Academy of Sciences, 100049 Beijing, People's Republic of China\\
$^3$ Uppsala University, Box 516, SE-75120 Uppsala, Sweden\\
$^4$ Guangxi Normal University, Guilin 541004, People's Republic of China

}

\begin{abstract}

The world's largest sample of 1.31 billion $J/\psi$ events accumulated
at the BESIII detector, provides a unique opportunity to investigate
$\eta$ and $\eta^\prime$ physics via two body $J/\psi$ radiative
or hadronic decays. For many $\eta^\prime$ decay channels the
low background data samples are up to three orders of magnitude larger
than collected in any previous experiment.
Here we review the most significant results on $\eta$ and
$\eta^\prime$ obtained at BESIII so far.  The analyses
range from detailed studies of the common decays dynamics, observations of new
radiative and Dalitz decays,  and search for rare/forbidden decays with
sensitivity up to  $\BR\sim 10^{-5}$. Finally,
prospects of the forthcoming runs at $J/\psi$ peak for the $\eta$ and
$\eta^\prime$ physics are discussed.

\end{abstract}



\footnotetext[0]{\hspace*{-3mm}\raisebox{0.3ex}{$\scriptstyle\copyright$}2013
Chinese Physical Society and the Institute of High Energy Physics of
the Chinese Academy of Sciences and the Institute
of Modern Physics of the Chinese Academy of Sciences and IOP Publishing Ltd}%

\begin{multicols}{2}

\section{Introduction}
\label{intro}


More than half a century after $\eta$ \cite{Pevsner:1961pa} and
$\eta^\prime$ \cite{Kalbfleisch:1964zz,Goldberg:1964zza} discoveries,
the mesons still attract attention of both theory and experiment.
As the neutral members of the ground state pseudoscalar nonet, they
play an important role in understanding low energy Quantum
Chromodynamics (QCD).  The main properties of $\eta$ and $\eta^\prime$
mesons are firmly established and their main decay modes are fairly
well known. Decays of the $\eta/\eta^\prime$ probe a wide variety of
physics issues {\it e.g.} $\pi^0-\eta$ mixing, light quark masses
and pion-pion  scattering.
 In particular the $\eta^\prime$ meson, much heavier than the
Goldstone bosons of broken chiral symmetry, plays a special role
as predominantly the singlet state arising from
the strong axial $U(1)$ anomaly. In addition the decays of
both mesons are
used to search for
processes beyond any considered extension of
the Standard Model (SM) and to test fundamental
discrete symmetries.

The main decays of the $\eta/{\eta}^{\prime}$ meson
are hadronic and radiative processes.
Alternatively one can divide the decays
into two following
classes.  The first class consists of hadronic decays into three
pseudoscalar mesons, such as ${\eta}^{\prime}$ ${\to}$
${\eta}{\pi}{\pi}$. Those processes are already included
in the lowest order, ${\cal O}(p^2)$, of chiral perturbation
theory  (ChPT)~\cite{Gasser:1983yg}.
The second class includes anomalous processes involving
odd number of pseudoscalar mesons,
such as ${\eta}^{\prime} \to \rho^0 \gamma$ and ${\eta}^{\prime} \to \pip\pim\pip\pim$.
They are
driven by the Wess-Zumino-Witten (WZW) term \cite{Wess:1971yu,Witten:1983tw}
which enters at  ${\cal O}(p^4)$ order~\cite{Bijnens:1989jb}.
Dynamics of $\eta$ decays remains a subject of extensive
studies aiming at precision tests of ChPT in $SU_L(3)\times SU_R(3)$ sector
({\it i.e.} involving $s$ quark).
Model-dependent approaches for describing low energy meson
interactions, such as Vector Meson Dominance
(VMD)~\cite{Sakurai:1960ju,Landsberg:1986fd},
and the large number of colors, $N_C$, extensions of ChPT~\cite{Kaiser:2000gs}
together with dispersive methods could be extensively
tested in ${\eta}^{\prime}$ decays.

\begin{tablehere}
\begin{center}
 \caption{\label{tab:eta} The available $\eta/\eta^\prime$ decays calculated with the $1.31\times 10^9$ $J/\psi$ events at BESIII.}
 \begin{tabular}{l c c c }\hline\hline
        Decay Mode    &       $\mathcal{B}$ ($\times 10^{-4}$) ~\cite{PDG}   & $\eta/\eta^\prime$ events \\ \hline
      $J/\psi\rightarrow\gamma\eta^\prime$ &$51.5\pm1.6$ &    $6.7\times 10^6$ \\ \hline
          $J/\psi\rightarrow\gamma\eta$ &$11.04\pm0.34$ &$1.4\times 10^6$\\  \hline
       $J/\psi\rightarrow\phi\eta^\prime$ & $7.5\pm0.8$ &    $9.8\times 10^5$ \\ \hline
          $J/\psi\rightarrow\phi\eta$ & $4.5\pm0.5$&   $5.9\times 10^5$ \\  \hline
                  \end{tabular}
                  \end{center}
\end{tablehere}

\begin{table*}[!htbp]
\centering
 \caption{\label{tab:BIIIres} Some of the BESIII results on $\eta'$ branching fractions, $\BR$, based on the sample of $1.31\times 10^{9}$ $J/\psi$ events.
Extracted yields with statistical errors,
 detection efficiency and branching fractions for the studied $\etap$ decay modes,
 where the first error is statistical, the second systematic, and the third from model dependence. The last column gives the status
before BESIII experiment.}
 \begin{tabular}{l c c c c c c }\hline\hline
        Decay Mode    &               Yield       & $\varepsilon$ (\%)&$\mathcal{B}$ ($\times 10^{-4}$)  &Ref.    &Comment \\ \hline
     $\eta'$ $\to$ $\pip\pim\pio$&  6067 $\pm$ 91  &      25.3           &  $35.91$ $\pm$$ 0.54\pm 1.74$& \cite{Ablikim:2016frj}&previously 20 events\\
  ~~~~~~~$(\pip\pim\pio)_S$                &  6580 $\pm$  130 &      26.2           &  $37.63$$\pm$$ 0.77\pm 2.22\pm 4.48$& \cite{Ablikim:2016frj}& first measurement\\
    ~~~~~~~$\rho^\pm\pi^\mp$                &  1231 $\pm$  98 &      24.8           &  $7.44$$\pm$$ 0.60\pm 1.26\pm 1.84$& \cite{Ablikim:2016frj}& first measurement\\

 $\eta'$$\to$$\pio\pio\pio$ &  2015 $\pm$ 47  &      8.8            &  $35.22$$\pm$$ 0.82\pm 2.60$& \cite{Ablikim:2016frj}&previously 235 events\\
$\eta'$$\to$$ e^+e^-\gamma$&  864$\pm$ 36 &      24.5           &  $4.69$$\pm$$ 0.20\pm 0.23$&\cite{Ablikim:2015wnx}& first measurement\\
$\eta'$$\to$$ e^+e^-\omega$&   66$\pm$11 &       5.45          &  $1.97$$\pm$ $ 0.34\pm 0.17$&\cite{Ablikim:2015eos}&  first measurement\\
$\eta'$$\to$$\gamma\omega$&  33187 $\pm$ 351 &      21.9           &  $255.00$$\pm$$ 3.00\pm 16.00$& \cite{Ablikim:2015eos}& \\
$\eta'$$\to$$\gamma\gamma\pi^0$&  655 $\pm$ 68 &      15.9           &  $6.16$$\pm$$ 0.64\pm 0.67$& \cite{Ablikim:2016tuo}&  first measurement \\
$\eta'$$\to$$\pi^+\pi^-\pi^+\pi^-$& 199  $\pm$16  &34.5                 & $0.853$ $\pm$$0.069 \pm 0.069$&  \cite{Ablikim:2014eoc}&first measurement \\
$\eta'$$\to$$\pi^+\pi^-\pi^0\pi^0$&  84 $\pm$16  &7.0                 &  $1.82$$\pm$$0.35 \pm 0.18$&  \cite{Ablikim:2014eoc}&first measurement \\
  \hline\hline
        \end{tabular}
\end{table*}

The BESIII detector~\cite{Ablikim:2009aa}, operating at the Beijing
Electron Positron Collider (BEPCII), is a general purpose facility
designed for $\tau$-charm physics studies in $e^+e^-$ annihilation
with high precision.  Since its commissioning in 2008, a series of
important results have been achieved, including charmonium decays,
light hadron spectroscopy and charm meson decay, with the world's
largest data samples in the $\tau$-charm region.  Due to high
production rate of light mesons in the charmonium, {\it e.g.},
$J/\psi$, decays, the BESIII experiment also offers a unique
possibility to investigate the light meson decays. Radiative decays
$J/\psi\to\eta\gamma$ and
$J/\psi\to\etap\gamma$
provide clean and efficient sources of $\eta/\etap$ mesons
for the decay studies. The accompanying radiative photon,
with energy of 1.5 GeV/$c^2$ and 1.4 GeV/$c^2$ respectively,
is well separated from the decay products.
An alternative source of the $\eta$ ($\eta^\prime$) is the hadronic two-body
process of  $J/\psi\to\phi\eta$
($J/\psi\to\phi\etap$)
where $\phi$ is identified via $\phi\to K^+K^-$ decay
could be used to tag $\eta$($\etap$) decays where not all decay
products are reconstructed.

With two runs in 2009 and in 2012 a total data sample
of $1.31\times 10^{9}$ $J/\psi$ events~\cite{Ablikim:2012cn,Ablikim:2016fal} was collected at the BESIII detector,
the available $\eta$ and $\eta^\prime$ events are summarized in Table.~\ref{tab:eta}
from radiative decays of $J/\psi\to \gamma\eta$, $\gamma\eta^\prime$, and hadronic decays of $J/\psi\rightarrow\phi\eta$, $\phi\eta^\prime$.

 The review presents a recent progress on $\eta/\eta^\prime$ decays
at the BESIII experiment. Unless specifically mentioned the analyses
are based on the full data sample of $1.31\times 10^{9}$ $J/\psi$
events. However, some earlier analyses use data from 2009 run only with
$225.3\times 10^{6}$ $J/\psi$ events.
A summary of some branching fractions measured by BESIII and the
collected data samples using full data set is presented in Table~\ref{tab:BIIIres}.  For
the common three body $\eta$ and $\etap$ processes results on
the decay distributions are reported.  In addition the upper limits
at 90\% confidence level (C.L.) for
rare and forbidden decay modes are presented. Finally the
prospects for the analyses based on the $10^{10}$ $J/\psi$ events to
be collected at BESIII in the near future are discussed.

\section[]{\boldmath$\eta/\eta^\prime$ hadronic decays}
\subsection[] {\boldmath  $\eta \ra \pi^{+}\pi^{-}\pi^0$ and $\eta\ra\pi^0\pi^0\pi^0$ \cite{Ablikim:2015cmz}}
Decays of the $\eta$ meson into 3$\pi$ violate isospin symmetry and were
first considered to be electromagnetic transitions.  However, it turns
out the electromagnetic contribution is strongly suppressed
\cite{Sutherland:1966zz,Bell:1996mi,Baur:1995gc,Ditsche:2008cq}.
Therefore the decays provide a unique opportunity for a precision
determination of $m_u/m_d$ quark mass ratio in a strong process
\cite{Leutwyler:1996qg}.  The challenge for the theory is to provide a
model independent description of the process based on ChPT,
supplemented by general analytic properties of the amplitudes
(dispersive methods).  This approach keeps the promise to finally
resolve long standing discrepancy between the lowest order ChPT prediction for the decay
width of $\eta\rightarrow\pi^+\pi^-\pi^0$ of 66 eV~\cite{Osborn:1970nn} and the experimental value
of $300\pm11$ eV~\cite{PDG}. This would conclude several years of efforts
put by several theory groups to understand the problem, see {\it e.g.}
\cite{
  Gasser:1984pr,Kambor:1995yc,Anisovich:1996tx,Bijnens:2007pr,Schneider:2010hs,Kampf:2011wr,Colangelo:2011zz,Colangelo:2016jmc,Guo:2015zqa}.
However, now there is a need for high statistics
Dalitz plot distributions of $\eta\to\pi^+\pi^-\pi^0$ to test
and/or constrain the theory predictions.

\begin{figure*}[htbp]
     \includegraphics[width=0.35\textwidth]{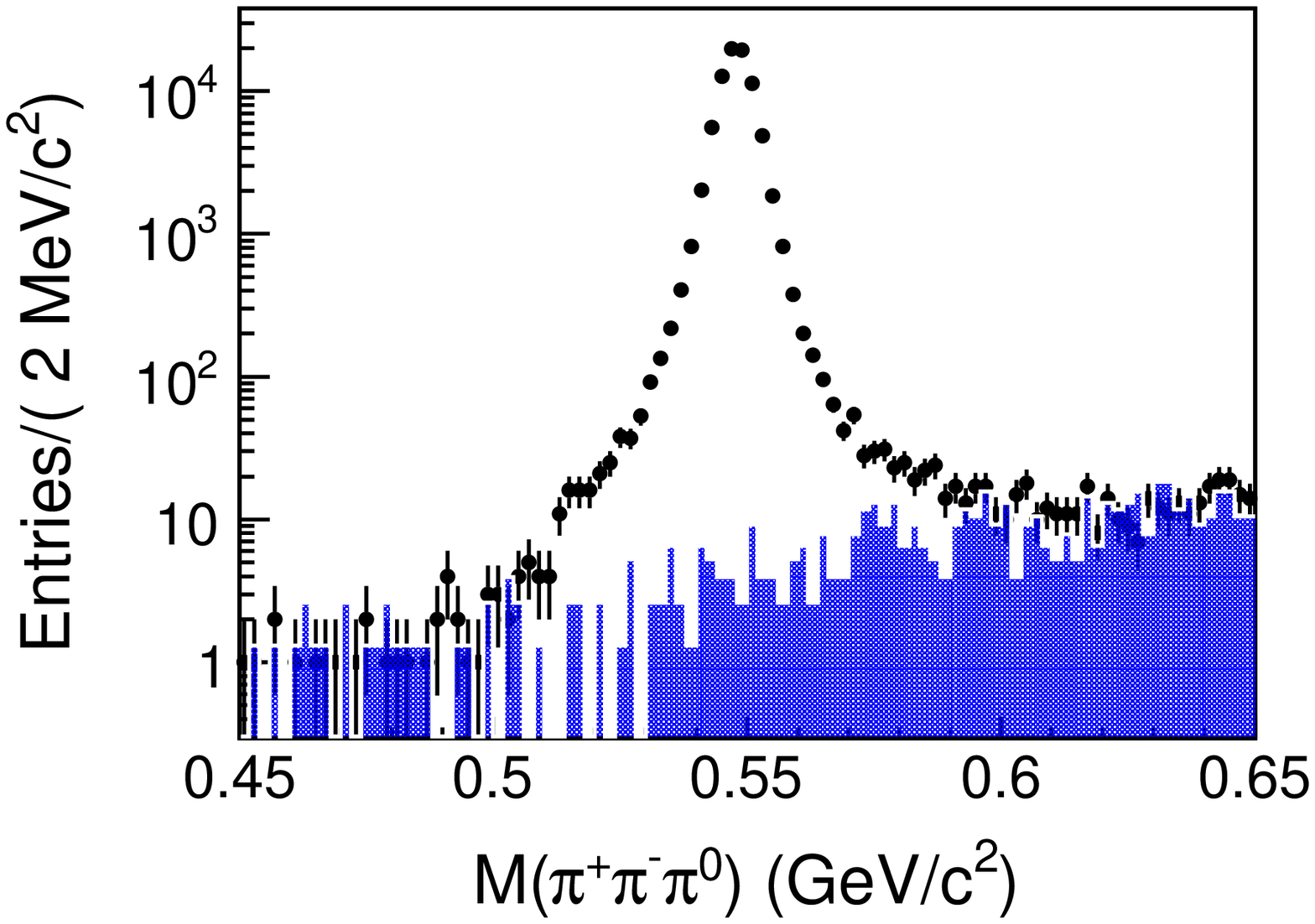}\put(-35,100){\bf
(a)}
\includegraphics[width=0.35\textwidth]{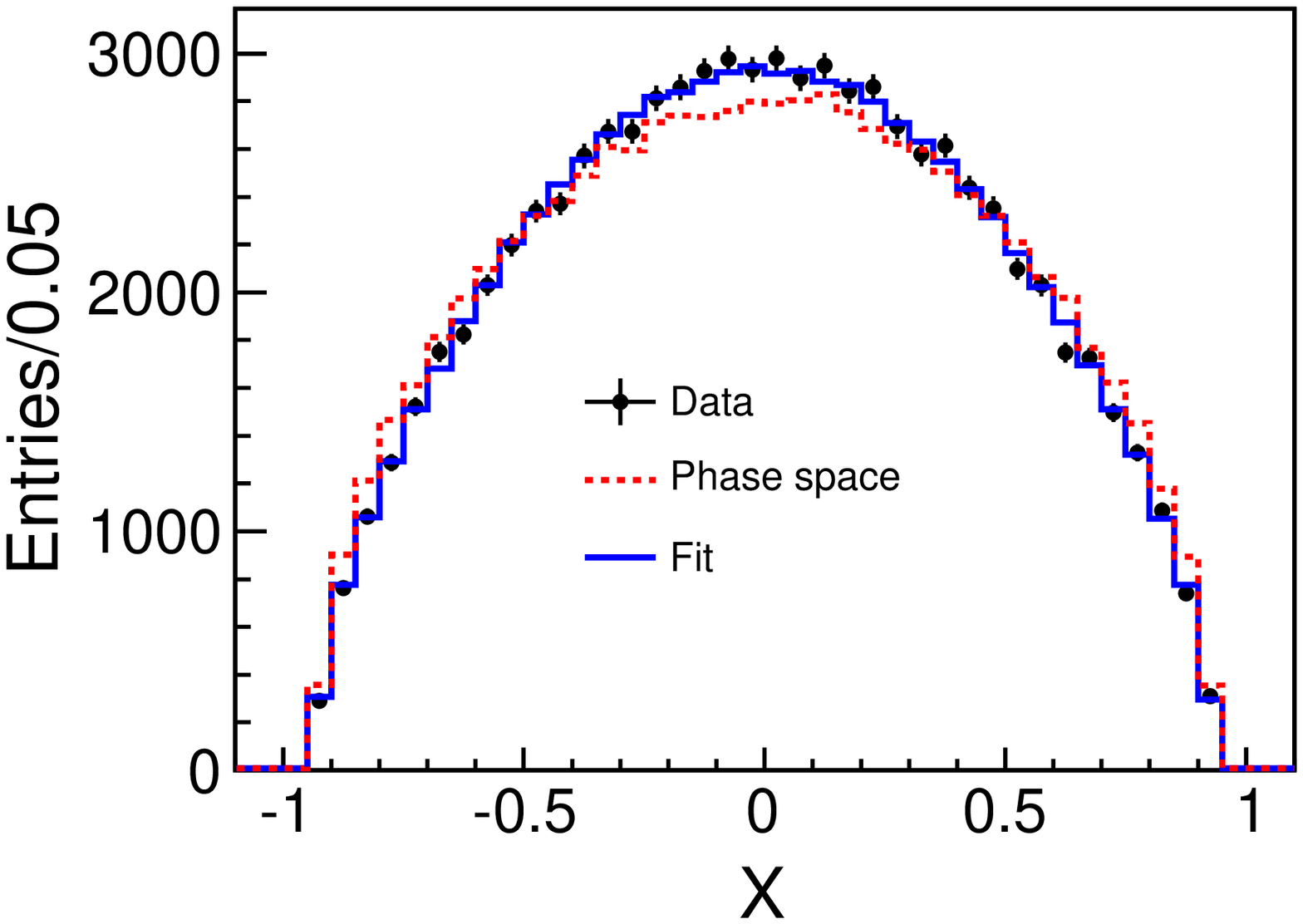}\put(-35,100){\bf
(b)}
\includegraphics[width=0.35\textwidth]{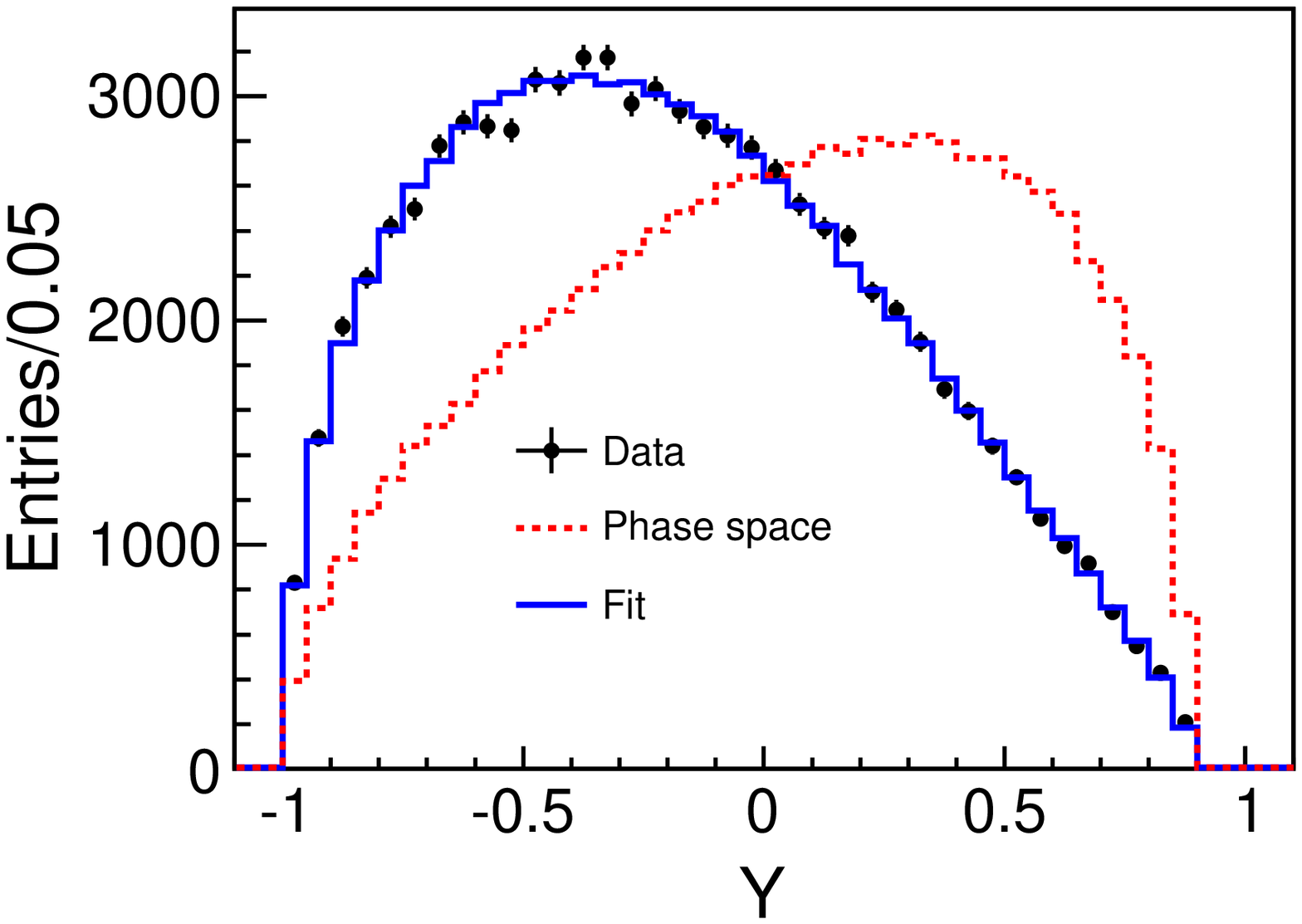}\put(-35,100){\bf
(c)}
    \caption{\label{fig:etacha_dalXY_fit}  (a) Distribution of $\pip\pim\pio$ invariant mass.
Projections of the Dalitz plot as a function of (b) $X$ and (c) $Y$
                        for $\eta\rightarrow \pip\pim\pio$ obtained from data (dots with error bars), the fit projections (solid
                        line) and phase space distributed MC events (dashed line).
                      }
\end{figure*}

With the radiative decay $J/\psi\rightarrow\gamma\eta$,
a clean sample of  $8\times 10^4$ $\eta\rightarrow \pi^+\pi^-\pi^0$ candidate events was selected at BESIII.
Figure~\ref{fig:etacha_dalXY_fit}(a) shows the $\pip\pin\pio$ invariant mass, with the pronounced $\eta$ peak and $\sim0.1$\% background.

The two Dalitz plot variables are
defined as $X={\sqrt{3}}(T_{\pi^+}-T_{\pi^-})/{Q}$ and
$Y= 3{T_{\pi^0}}/{Q}-1$,  where
$T_{\pi}$ denotes the kinetic energy of a pion in the $\eta$ rest
frame and  $Q=m_{\eta}-m_{\pip}-m_{\pim}-m_{\pio}$ is the excess energy of the
reaction. The distributions of $X$ and $Y$ are shown in Figs.~\ref{fig:etacha_dalXY_fit}
(b) and (c).   Using the same parameterization as in
Ref.~\cite{Ambrosino:2008ht},  the decay amplitude squared is expressed as
\begin{equation}\label{eq:etacha_amp}
\begin{gathered}
        |A(X,Y)|^{2} \propto 1 + aY + bY^2 + cX + dX^2 \\
        						     + eXY + fY^{3} + \ldots,
\end{gathered}
\end{equation}
where the coefficients $a, b, c, \ldots$ are the Dalitz plot
parameters.  Terms with odd powers of $X$ ($c$ and $e$
parameters) are included only to test charge conjugation
($C$)
conservation.  An unbinned maximum likelihood fit to data gives the
Dalitz plot parameters shown in Table.~\ref{tab:otherexp} where they
are compared to the results from previous measurements and theoretical
calculations.
The effect of including $c$ and $e$ parameters was
tested in an alternative fit. The $a$, $b$, $d$
and $f$ parameters are almost unchanged, while the parameters $c$ and $e$ are
consistent with zero within one standard deviation.

\begin{table*}[!htbp]
 \begin{center}
 \caption{\label{tab:otherexp} Theoretical and experimental values
for the $\eta\to\pip\pim\pio$ Dalitz plot parameters.}
 \begin{tabular}{lcccc}\hline\hline
  	Theory/Exp. & $a$ 	& $b$  & 	$d$	 &	 $f$	 \\\hline

   	ChPT NLO\cite{Gasser:1984pr}  & $-1.33 $ & $0.42$ & $0.08$ & $0$ \\
   	ChPT NNLO\cite{Bijnens:2007pr}  & $-1.271 \pm 0.075$ & $0.394 \pm 0.102$ & $0.055 \pm 0.057$ & $0.025 \pm 0.160$ \\
   	Dispersive Theory\cite{Kambor:1995yc}  & $-1.16$ & $0.26$ & $0.10$ & $0$ \\
   	Absolute Dispersive\cite{Bijnens:2002qy}  & $-1.21$ & $0.33$ & $0.04$ & $0$ \\
   	UA\cite{Borasoy:2005du}  & $-1.049 \pm 0.025$ & $0.178 \pm 0.019$ & $0.079 \pm 0.028$ & $0.064 \pm 0.012$ \\
   	NREFT\cite{Schneider:2010hs}  & $-1.218 \pm 0.013$ & $0.314 \pm 0.023$ & $0.051 \pm 0.003$ & $0.084 \pm 0.019$ \\
\hline
   	Layter\cite{Layter:1973ti}  &$ -1.08\pm0.014 $&$ 0.03\pm0.03 $&$ 0.05\pm0.03 $&$ - $\\
   	CBarrel\cite{Abele:1998yj}  &$ -1.22\pm0.07 $&$ 0.22\pm0.11 $&$ 0.06(fixed) $&$ - $\\
   	KLOE08\cite{Ambrosino:2008ht}  &$ -1.09^{+0.013}_{- 0.024} $&$ 0.124\pm0.016 $&$ 0.057^{+0.016}_{-0.022} $&$ 0.14\pm0.03$\\
   	WASA-at-COSY\cite{Adlarson:2014aks}  &$ -1.144\pm 0.018$&$ 0.219\pm0.019\pm0.047 $&$ 0.086\pm0.018\pm0.015 $&$ 0.115\pm0.037$\\
    BESIII\cite{Ablikim:2015cmz}  &$ -1.128\pm0.015\pm0.008 $&$ 0.153\pm0.017\pm0.004 $&$ 0.085\pm0.016\pm0.009 $&$ 0.173\pm0.028\pm0.021$\\
    KLOE16\cite{Anastasi:2016cdz} & $-1.104\pm0.003\pm0.002$ & $0.142\pm0.003^{+0.005}_{-0.004}$ & $0.073\pm0.003^{+0.004}_{-0.003}$ & $0.154\pm0.006^{+0.004}_{-0.005}$ \\
\hline\hline
\end{tabular}
\end{center}
\end{table*}

For $\eta\rightarrow\pio\pio\pio$, the amplitude squared is nearly
constant and the deviation can be parameterized in the lowest order
using just one variable
$Z=\frac{2}{3}\sum^{3}_{i=1}({3T_{i}}/{Q}-1)^{2}$, where $Q =
m_{\eta}-3m_{\pio}$ and  $T_{i}$ denotes the kinetic energy of each
$\pio$ in the $\eta$ rest frame.  The Dalitz plot density
distribution could be parameterized using a linear term,
\begin{equation}\label{eq:etaneu_amp}
    |A(Z)|^2\propto 1 + 2\alpha Z + \ldots,
\end{equation}
where $\alpha$ is the slope parameter.

The $\pio\pio\pio$ mass spectrum is shown in Fig.~\ref{fig:M3pieta}(a), with the $\eta$ peak and the background estimated to be less than 1\%. The  distribution of the variable $Z$
is displayed in Fig.~\ref{fig:M3pieta}(b). Due to the kinematic boundaries
and the cusp at the $\pio\pio\to\pip\pim$ threshold \cite{Gullstrom:2008sy,Schneider:2010hs}, only the interval of $0<Z<0.7$
 is used to extract the slope parameter $\alpha$ from the data.
In analogy to the  $\eta\ra\pip\pim\pio$ measurement
an unbinned maximum likelihood fit, as displayed in the inset of
Fig.~\ref{fig:M3pieta}(b), yields the Dalitz plot slope parameter $\alpha = -0.055 \pm
0.014\pm 0.004$, which is compatible with the recent results from other experiments.

\begin{figurehere}
\begin{center}
    \includegraphics[width=8.3cm, height=7cm]{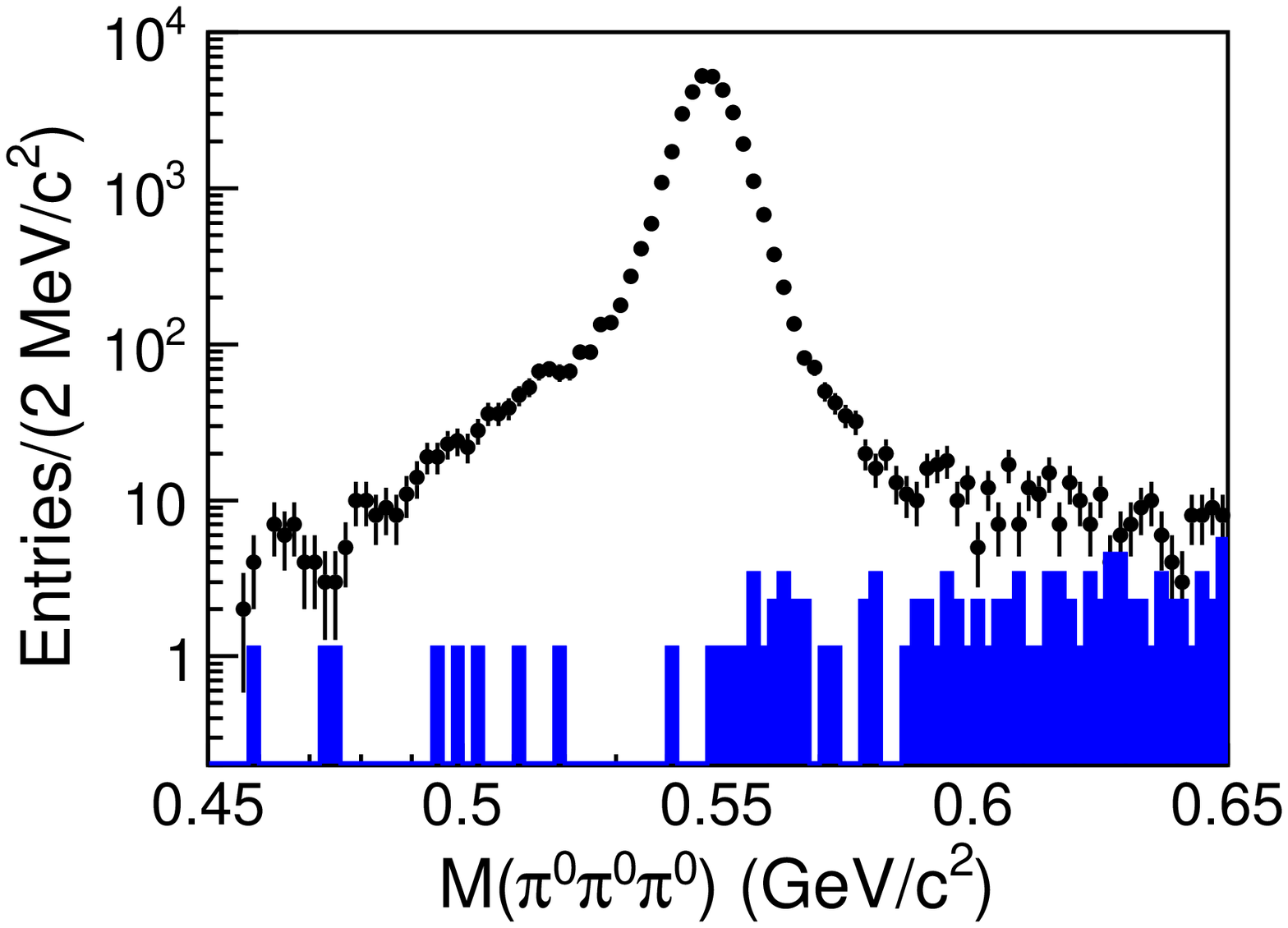}\put(-40,160){\bf (a)}

    \includegraphics[width=8.3cm, height=7.cm]{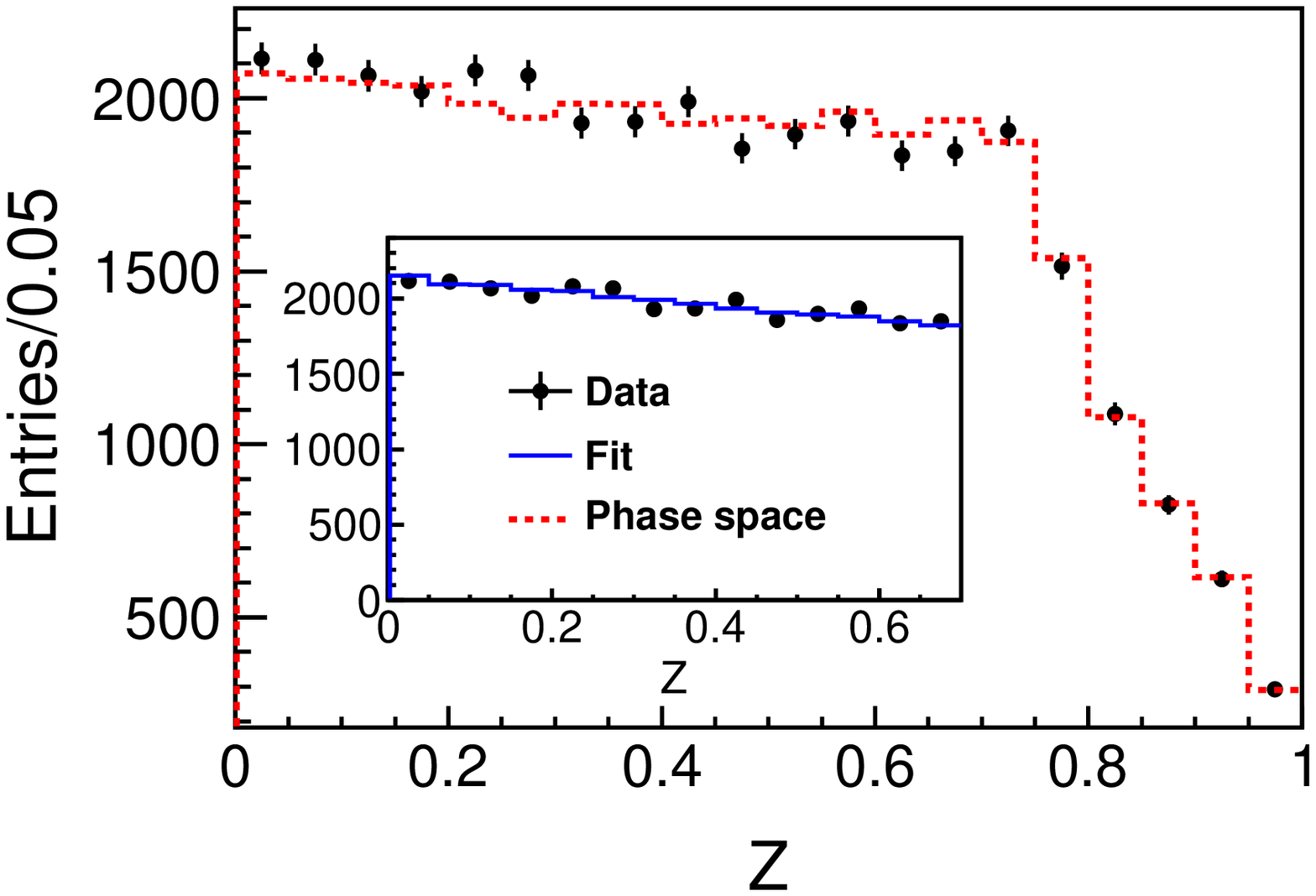}\put(-40,160){\bf
(b)}

    \caption{\label{fig:M3pieta} (a) Distribution of $M(\pio\pio\pio)$
in the $\eta$ mass region.
                (b) Distribution of the variable $Z$ for
$\eta\ra\pio\pio\pio$. Dots with error bars are for data,
                histograms for background contributions, dashed histograms
for phase space distributed MC events and the solid lines in the inset are the
results of the fit.}
\end{center}
\end{figurehere}

\begin{tablehere}
\begin{center}
\begin{small}
\caption{\label{tab:expresults_etanue}Theoretical and experimental values
for  $\eta\to\pio\pio\pio$ Dalitz plot slope parameter $\alpha$.}
 \begin{tabular}{lc}\hline\hline
  	Theory/Exp. & $\alpha$ \\\hline
  	ChPT/NLO\cite{Gasser:1984pr} & 0.015\\
  	dispersive\cite{Kambor:1995yc} & (-0.014)-(-0.007)\\	
  	UA\cite{Borasoy:2005du} & $-0.031 \pm 0.003$\\
  	ChPT/NNLO\cite{Bijnens:2007pr} & $0.013 \pm 0.032$\\
        \hline
   	KLOE\cite{Ambrosinod:2010mj}  & $-0.0301 \pm 0.0035^{+0.0022}_{-0.0035}$\\
   	WASA-at-COSY\cite{Adolph:2008vn}  & $-0.027 \pm 0.008 \pm 0.005$\\
   	CBall\cite{Prakhov:2008ff}  & $-0.0322 \pm 0.0012 \pm 0.0022$\\
   	SND\cite{Achasov:2001xi}  & $-0.010 \pm 0.021 \pm 0.010 $\\
   	CBarrel\cite{Abele:1998yi}  & $-0.052 \pm 0.017 \pm 0.010$\\
   	GAM2\cite{Alde:1984wj}  & $-0.022 \pm 0.023$\\
   	BESIII\cite{Ablikim:2015cmz}  & $-0.055 \pm 0.014 \pm 0.004$\\
\hline\hline
\end{tabular}
\end{small}
\end{center}
\end{tablehere}

\subsection[]{\boldmath $\etap\to\pi^+\pi^-\eta$~\cite{Ablikim:2010kp} and
$\etap\to\pi^{+(0)}\pi^{-(0)}\eta$ ~\cite{Ablikim:2017kp}}

The combined branching fraction of the two main hadronic decays of
$\etap$: $\etap\to\pi^{+}\pi^{-}\eta$ and $\etap\to\pi^{0}\pi^{0}\eta$
is nearly $2/3$.  The ratio
$\BR(\etap\to\pi^+\pi^-\eta)/\BR(\etap\to\pi^0\pi^0\eta)$ should be
exactly two in the isospin limit.  The decays involve both $\eta$ and
pions in the final state and therefore allows to extract information
about $\pi\eta$ interactions. However, the excess energy of the
processes is relatively small: 130 MeV and 140 MeV for $\pi^+\pi^-\eta$
and $\pi^0\pi^0\eta$ respectively. This means precision high statistics
experimental studies of the Dalitz plots together with an
appropriate theory
framework for extraction of the $\pi\eta$ phase shifts are needed.

 The two Dalitz plot variables, $X$ and $Y$, are usually defined as
 $X=\frac{\sqrt{3}}{Q}(T_{\pi^+}-T_{\pi^-})$ and $Y=\frac{m_{\eta}+2m_{\pi}}{m_{\pi}}
\frac{T_{\eta}}{Q}-1$,  where $T_{\pi,\eta}$  denote the kinetic energies of the
mesons in the  $\eta^\prime$ rest frame and
$Q= T_\eta+T_{\pi^+} +T_{\pi^-}$ = $m_{\eta^\prime}-m_{\eta}-2m_{\pi}$.

Two different parametrizations of the Dalitz plot distribution are used.
The historically first one assumes a linear amplitude in $Y$ variable:
 \begin{equation}\label{eq:etapcha_ampl}
\begin{gathered}
        |A(X,Y)|^{2} \propto |1 + \alpha Y|^2+ cX + dX^2,\\
\end{gathered}
\end{equation}
the other representation is just a general polynomial expansion:
 \begin{equation}\label{eq:etapcha_amp}
\begin{gathered}
        |A(X,Y)|^{2} \propto  1 + aY + b Y^2+ cX + dX^2,\\
\end{gathered}
\end{equation}
where, $\alpha$ is complex and $a$, $b$, $c$, $d$ are real parameters.
These two representations are equivalent in case of  $b>a^2/4$.

\begin{figure*}[htbp]
\begin{center}
\includegraphics[height=5cm, width=5.2cm]{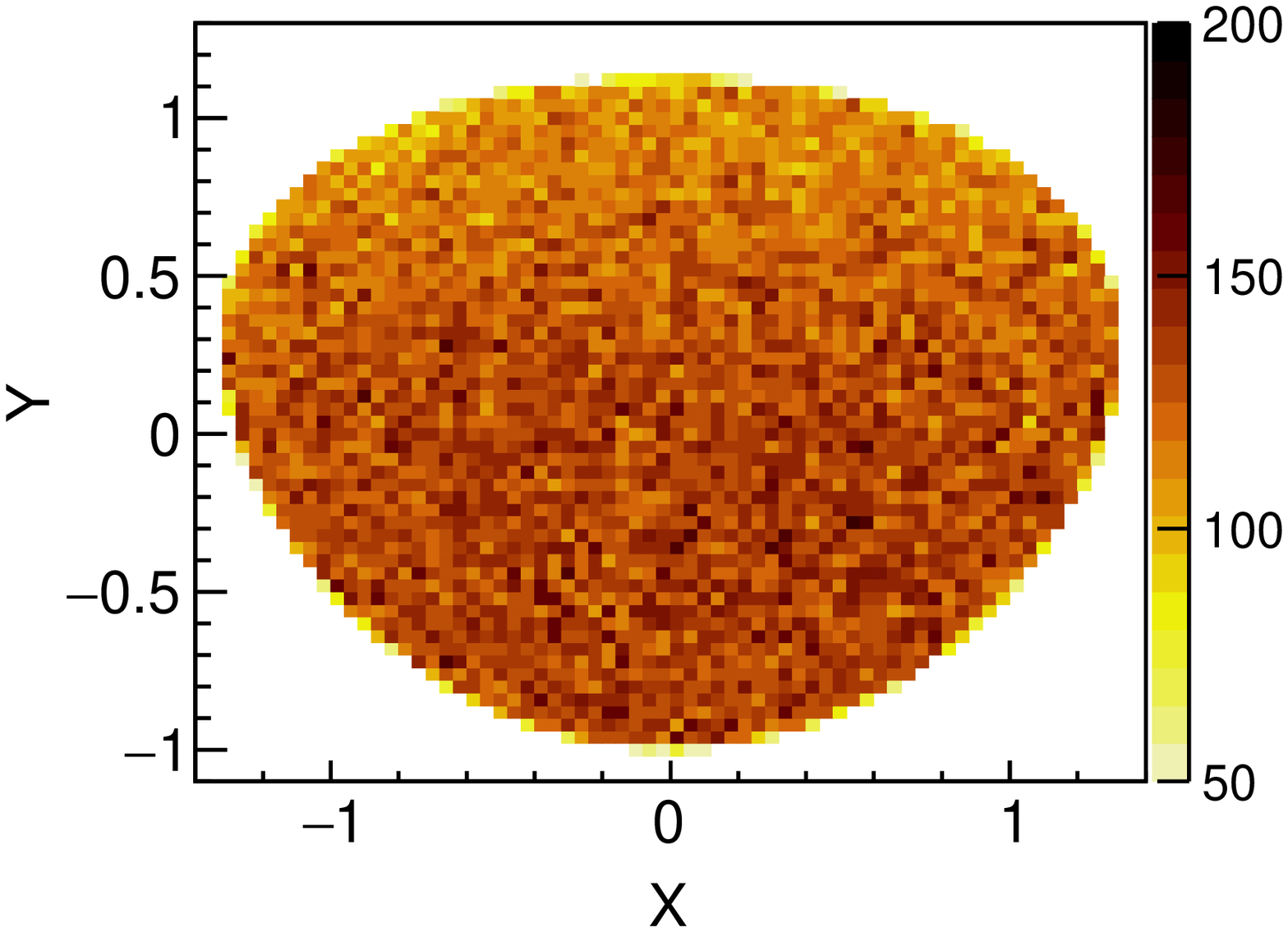}
\put(-35,115){(a)}
\includegraphics[width=5.2cm]{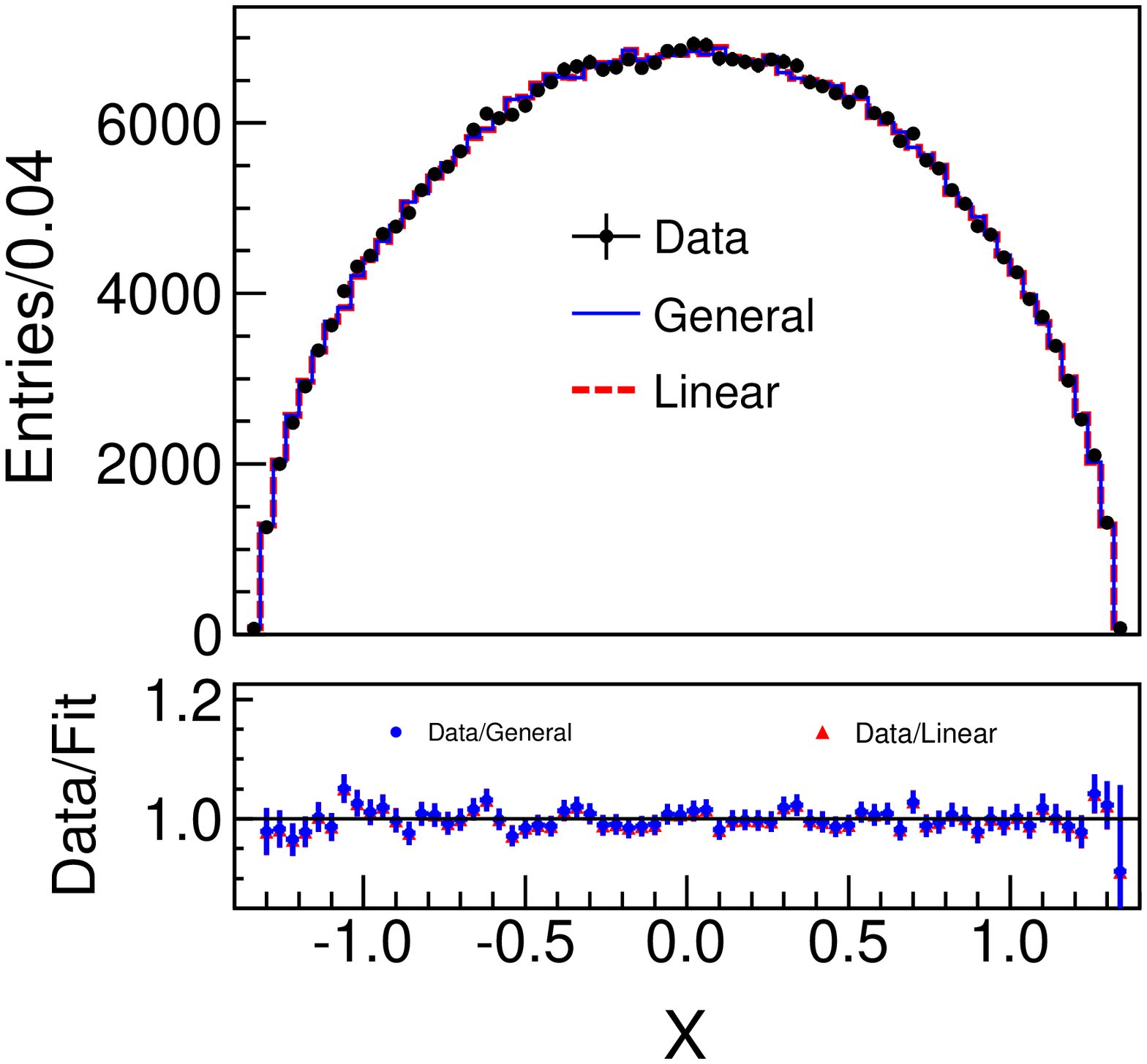}
\put(-35,115){(b)}
\includegraphics[width=5.2cm]{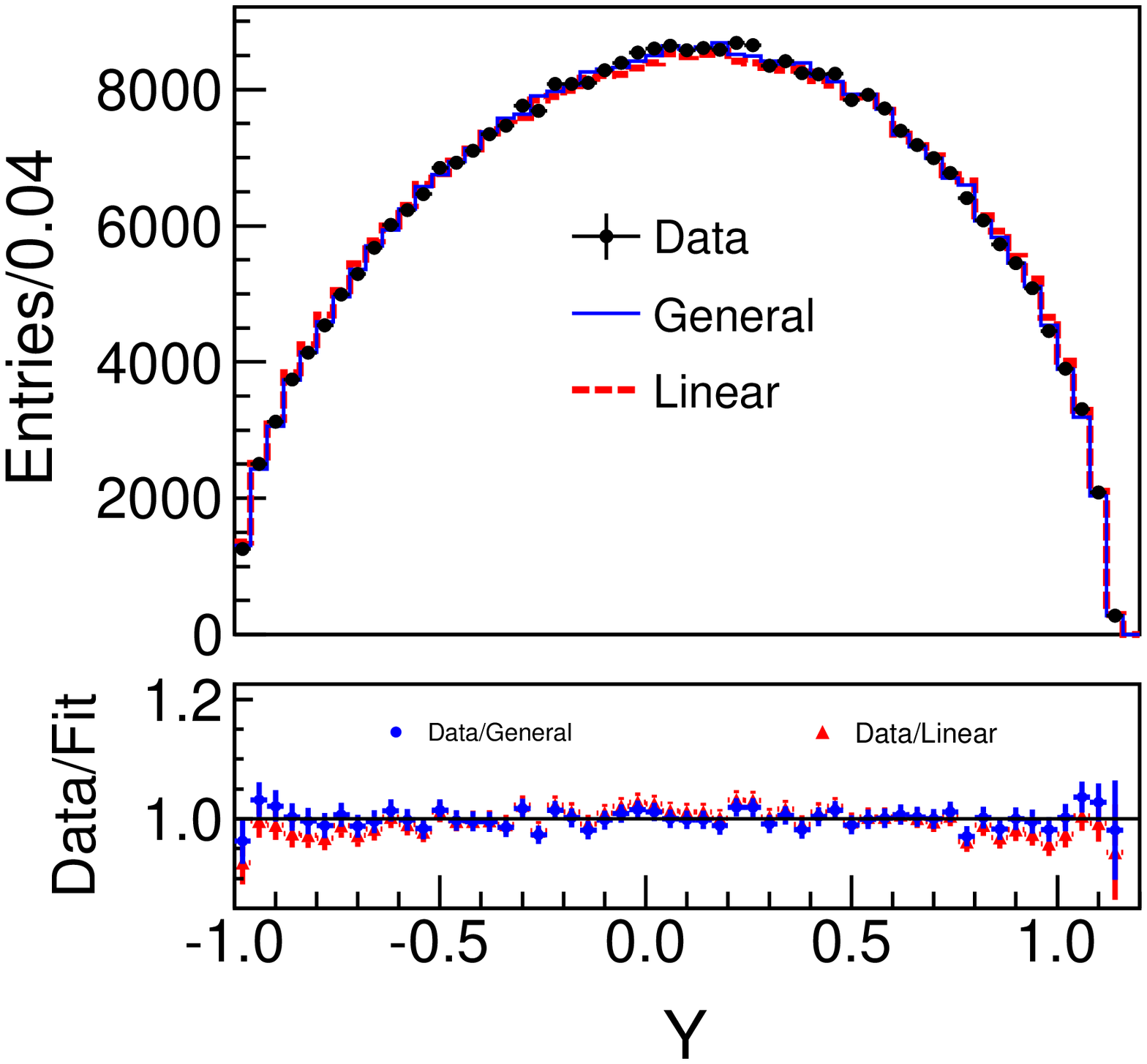}
\put(-35,115){(c)}
\caption{(a) The
experimental Dalitz plot for the decay $\eta^\prime \rightarrow \pi^+\pi^-\eta$ in
terms of the variables $X$ and $Y$ with the $\pi^+\pi^-\eta$ mass in the
$\eta^\prime$ mass region. The corresponding projections on variables $X$
and $Y$ are shown in (b) and (c), respectively, where the dashed
histograms are from MC of $\eta^\prime \rightarrow \pi^+\pi^-\eta$ events
generated with phase space. The solid histograms are the fit
results described in the text.} \label{xy}
\end{center}
\end{figure*}

\begin{figure*}[htbp]
\begin{center}
\includegraphics[height=5cm, width=5.2cm]{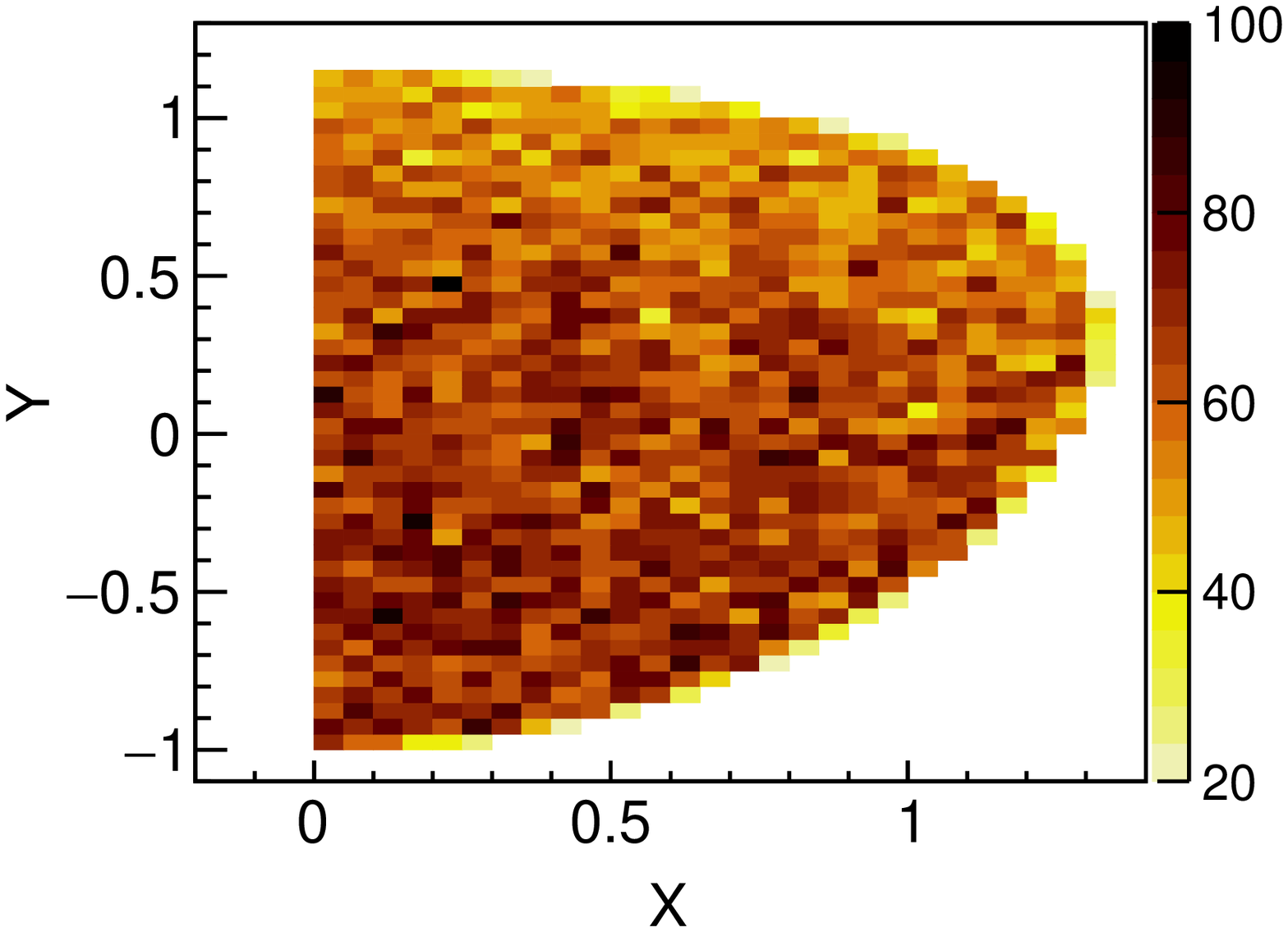}
\put(-35,115){(a)}
\includegraphics[width=5.2cm]{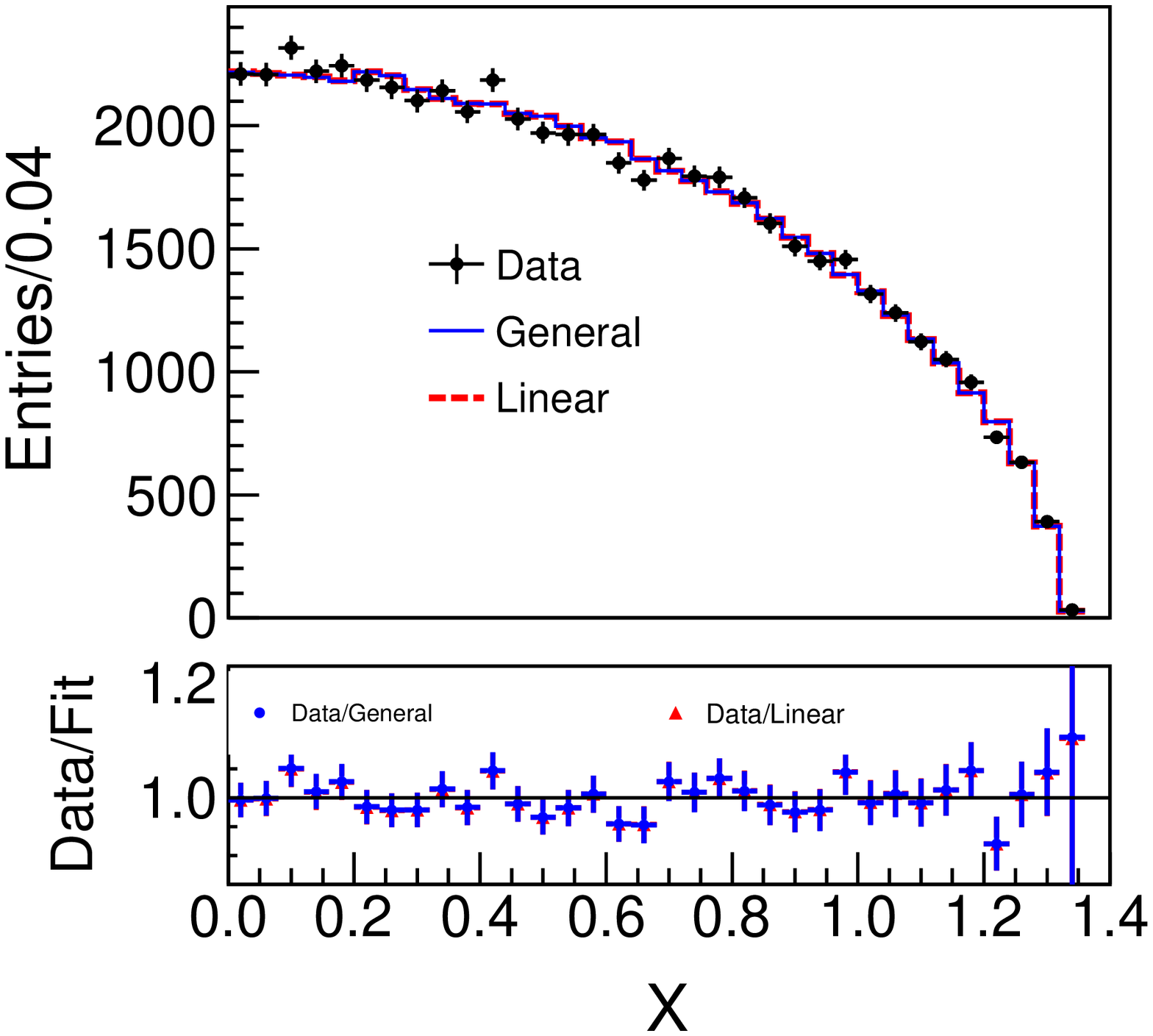}
\put(-35,115){(b)}
\includegraphics[width=5.2cm]{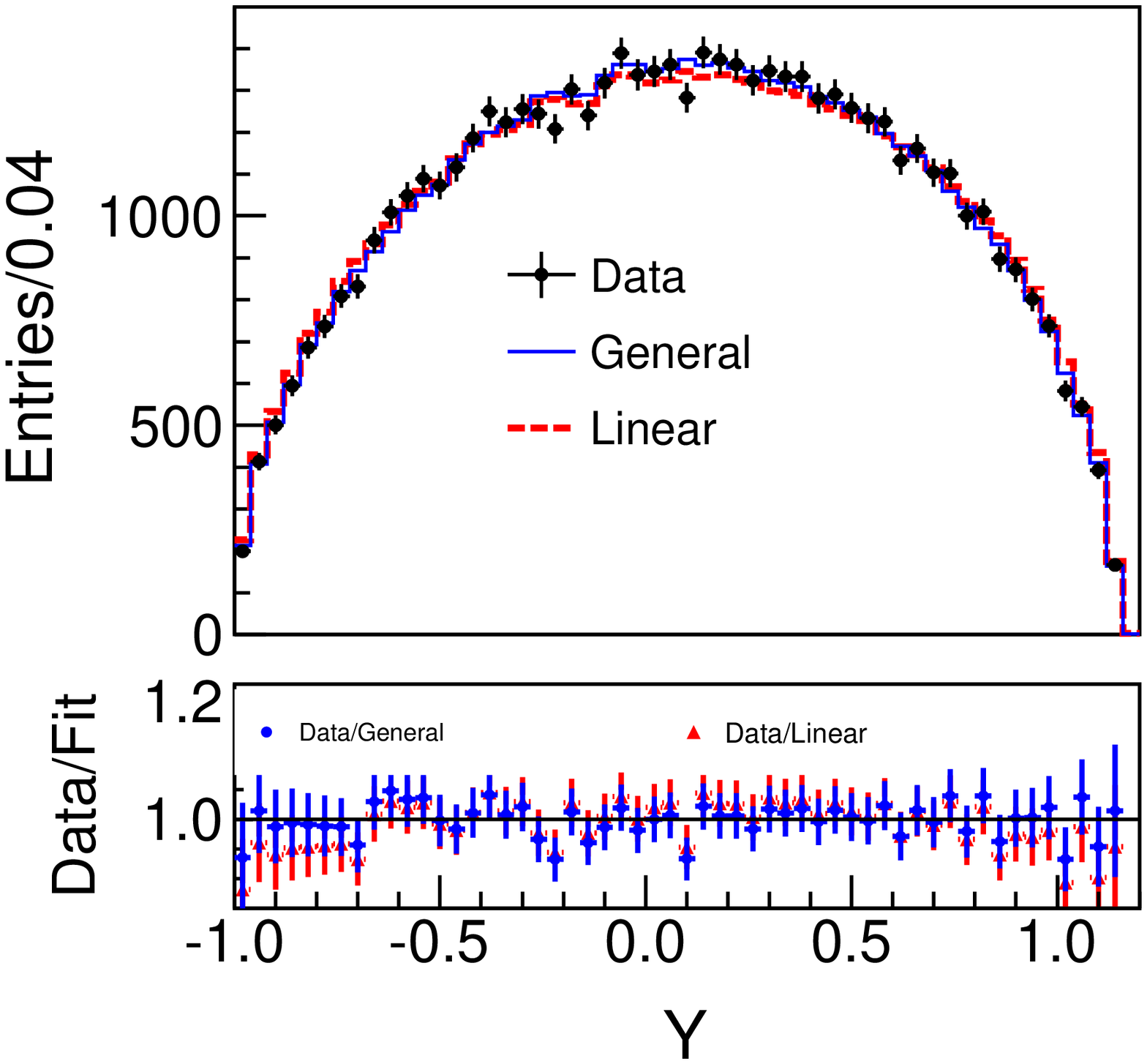}
\put(-35,115){(c)}
\caption{(a) The
experimental Dalitz plot for the decay $\eta^\prime \rightarrow \pi^0\pi^0\eta$ in
terms of the variables $X$ and $Y$ with the $\pi^+\pi^-\eta$ mass in the
$\eta^\prime$ mass region. The corresponding projections on variables $X$
and $Y$ are shown in (b) and (c), respectively, where the dashed
histograms are from MC of $\eta^\prime \rightarrow \pi^0\pi^0\eta$ events
generated with phase space. The solid histograms are the fit
results described in the text.} \label{xyetap}
\end{center}
\end{figure*}

Initial  BESIII  $\etap\rightarrow \pi^+\pi^-\eta$
Dalitz plot analysis~\cite{Ablikim:2010kp} was based on  2009 data and
the above two
representations was used.  The extracted parameters are generally consistent
with the previous measurements and theoretical predictions.
The negative value of $b$ parameter indicates, with an uncertainty of
30\%, that the two
representations may not be equivalent.  The most recent BESIII analysis~\cite{Ablikim:2017kp} uses
nearly background free samples of $3.5\times 10^5$
$\eta^\prime\rightarrow\eta\pi^+\pi^-$ events and $5.6\times 10^4$
$\eta^\prime\rightarrow\eta\pi^0\pi^0$ events from $1.31\times 10^9$
$J/\psi$.  The goal was determination of the Dalitz plot parameters for the two
decay modes and a search for the cusp at the $\pio\pio\to\pip\pim$
threshold in $\eta^\prime\rightarrow\eta\pi^0\pi^0$.
And the fit results for the above two representations, which are shown in Fig.~\ref{xy} and Fig.~\ref{xyetap}, and the  corresponding fitted  parameters are summarized in Table~\ref{fitres}.

\begin{table*}
\centering
 \caption{\label{fitres} Experimental and theoretical values of the Dalitz plot
 parameters for $\eta^\prime\rightarrow\eta\pi^+\pi^-$ and $\eta^\prime\rightarrow\eta\pi^0\pi^0$.
 The values for parameter $c$ and $\Im(\alpha)$ are given only for comparison with previous experiments.}
 \resizebox{\textwidth}{!}{%
 \begin{tabular}{c|cccc|ccc}\hline\hline
  \multirow{2}{*}{Para.}& \multicolumn{4}{|c|}{$\eta^\prime\rightarrow\eta\pi^+\pi^-$ } & \multicolumn{3}{|c}{$\eta^\prime\rightarrow\eta\pi^0\pi^0$} \\
   & ChUA~\cite{Borasoy:2005du} & Large N$_C$~\cite{Escribano:2010wt} & VES~\cite{Dorofeev:2006fb} & BESIII~\cite{Ablikim:2017kp} &
	 ChUA~\cite{Borasoy:2005du} & GAMS-4$\pi$~\cite{Blik:2008zz} & BESIII~\cite{Ablikim:2017kp} \\hline

  $a$ & $-0.116\pm0.011$ & $-0.098\pm0.048$ & $-0.127\pm0.018$ &$-0.056\pm0.004\pm0.003$ & $-0.127\pm0.009$ & $-0.067\pm0.016$ & $-0.087\pm0.009\pm0.006$ \\
  $b$ & $-0.042\pm0.034$ & $-0.050\pm0.001$ & $-0.106\pm0.032$ &$-0.049\pm0.006\pm0.006$ & $-0.049\pm0.036$ & $-0.064\pm0.029$ & $-0.073\pm0.014\pm0.005$ \\
  $c$ & $...$ & $...$    & $+0.015\pm0.018$ &  $(2.7\pm2.4\pm1.8)\times10^{-3}$ & $...$ & $...$ & $...$ \\
  $d$ & $+0.010\pm0.019$ & $-0.092\pm0.008$ & $-0.082\pm0.019$ & $-0.063\pm0.004\pm0.004$ & $+0.011\pm0.021$ & $-0.067\pm0.020$ & $-0.074\pm0.009\pm0.004$ \\\hline
  $\Re(\alpha)$ & $...$ & $...$ & $-0.072\pm0.014$ & $-0.034\pm0.002\pm0.002$ & $...$ & $-0.042\pm0.008$ & $-0.054\pm0.004\pm0.001$ \\
  $\Im(\alpha)$ & $...$ & $...$ & $~0.000\pm0.100$ & $~0.000\pm0.019\pm0.001$ & $...$ & $~0.000\pm0.070$ & $~0.000\pm0.038\pm0.002$ \\
  $c$           & $...$ & $...$ & $+0.020\pm0.019$ & $(2.7\pm2.4\pm1.5)\times10^{-3}$ & $...$ &  $...$  &  $...$  \\
  $d$           & $...$ & $...$ & $-0.066\pm0.034$ & $-0.053\pm0.004\pm0.004$ & $...$ & $-0.054\pm0.019$ & $-0.061\pm0.009\pm0.006$ \\
  \hline\hline

	\end{tabular}
	}%
\end{table*}

For the  $\eta^\prime\rightarrow\eta\pi^+\pi^-$ decay,  the  results,  superseding  the previous BESIII measurement~\cite{Ablikim:2010kp},   are not consistent
with the measurement from VES and the theoretical predictions within the framework of $U(3)$ chiral effective
field theory in combination with a relativistic coupled-channels method
(Chiral Unitary Approach -- ChUA)~\cite{Borasoy:2005du}. In particular for the cofficient $a$,  the discrepancies are about four standard deviations.
On the other hand large-$N_C$ ChPT prediction
at next-to-next-to-leading order~\cite{Escribano:2010wt} is consistent with the measured $a$ value due
to the large theoretical uncertainty. For the cofficient $c$ violating charge conjugation, the fitted values
are consistent with zero within one standard deviation for both representations.

In case of  $\etap\ra\eta\pio\pio$, the results are in general consistent with the previous measurements and theoretical
predictions within the uncertainties from
both sides. The latest results~\cite{Adlarson:2017} reported by the A2 experiment are also in agreement with those obtained from BESIII.
 We notice  a discrepancy of 2.6 standard deviations  for parameter $a$
between $\etap\ra\eta\pip\pim$ and $\etap\ra\eta\pio\pio$ modes.
The present results are not precise enough to firmly establish
isospin violation and
additional effects, {\it e.g.}, radiative corrections~\cite{Kubis:2009sb} and $\pi^+/\pi^0$
mass difference
should be considered in the
future experimental and theoretical studies.

It was also found that the linear representation could not
describe data.  The discrepancies  between data and the fit are evident in the $Y$ projection
for the both decay modes, which is yet another
indication that  the linear and general representations are not equivalent.
In addition,  a search for the cusp in
$\etap\ra\eta\pi^0\pi^0$ performed by inspecting  the $\pi^0\pi^0$ mass spectrum
close to $\pip\pim$ mass threshold, reveals no
statistically significant effect.
Most recent theoretical
dispersive analysis of the cusp
in the $\etap\ra\eta\pi^0\pi^0$ \cite{Isken:2017dkw} uses Dalitz plot parameters from
VES and 2009 BESIII\cite{Ablikim:2010kp} $\etap\ra\eta\pip\pim$ data.
However, the amplitudes from Ref.~\cite{Isken:2017dkw}   should be preferably fitted
directly to the Dalitz plot data for the two decay modes.

\subsection[] {\boldmath $\etap\ra\pi^0\pi^0\pi^0$~\cite{Ablikim:2015cmz}}

The isospin violating decay of $\eta^\prime\ra\pi^0\pi^0\pi^0$ was first observed in $\pi^- p \rightarrow \eta n$~\cite{Binon:1984fe}. In a later
experiment the Dalitz plot slope parameter was
extracted to be   $\alpha = -0.1\pm0.3$  with limited statistics of around 60 events~\cite{Alde:1987jt}, and in 2008 GAMS-4$\pi$ analysis updated to
$\alpha = - 0.59 \pm 0.18$ using $235\pm 45$ events~\cite{Blik:2008zz}.
Using the 2009 $J/\psi$ data sample BESIII has reported the
branching fraction  of the decay which is about two times larger
than Ref.~\cite{Binon:1984fe,Alde:1987jt,Blik:2008zz} (average
$\BR=(1.77\pm 0.23)\times10^{-3}$ for the three experiments).
However, in this first BESIII analysis the Dalitz plot
slope parameter was not reported~\cite{BESIII:2012aa}.

With the full $J/\psi$ data set a determination of
the Dalitz plot slope was possible~\cite{Ablikim:2015cmz}.
The $\eta^\prime$ signal is clearly observed in $\pi^0\pi^0\pi^0$ mass spectrum,
Fig.~\ref{fig:M3pi}(a), where the hatched and shaded histograms show the background contributions from the inclusive $J/\psi$ decays and $\eta^\prime\ar\pi^0\pi^0\eta$, respectively.
As shown in Fig.~\ref{fig:M3pi}(b), a maximum-likelihood fit to the  events
with $Z$ in the  region of $0.2<Z<0.6$ gives
the Dalitz plot slope parameter:
$\alpha = -0.640 \pm 0.046 \pm 0.047$,
much more precise than previous measurements as summarized in Table.~\ref{tab:expresults_etapnue}.
The value deviates significantly from zero, which  implies that final state
interactions play an important role.
Up to now, there are only few theory predictions to compare the parameter
value.
One exception are the ChUA calculations of Ref.~\cite{Borasoy:2005du}.
The predicted value of the $\alpha$ coefficient
is in the  $-2.7$ to
$0.1$ range, consistent with the BESIII measurement.
\begin{tablehere}
\begin{center}
\caption{\label{tab:expresults_etapnue}Theoretical and experimental values
for $\eta'\to\pio\pio\pio$ Dalitz plot slope parameter $\alpha$.}
\begin{tabular}{lc}\hline\hline
  	Theory/Exp. & $\alpha$ \\\hline
  	ChUA\cite{Borasoy:2005du} &  $-2.7\sim 0.1$\\
   	GAMS\cite{Blik:2008zz} & $-0.59 \pm 0.18$\\
   	GAM2\cite{Alde:1987jt} & $-0.1 \pm 0.3$ \\
   	BESIII\cite{Ablikim:2015cmz}  & $-0.640 \pm 0.046 \pm 0.047$\\
\hline\hline
\end{tabular}
\end{center}
\end{tablehere}

\begin{figurehere}
\begin{center}
    \includegraphics[width=6.3cm, height=5cm]{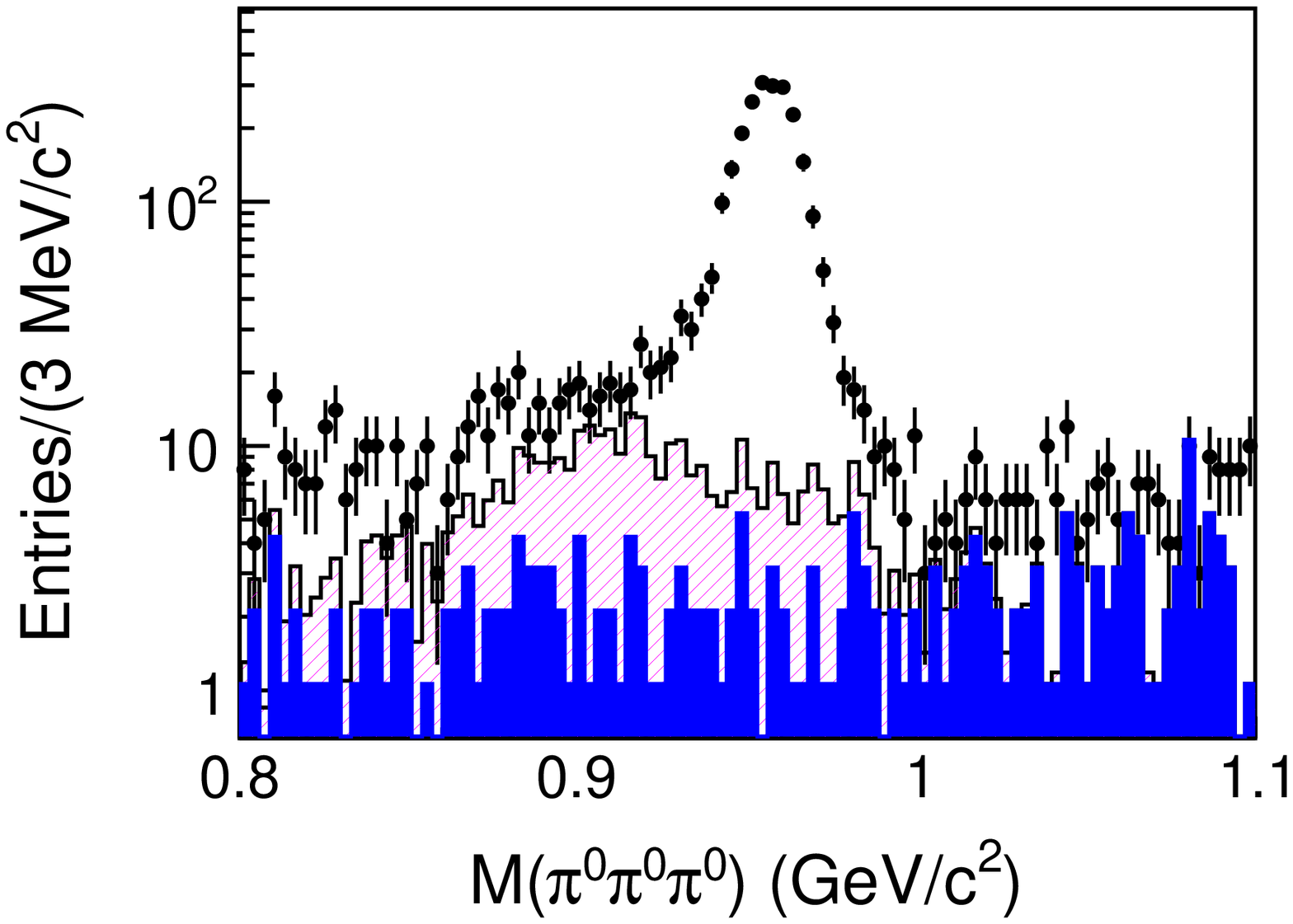}\put(-40,120){\bf
(a)}

    \includegraphics[width=6.3cm,
height=5cm]{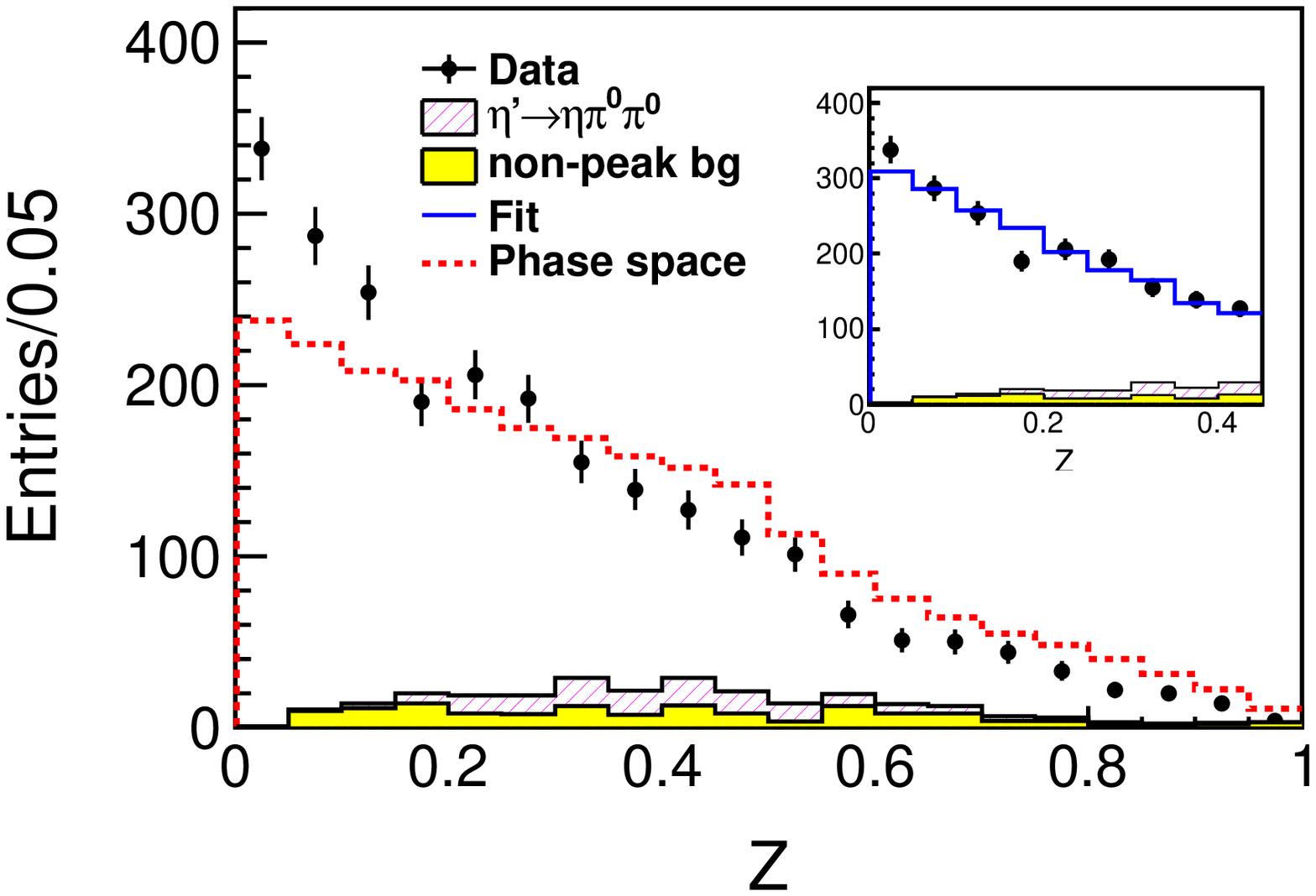}\put(-40,120){\bf (b)}

    \caption{\label{fig:M3pi}
 (a) Distribution of $M(\pio\pio\pio)$
in the $\etap$ mass region.
                (b) Distribution of the variable $Z$ for
$\etap\ra\pio\pio\pio$. Dots with error bars are for data,
                histograms for background contributions, dashed histograms
for phase space distributed MC events and the solid lines in the inset are the
results of the fit.}
\end{center}
\end{figurehere}

\subsection[]{\boldmath   Amplitude analysis of $\etap \ra \pi^{+(0)}\pi^{-(0)}\pi^0$ \cite{Ablikim:2016frj}}
At first, the low intensity process $\eta^\prime\rightarrow \pi^+\pi^-\pi^0$ may be considered
to come from $\pi^0-\eta$ mixing  in the
dominating decay $\eta^\prime\rightarrow\pi^+\pi^-\eta$ \cite{Gross:1979ur}. This
would offer a possibility to determine precisely  $u-d$ quark mass difference
from the branching fraction ratio of the two processes.
However, a recent analysis
shows that even at tree level other terms are needed \cite{Borasoy:2006uv}.
In addition the decay amplitudes are strongly
affected by the intermediate resonances.
Therefore the mixing of $\pi^0-\eta$ and $u-d$ quark mass difference cannot be extracted in a simple way.

The decay $\etap\to \pi^+\pi^-\pi^0$ was first observed  by the
CLEO experiment~\cite{Naik:2008aa} in 2008. BESIII has reported the branching
fraction measurement using 2009 $J/\psi$ data set \cite{BESIII:2012aa}
but the amplitude analysis was only possible with the full data
set. In particular it was expected that contribution of
$\eta'\to\rho^\pm\pi^\mp$ could be identified.  This expectation is
supported by the experimental distributions shown in the Dalitz plot
of $M^2(\pi^+\pi^-)$ versus $M^2(\pi^-\pi^0)$ in Fig.~\ref{m4pidalitz},
where the two clusters corresponding to the contribution
$\eta^\prime\rightarrow\rho^\pm\pi^\mp$ are seen.

\begin{figure*}
\includegraphics[height=2.in,width=2.2 in]{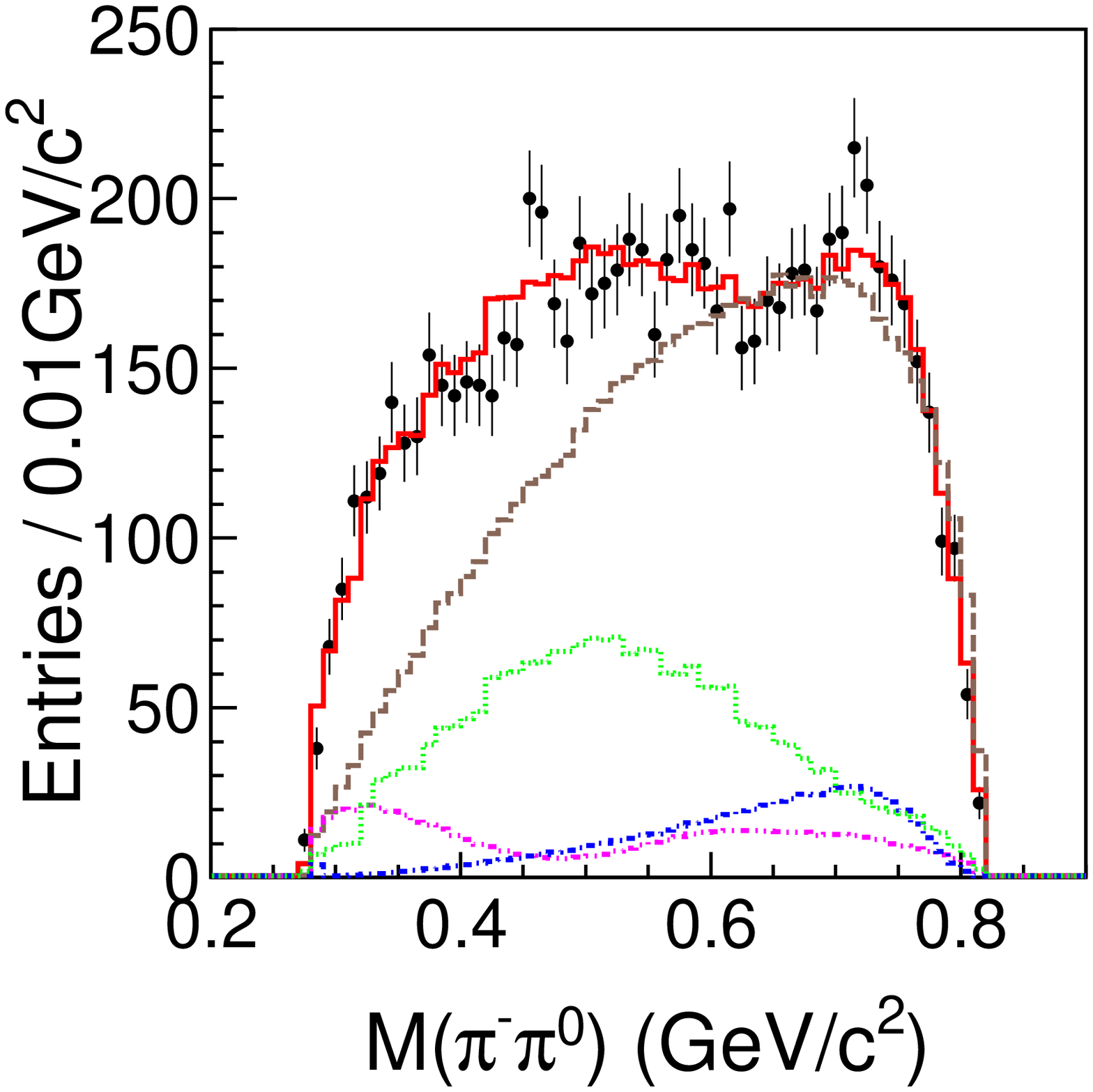}%
  \put(-30,117){\bf (a)}
    \includegraphics[height=2.in,width=2.2 in]{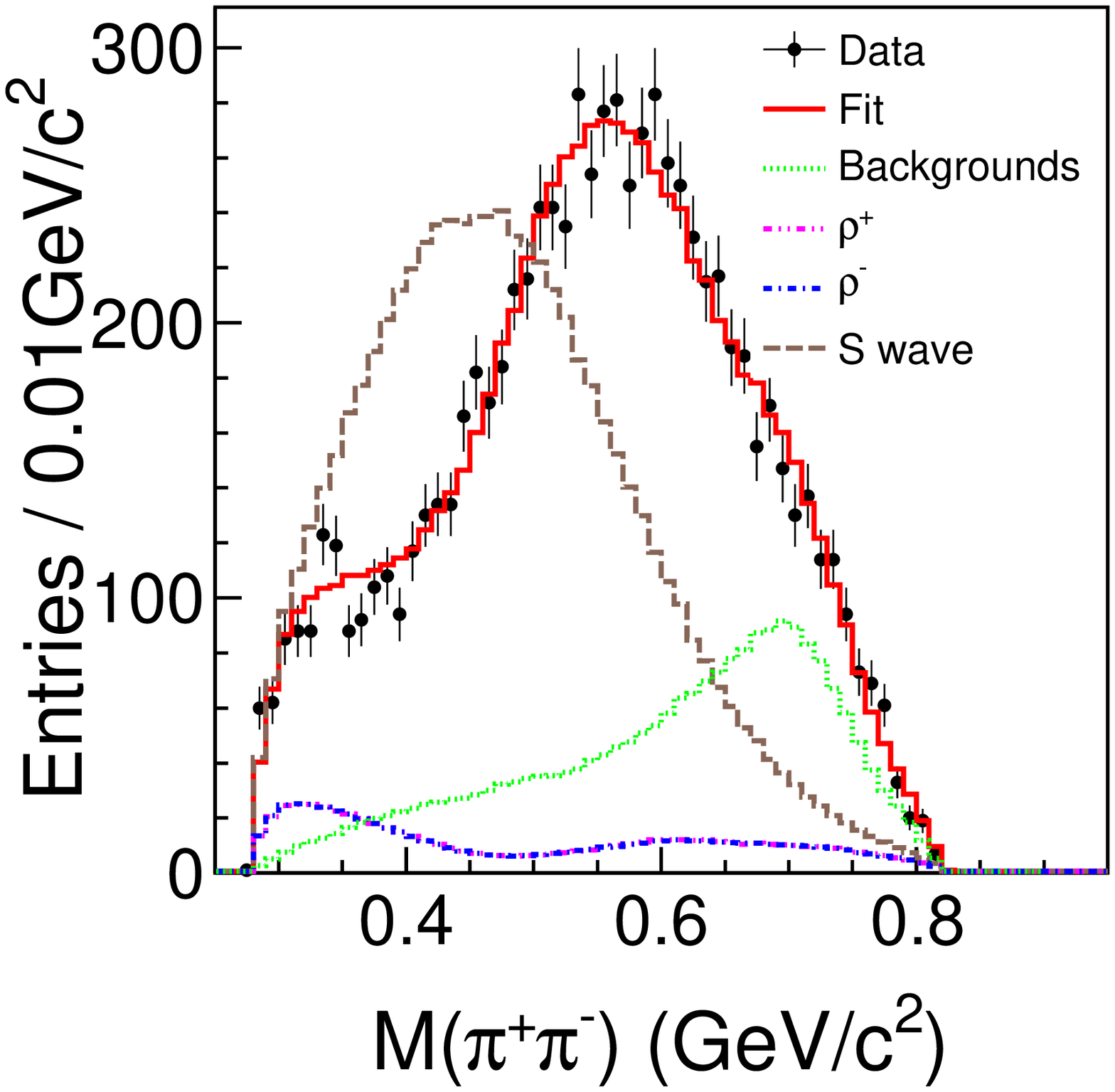}%
\put(-30,117){\bf (b)}
    \includegraphics[height=2.in,width=2.2 in]{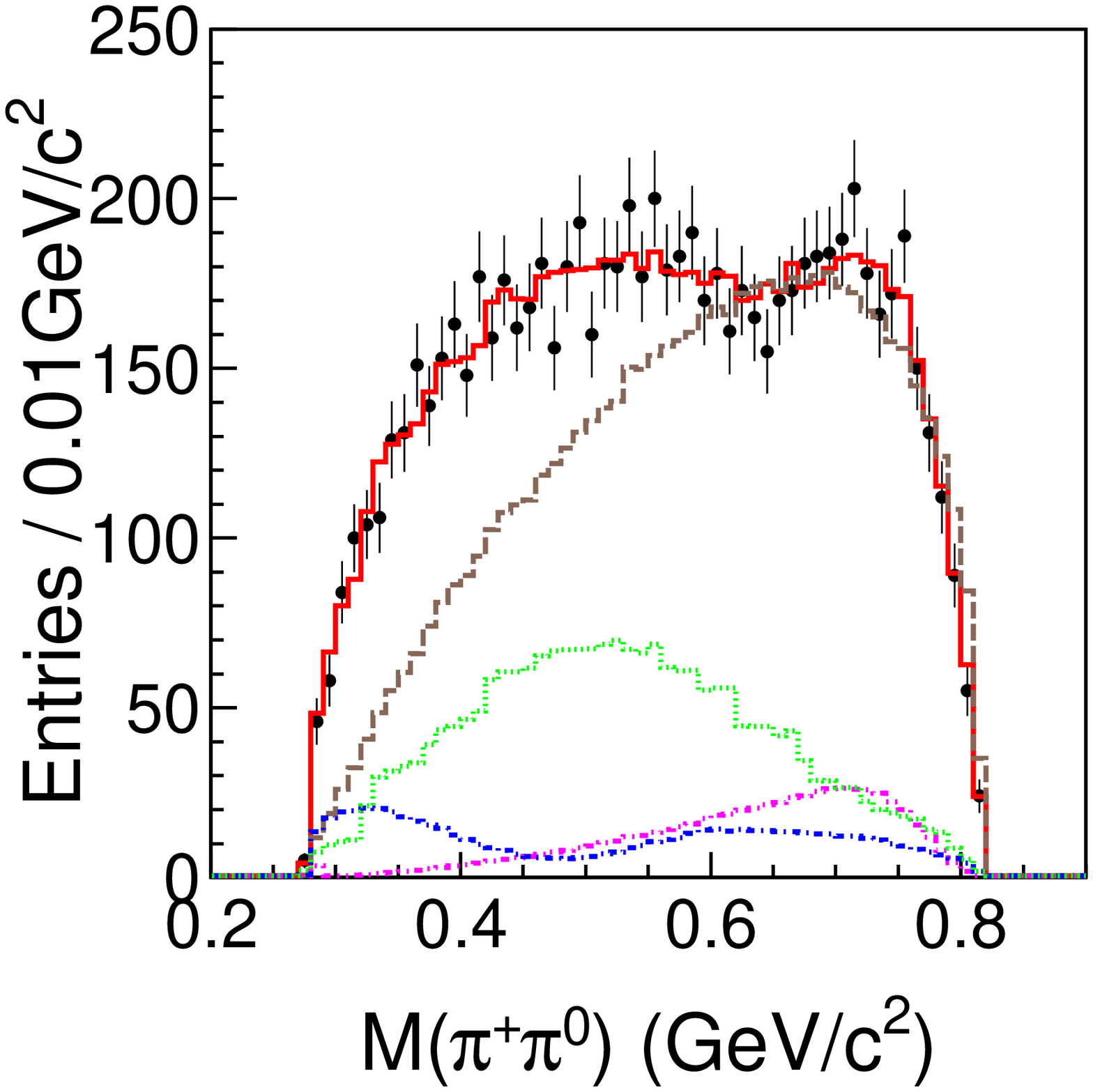}%
  \put(-30,117){\bf(c)}
\caption{  Comparison of the invariant mass distributions of (a) $\pim\pio$,
	 (b) $\pip\pim$, (c) $\pip\pio$ (dots with error bars)
	 and the fit projections (solid histograms).
	 The dotted, dashed, dash-dotted, and dash-dot-dotted histograms show
	 the contributions from background, $S$ wave, $\rho^-$, and $\rho^+$, respectively.
\label{mch-3pi}}
\end{figure*}

The common amplitude analysis of the decays
 $\etap\ra\pi^+\pi^-\pi^0$ and $\etap\ra\pi^0\pi^0\pi^0$ is performed using isobar model.
The fit results illustrated by
the invariant mass spectra of $\pip\pin$, $\pip\pio$ and $\pim\pio$ (Fig.~\ref{mch-3pi})
show significant $P$-wave contribution from $\etap\ra\rho^{\pm}\pi^{\mp}$
in $\etap\ra\pip\pim\pio$. The branching fraction ${\mathcal B}(\etap\ra\rho^{\pm}\pi^{\mp})$
is determined to be $(7.44\pm0.60\pm1.26\pm1.84_{\mathrm{model}})\times 10^{-4}$.
In addition to the non-resonant $S$-wave, the resonant $\pi$-$\pi$
$S$-wave with a pole at $(512\pm15)-i(188\pm12)$ MeV,
interpreted as the broad $\sigma$ meson, plays  essential role
in the $\etap\ra\pi\pi\pi$ decays.
Due to the large interference between
non-resonant and resonant $S$-waves, only the sum is used to describe the $S$-wave
contribution, and the branching fraction is  determined to be
${\mathcal B}(\etap\ra\pip\pim\pio)_S=(37.63\pm 0.77\pm2.22\pm4.48_{\mathrm{model}})\times10^{-4}$.

For  $\etap\ra\pio\pio\pio$, the $P$-wave contribution in
two-body rescattering is forbidden by Bose symmetry. The Dalitz plot for $\etap\ra\pio\pio\pio$ is shown
in Fig.~\ref{m4pi} (a) and the amplitude fit is displayed in Fig.~\ref{m4pi} (b). The corresponding branching fraction is measured to be
${\mathcal B}(\etap\ra\pio\pio\pio)=(35.22\pm0.82\pm2.60)\times10^{-4}$.

The branching fractions of $\etap\ra\pip\pim\pio$ and $\etap\ra\pio\pio\pio$ are in good agreement with
and supersede the previous BESIII measurements~\cite{BESIII:2012aa}.
The value for ${\mathcal B}(\etap\ra\pio\pio\pio)$ is two times larger than
GAMS measurement of $(16\pm3.2)\times10^{-4}$~\cite{Alde:1987jt}.
The significant resonant $S$-wave contribution also provides a
reasonable explanation for the negative slope parameter of the
$\etap\ra\pio\pio\pio$ Dalitz plot~\cite{Ablikim:2015cmz}.  The ratio
between the $S$-wave components of the two decay modes,
$\mathcal{B}(\etap\ra\pio\pio\pio)/\mathcal{B}(\etap\ra\pip\pim\pio)_S$,
is determined to be $0.94\pm0.029\pm0.13$, where the common systematic
cancels.  With the branching fractions of $\etap\ra\pi\pi\eta$ taken
from Particle Data Group (PDG)~\cite{PDG},
$r_{\pm}={{\BR}(\etap\ra\pip\pim\pio)}/{\mathcal{B}(\etap\ra\pip\pim\eta)}$
and
$r_{0}={{\BR}(\etap\ra\pio\pio\pio)}/{\mathcal{B}(\etap\ra\pio\pio\eta)}$
are calculated to be $(8.8\pm1.2)\times10^{-3}$ and
$(16.9\pm1.4)\times10^{-3}$, respectively.

\begin{figurehere}

\includegraphics[height=2.89 in,width=3.0in]{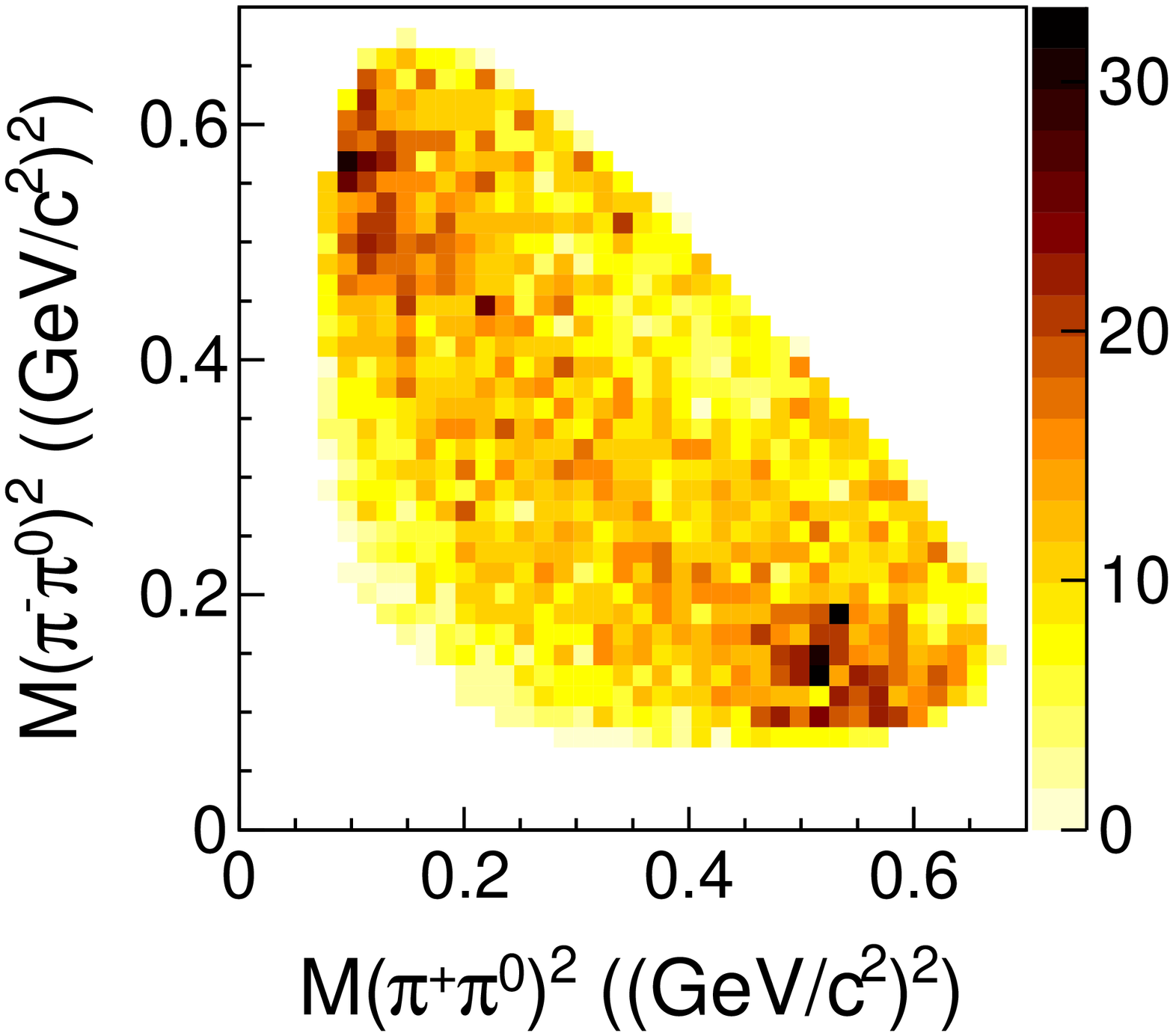}%
\caption{
Dalitz plot of $M^2(\pi^+\pi^0)$ versus $M^2(\pi^-\pi^0)$ for
$\etap\ra\pip\pim\pio$  candidate
    events selected from data.}
\label{m4pidalitz}
\end{figurehere}

\begin{figurehere}
    \includegraphics[height=2.6in,width=2.8 in]{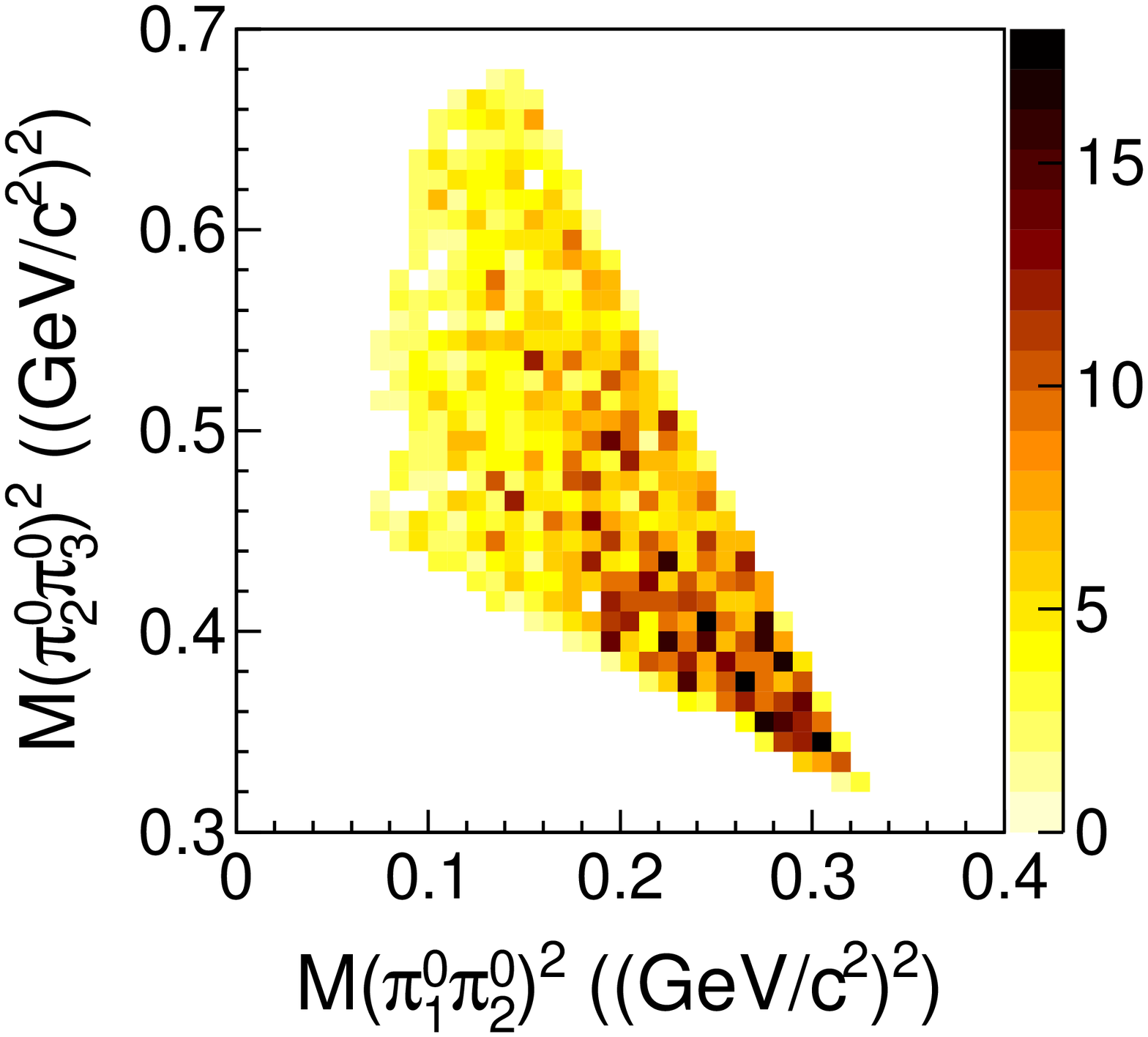}%
\put(-45,150){\bf(a)}

    \includegraphics[height=2.6in,width=2.8 in]{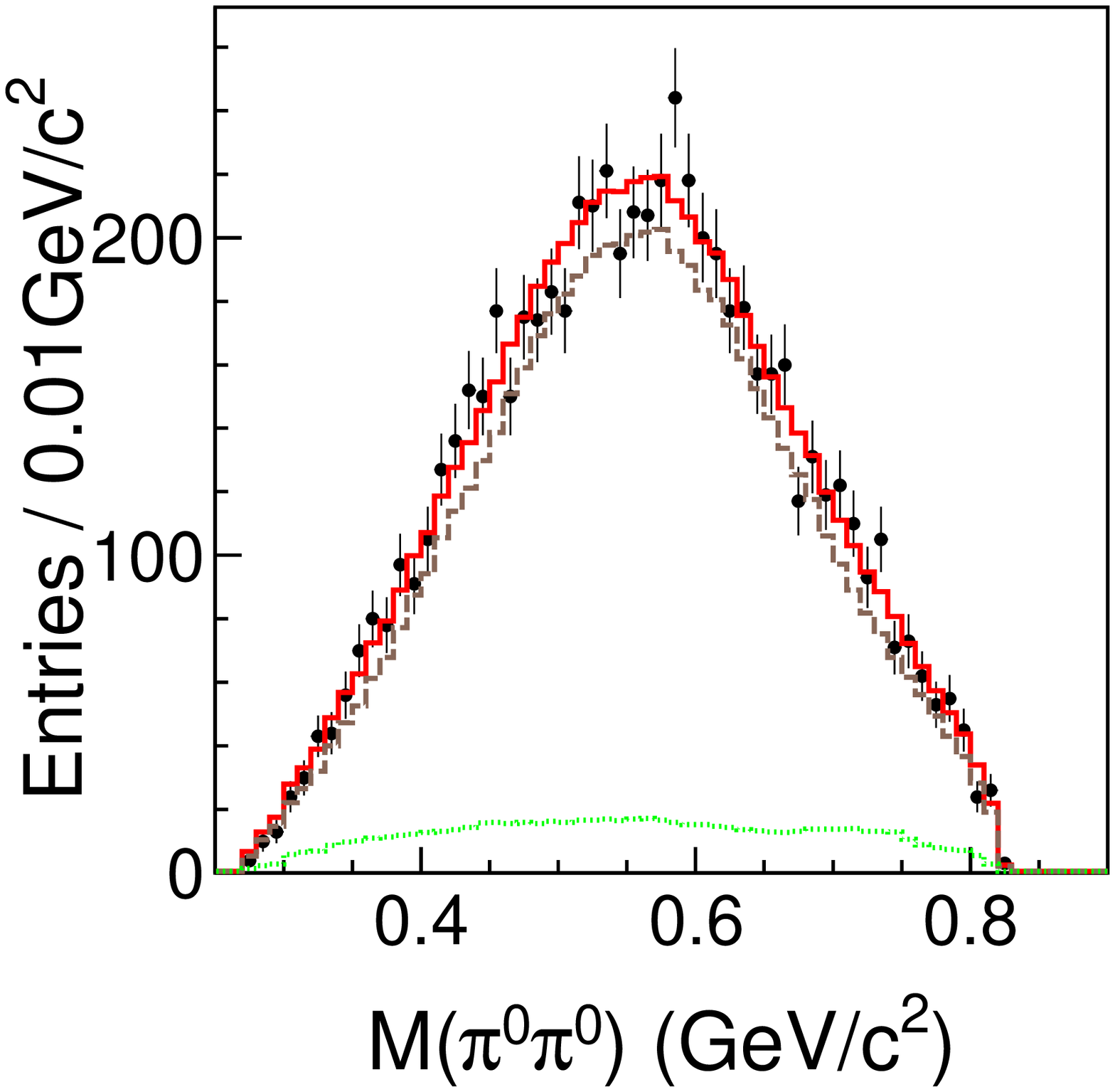}%
  \put(-45,150){\bf(b)}
\caption{
 (a)
    $\etap\ra\pio\pio\pio$ Dalitz plot for candidate
    events selected from data. (b) Comparison of $\pio\pio$ mass spectrum between data (dots with error bars)
	 and the fit projections (solid histograms).
\label{m4pi}}
\end{figurehere}


\subsection[]{\boldmath $\eta^\prime\rightarrow
\pi^+\pi^-\pi^+\pi^-,\pi^+\pi^-\pi^0\pi^0$~\cite{Ablikim:2014eoc}}

In ChPT anomalous hadronic decays
$\eta^\prime \to \pi^+\pi^- \pi^{+(0)}\pi^{-(0)}$ are
related to WZW pentagon contribution.
In the VMD model $\rho^0\rho^0$ or $\rho^+\rho^-$ intermediate
state should provide dominant contribution.
Using a combination of ChPT and VMD
the branching fractions were calculated to be
$\mathcal{B}(\eta^{\prime}\to\pi^{+}\pi^{-}\pi^{+}\pi^{-})$ $ = (1.0
\pm 0.3)\times10^{-4}$ and
$\mathcal{B}(\eta^{\prime}\to\pi^{+}\pi^{-}\pi^{0}\pi^{0}) = (2.4
\pm 0.7)\times10^{-4}$~\cite{Guo:2011ir}.

\begin{figure*}
\begin{center}
    \includegraphics[width=5.6cm,height=5.5cm]{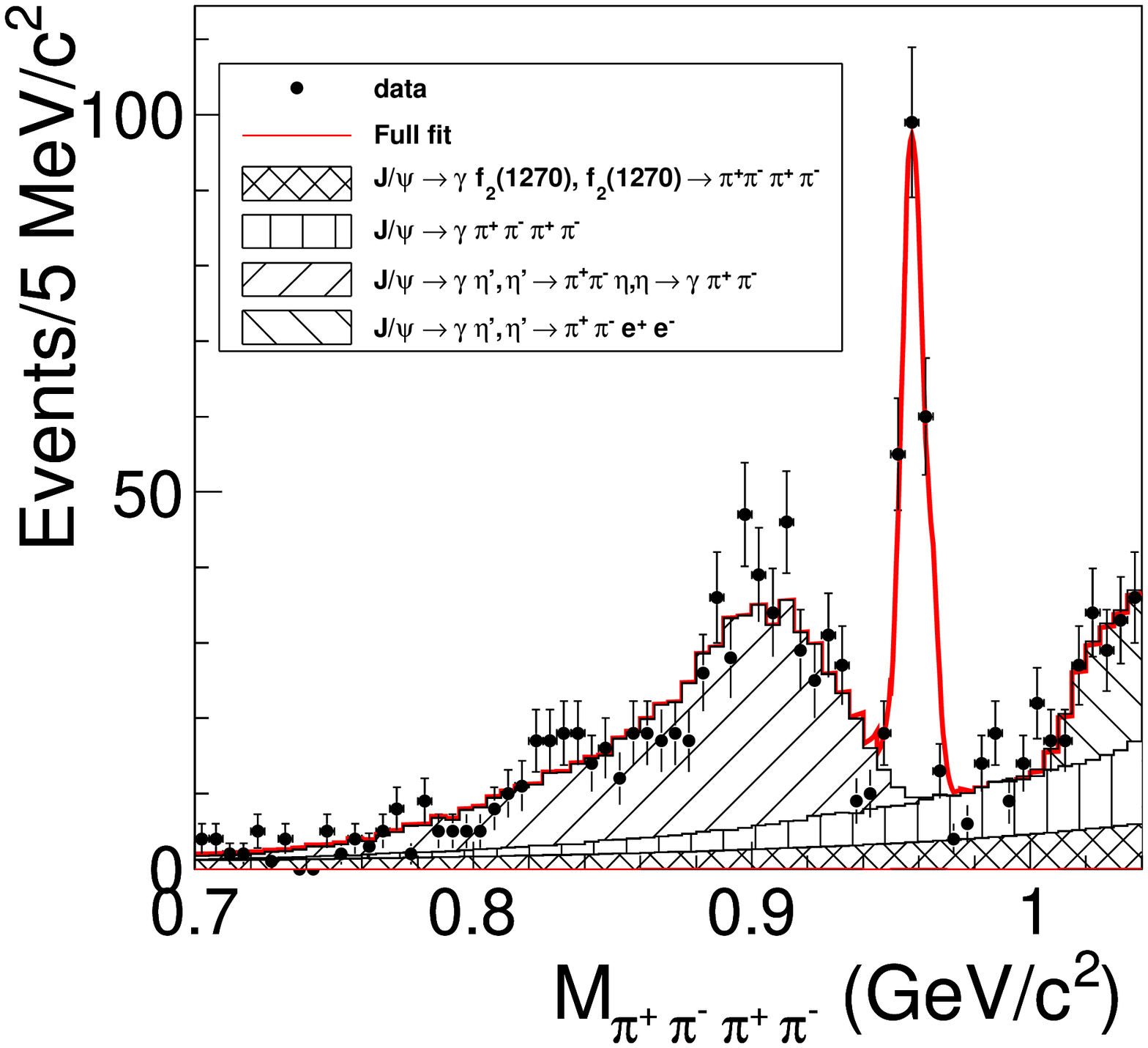}
\put(-30,135){\bf (a)}
    \includegraphics[width=5.6cm,height=5.5cm]{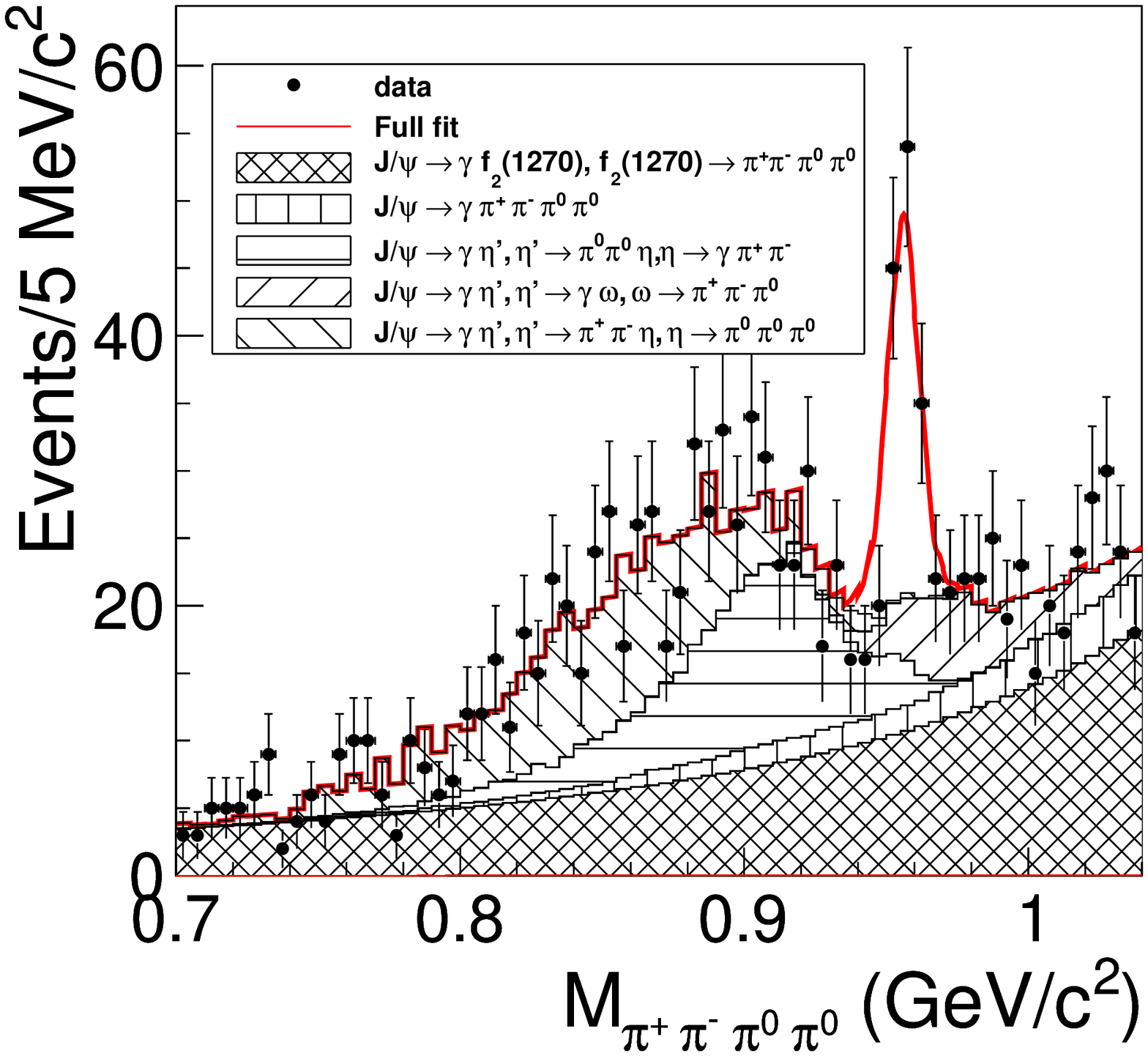}
\put(-30,135){\bf(b)}
    \includegraphics[width=5.6cm,height=5.5cm]{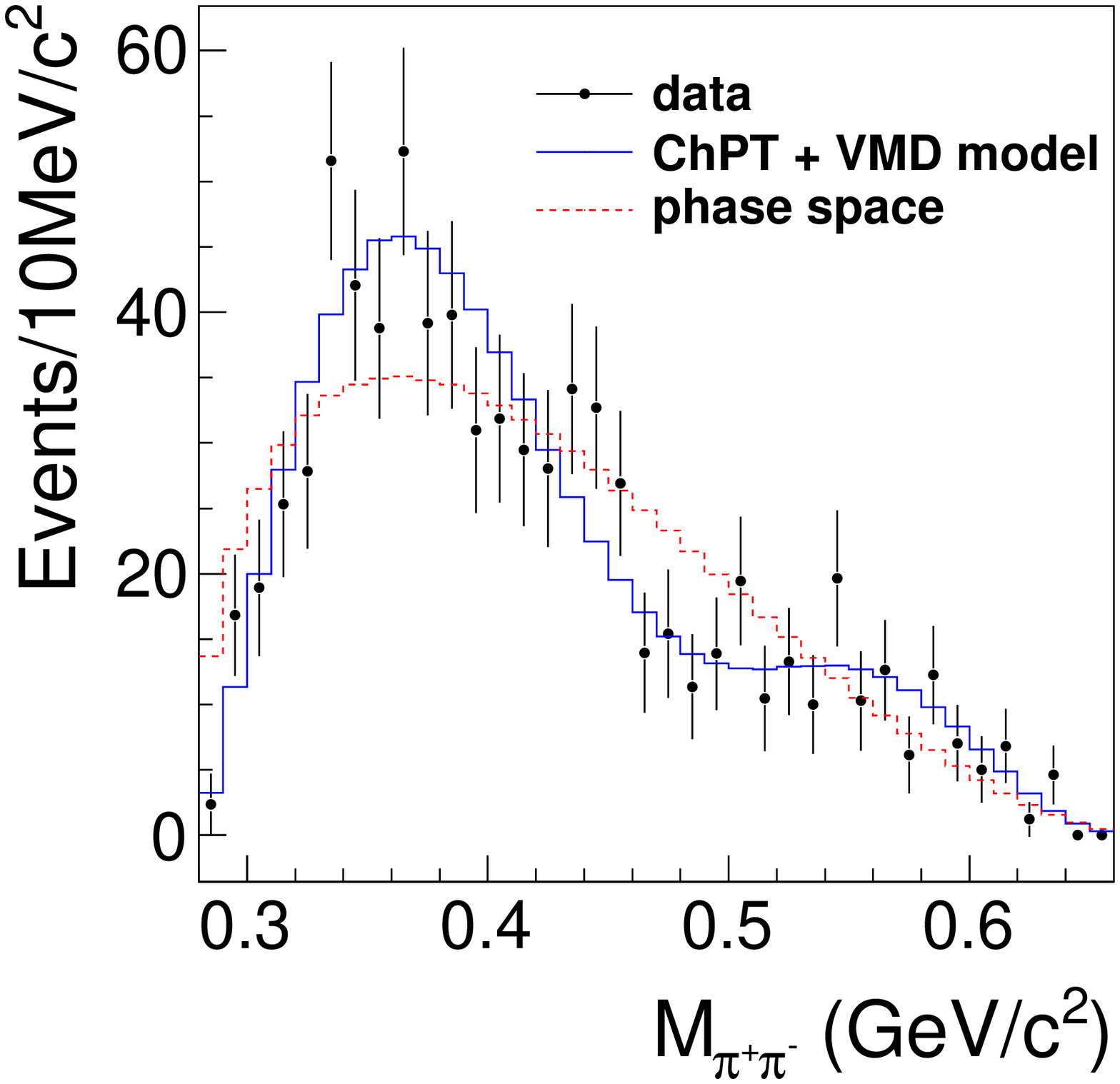}
\put(-30,135){\bf(c)}
    \caption{ Results of the fits to (a) $M(\pi^+\pi^-\pi^+\pi^-)$,
(b) $M(\pi^+\pi^-\pi^0\pi^0)$ and (c) $M(\pi^+\pi^-)$, where the background contributions are
displayed
as the hatched histograms.
\label{m4pi_fit}}
\end{center}
\end{figure*}

The $ \pi^+\pi^-\pi^{+(0)}\pi^{-(0)}$ invariant mass distributions for the BESIII
analysis are shown
in Figs.~\ref{m4pi_fit}(a) and (b), respectively, where the
$\eta^\prime$ peak is clearly seen.
The results of background simulations
are indicated by the hatched
histograms in Figs.~\ref{m4pi_fit}(a) and (b).
None of the background sources produces a peak in the
$\pi^+\pi^-\pi^{+(0)}\pi^{-(0)}$  invariant mass spectrum near  the $\eta^\prime$ mass.

In order to measure the branching fractions, the signal efficiency
was estimated using  a signal MC sample using two assumptions: the
flat phase space and the decay amplitudes from
Ref.~\cite{Guo:2011ir}. For
$\eta^\prime\rightarrow\pi^+\pi^-\pi^+\pi^-$, each of
the $M(\pi^+\pi^-)$ combinations was divided into 38 bins in the region of
[0.28, 0.66] GeV/$c^{2}$. With the procedure described above, the number
of the $\eta^\prime$ events in each bin is obtained by fitting
the $\pi^+\pi^-\pi^+\pi^-$ mass spectrum in this bin, and then the
background-subtracted $M(\pi^+\pi^-)$ is obtained as shown in Fig.~\ref{m4pi_fit} (c)
(four entries per event), where the errors are statistical only.
The comparison of the experimental $M(\pi^+\pi^-)$ distribution
and two
models is shown in Fig.~\ref{m4pi_fit}(c). The amplitude
of Ref.~\cite{Guo:2011ir} provides better description of the
data than the phase
space. Therefore this amplitude is used in the simulation
to determine the detection efficiency
for $\eta^\prime\rightarrow\pi^+\pi^-\pi^{+(0)}\pi^{-(0)}$ decays.

The signal yields are obtained from extended unbinned maximum
likelihood fits to the $ \pi^+\pi^-\pi^{+}\pi^{-}$ and $ \pi^+\pi^-\pi^{0}\pi^{0}$  invariant
mass distributions and the statistical significances for
$\eta^{\prime}\to\pi^{+}\pi^{-}\pi^{+}\pi^{-}$ and
$\eta^{\prime}\to\pi^{+}\pi^{-}\pi^{0}\pi^{0}$ are calculated to be
18$\sigma$ and 5$\sigma$, respectively.  The branching fractions of
$\eta^{\prime}\to\pi^{+}\pi^{-}\pi^{+(0)}\pi^{-(0)}$ are determined
to be $\mathcal{B}(\eta^{\prime}\to\pi^{+}\pi^{-}\pi^{+}\pi^{-}) =
(8.53\pm0.69\pm0.64)\times10^{-5}$ and
$\mathcal{B}(\eta^{\prime}\to\pi^{+}\pi^{-}\pi^{0}\pi^{0}) =
(1.82\pm0.35\pm0.18)\times10^{-4}$,  which are
 in agreement with the predictions in Ref.~\cite{Guo:2011ir},
but not with an older estimate based on broken-$SU(6)\times O(6)$ quark
model~\cite{Parashar:1979js}.

\section{Radiative and Dalitz decays}

\subsection[]{\boldmath $\eta'\to \pi^+\pi^-\gamma$ (preliminary results) }

The anomalous process $\etap \to \gamma\pi^+ \pi^-$ is the second most
probable decay of the $\etap$ meson (${\cal B}=29.1\pm0.5)$ \% \cite{PDG}) and frequently used for $\etap$
tagging.  In the VMD model the main
contribution to the decay comes from $\etap \to \gamma\rho^0$
\cite{GellMann:1962jt}. In the past the di-pion mass distribution was
studied by several experiments {\it e.g.} JADE~\cite{Bartel:1982qd},
CELLO~\cite{Behrend:1982ze}, PLUTO~\cite{Berger:1984xk}, TASSO~\cite{Althoff:1984jq},
TPC/$\gamma\gamma$~\cite{Aihara:1986sp}, and ARGUS~\cite{Albrecht:1987ed}. A peak shift of
about $+$20 MeV  with respect to the expected position from
the $\rho^0$ contribution was consistently observed.
Dedicated~\cite{Bityukov:1990db} analysis using $\sim$2000 $\etap
\to \gamma\pi^+ \pi^-$ events concluded that $\rho^0$ contribution is
not sufficient to describe the di-pion mass spectrum.  This
discrepancy could be attributed to the WZW box anomaly
contribution which should be included as an
extra non resonant term in the decay amplitude.  It was suggested that
the fits to the shape of the di-pion distribution will allow to
determine the ratio of the two contributions~\cite{Benayoun:1992ty}.
The evidence for the box anomaly with a significance of 4$\sigma$ was
reported in 1997 by the Crystal Barrel experiment~\cite{Abele:1997yi} using a
sample of 7490$\pm$180 $\etap$ events but this observation was not
confirmed by the subsequent measurement by the L3
Collaboration~\cite{Acciarri:1997yx} using 2123$\pm$53 events.  Recently
proposed model-independent approach, based on ChPT and a dispersion
theory, describes the $\eta/\etap \to \pi^+ \pi^- \gamma$ decay
amplitudes as a product of an universal and a reaction specific
part~\cite{Stollenwerk:2011zz}.
The universal part could be extracted from the pion vector form factor
measured precisely in $e^+e^-\rightarrow
\pi^+\pi^-$. The reaction specific part was
determined experimentally for the $\eta \to \pi^+ \pi^- \gamma$ decay
by WASA-at-COSY \cite{Adlarson:2011xb} and KLOE \cite{Babusci:2012ft}.
 It was shown the di-pion distribution for the $\eta$
decay can not be described by the pion vector form factor only.  In
Ref.~\cite{Hanhart:2013vba} it was hypothesized that the reaction
specific part could be similar for the $\eta$ and $\etap$ decays.

\begin{figure*}[hbtp]
\begin{center}

  \includegraphics[width=5.cm,height=4.6cm]{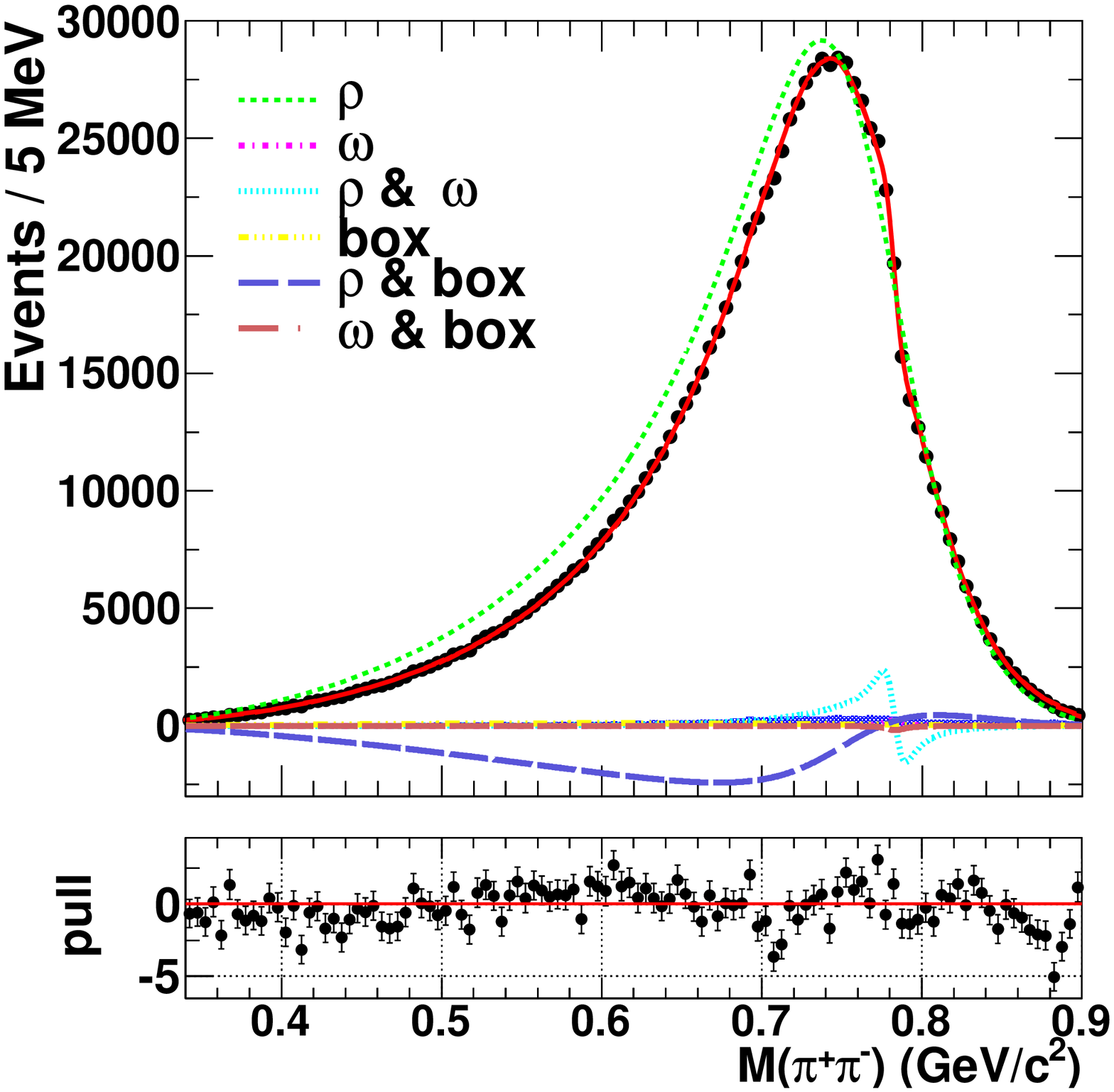}
  \put(-25,110){\bf (a)}
  \includegraphics[width=5.cm,height=4.6cm]{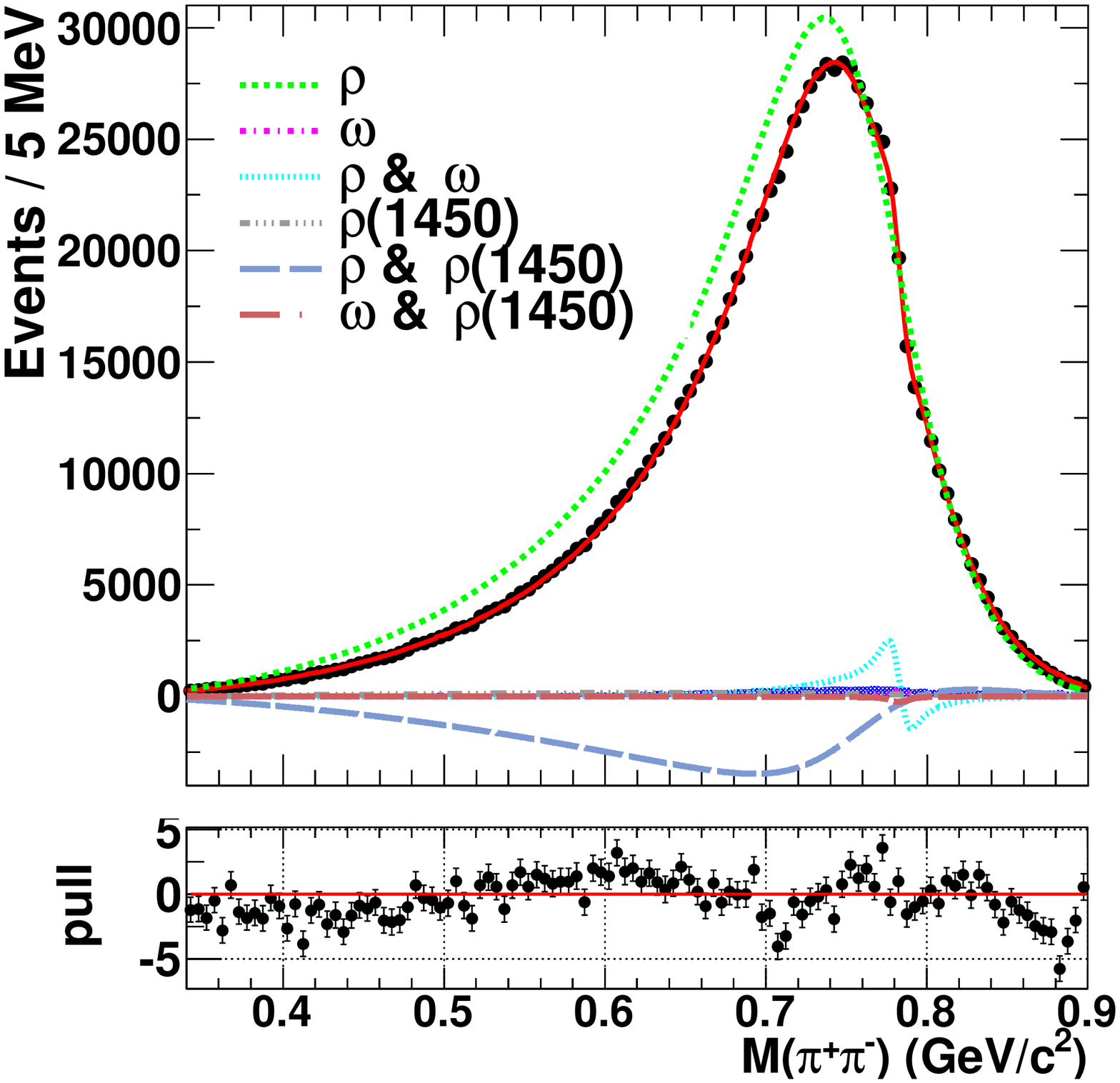}
  \put(-25,110){\bf (b)}
    \includegraphics[width=5.cm,height=4.6cm]{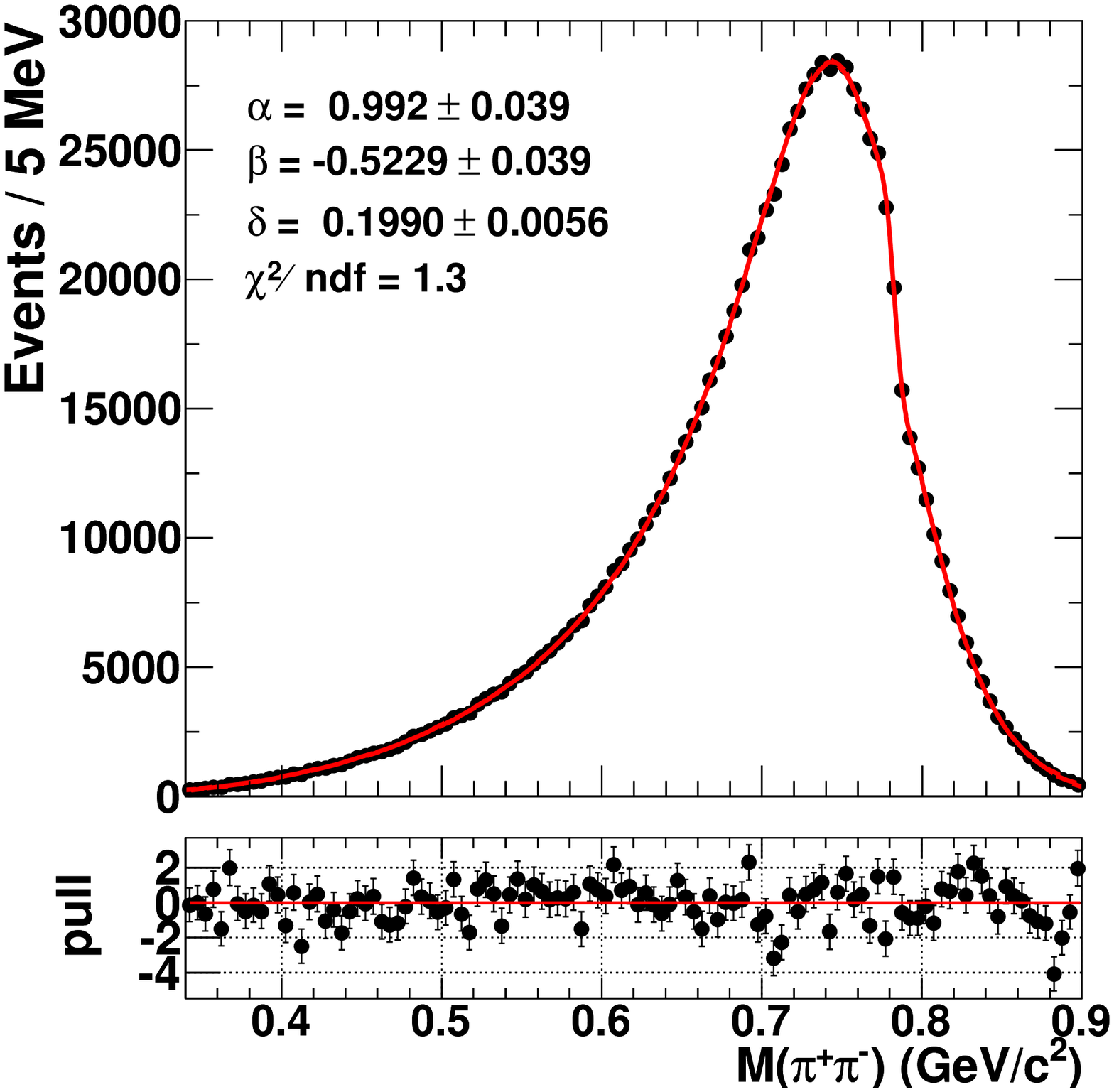}
    \put(-25,110){\bf (c)}
      \caption{The results of the model-dependent fits to $M(\pip\pim)$ with (a) $\rho^0-\omega-box~anomaly$ and (b)  $\rho^0-\omega-\rho^0(1450)$.    (c)
      The results of model-independent fit with $\omega$ interference.  }
  \label{etap-invidata}
\end{center}
\end{figure*}

For BESIII analysis a low background data sample of $9.7\times10^{5}$
$\etap\to\gamma\pi^+\pi^-$ decays candidates is selected. The
 distribution of the $\pi^+\pi^-$ invariant mass, $M(\pi^+\pi^-)$,
is displayed in Fig.~\ref{etap-invidata}.  The
$\rho^0-\omega$ interference is seen for first time in this decay.
In the model-dependent approach the data can
not be described with Gounaris-Sakurai parameterisation~\cite{Gounaris:1968mw} of the $\rho^0$
and the $\omega$ contributions including the interference.
The fit performance gets much better after including the box anomaly, Fig.~\ref{etap-invidata}(a), with a statistical
significance larger than 37$\sigma$.
An alternative fit was performed by replacing the box anomaly
with $\rho^0(1450)$,
Fig.~\ref{etap-invidata}(b), by
 fixing its mass the width to the world average values. The fit is slightly
worse but it still
provides a reasonable description of the data.

Using model-independent approach of Ref.~\cite{Stollenwerk:2011zz} and including
$\rho^0-\omega$ mixing
the pion vector form factor $F_{V}(s)$ (where $s=M^2(\pi^+\pi^-)$) and amplitudes for
$\eta/\etap\to\gamma\pip\pin$ decays are proportional to
$P(s)\cdot \Omega(s)$ where $P(s)$ is a reaction
specific term, $P(s) = 1 + \kappa s + \lambda
s^2+ \xi \cdot BW_{\omega}+\mathcal{O}(s^4)$, $\Omega (s)$ is the
Omnes function describing $\pi-\pi$ interactions with $L=1$
\cite{GarciaMartin:2011cn,Hanhart:2013vba}.
For $\eta\ar\gamma\pip\pin$
only the linear term $\kappa=1.32\pm0.13$~GeV$^{-2}$ \cite{Adlarson:2011xb,
  Babusci:2012ft} is needed.  The fit to the BESIII $\eta'\ar\gamma\pip\pin$  data is shown  in
Fig.~\ref{etap-invidata}(c), it
yields $\kappa=0.992~\pm~0.039$ GeV$^{-2}$, $\lambda=-0.523~\pm~0.039$GeV$^{-4}$,
$\xi=0.199~\pm~0.006$ with the fit goodness  $\chi^2/$ndf=145/109.
The  presence of the quadratic term is consistent  with
recent calculations including  intermediate $\pi^\pm a_2^\mp$ state \cite{Kubis:2015sga}.



\subsection[]{\boldmath $\eta^\prime\rightarrow \pi^+\pi^- l^+l^-$~\cite{Ablikim:2013wfg} }
\label{sec:llpp}

The first observation of the
conversion decay $\etap\ar\pip\pim\EE$ was reported in 2009 by the CLEO~\cite{Naik:2008aa} Collaboration.
The decay is directly related to $\etap\ar\pip\pim\gamma$
and involves virtual photon,
$\etap\ar\pip\pim \g^*\ar\pip\pim\EE$. The conversion process
provides a more stringent test of the models. Predictions for
 the decay are given within
VMD model and unitarized
ChPT~\cite{Faessler:1999de,Borasoy:2007dw,Petri:2010ea}.
In the $e^+e^-$ invariant mass  ($q\equiv M(e^+e^-)$) distribution
for a conversion decay the contribution of the
photon propagator translates to the pole-like $1/q$
dependence  close to  the lower kinematic boundary of $q=2m_{e}$
(see also Sec.~\ref{sub-etap-ggpio}).
For the $\etap\ar\pip\pim\EE$ decay a dominant $\rho^0$ contribution in
the $M(\pip\pim)$ is also expected.
The CLEO measurement based on just $7.9^{+3.9}_{-2.7}$ signal events was unable to explore these distributions. However, the measured branching fraction
$\BR(\etap\ar\pip\pim\EE)=(2.5^{+1.2}_{-0.9}\pm0.5)\times10^{-3}$~\cite{Naik:2008aa},
is consistent with the predicted value of $\sim 2\times 10^{-3}$.
The corresponding conversion decay with a $\MM$ pair
are suppressed by two orders of magnitude due
to $q=2m_\mu$ cutoff. In the CLEO analysis an upper limit
of
$\BR(\etap\ar\pip\pim\MM)<2.4\times10^{-4}$, at 90\% C.L. was
set.

The finished BESIII analysis is based only on 2009 data.
Figure~\ref{fig:eekinematics} displays the ${\EE}$ mass spectrum by
requiring $|M(\pip\pim\EE)-m_{\etap}|<0.02$ GeV/$c^2$, where the
background from $\g\pip\pim$ following the photon conversion
in the detector material could be clearly seen.
The peak close to $2m_e$ corresponds to the  $\etap\ar\pip\pim\EE$
signal and the second peak
around 0.015 GeV/$c^2$ comes from the
$\etap\ar\g\pip\pim$ background.
For the selected data sample any other background is negligible
and the efficiency for the signal is 16.9\%.
The number of $429\pm 24$ signal events is taken
from a fit of the two contributions to the $M(\EE)$ distribution.
The corresponding branching fraction  of
  $\BR(\etap\ar\pip\pim\EE)=(2.11\pm0.12\pm
  0.15)\times10^{-3}$ is in good agreement with
  theoretical predictions and with the CLEO result. The mass spectra of $\pip\pim$ and $e^+e^-$ are
consistent with the expected $\rho^0$ domination in
$M(\pip\pim)$ distribution and the peak in
the $M(\EE)$ distribution just above
2$m_{e}$ threshold with a long tail.

\begin{figurehere}\centering
\includegraphics[width=0.45\textwidth]{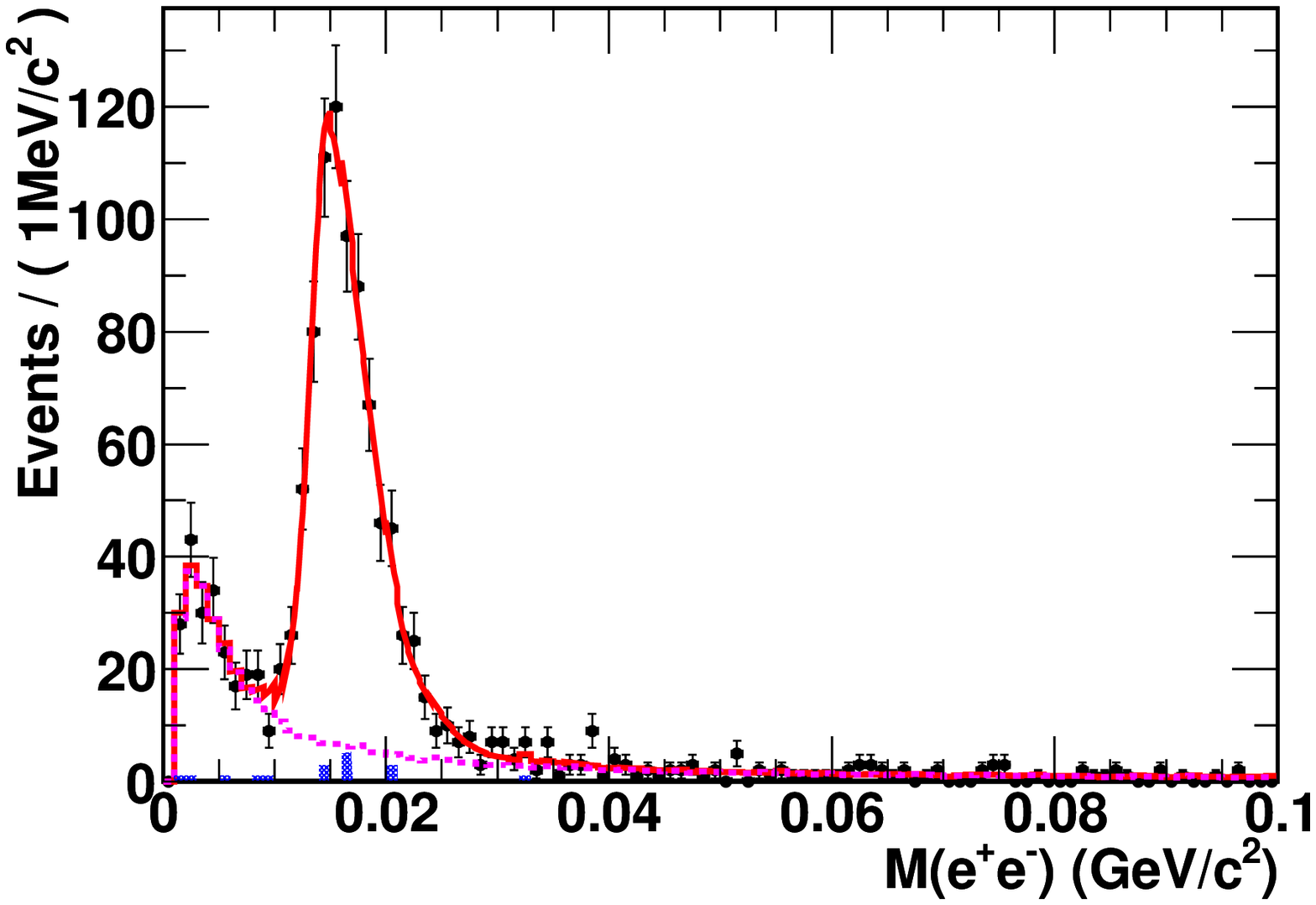}
\caption{The invariant mass spectrum of $\EE$ for data (dots with
error bars) after all the selection criteria are applied. The solid
line
    represents the fit result, the dotted histogram is MC signal shape and the shaded histogram is for backgrounds obtained from $\etap$ sideband events.}
    \label{fig:eekinematics}
    \end{figurehere}

  Figure~\ref{fig:dimukinematics} shows the invariant mass of
$\pip\pim\MM$, where no $\eta^\prime$ signal is observed. The remaining
events in the $\eta^\prime$ mass region are consistent with the
contributions from the background estimated with MC simulations.
The upper limit of $\BR(\etap\ar\pip\pim\MM)<2.9\times10^{-5}$ at the
  90\% C.L. is set.

\begin{figurehere}\centering
\includegraphics[width=0.45\textwidth]{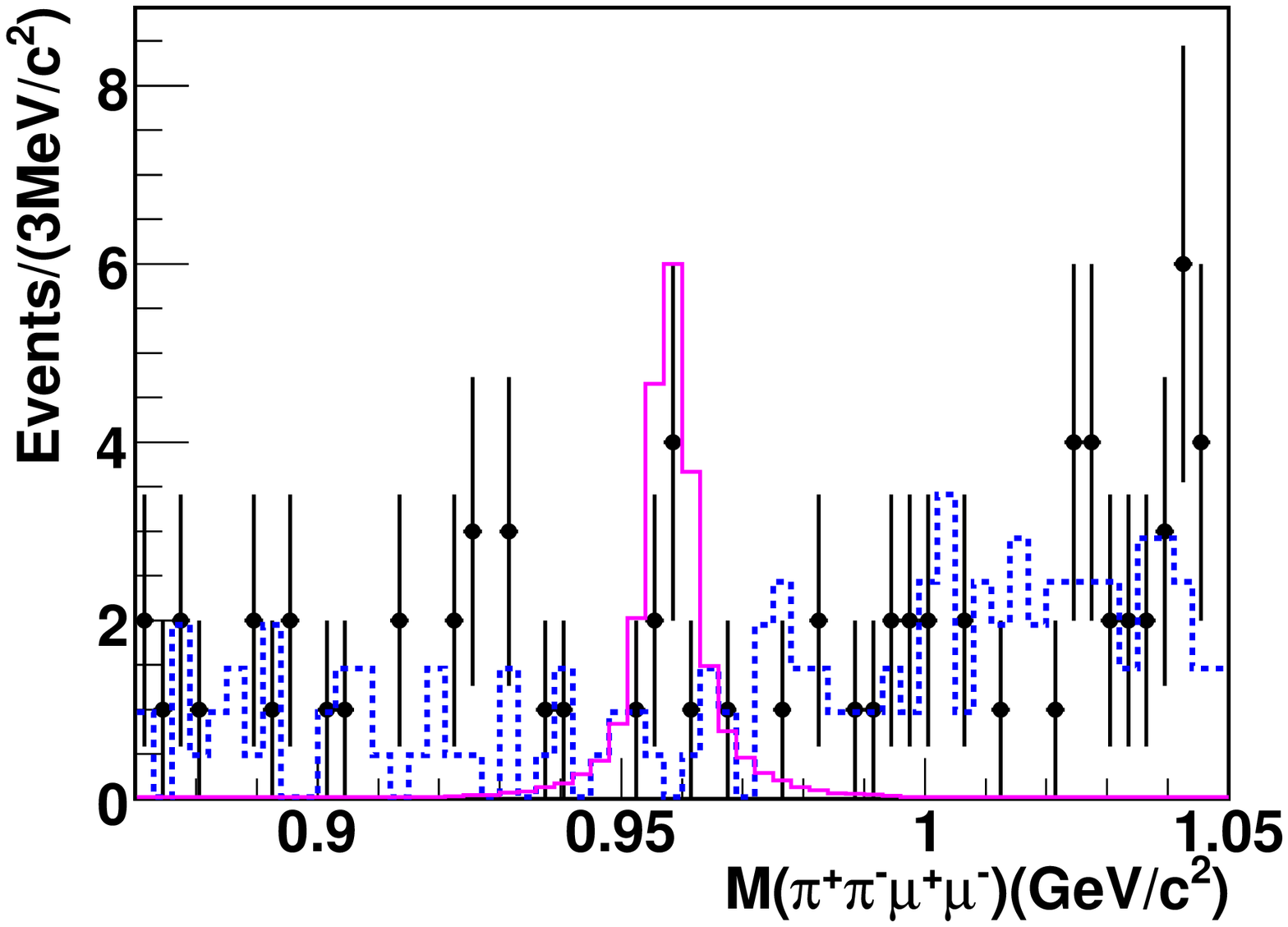}
\caption{Invariant mass spectrum of $\pi^+\pi^-\mu^+\mu^-$ for data (dots with
error bars) after all the selection criteria are applied. The solid
line
    represents the fit result, the dotted histogram is MC signal shape and the shaded histogram is for backgrounds obtained from $\etap$ sideband events.}
    \label{fig:dimukinematics}
    \end{figurehere}


\subsection[]{\boldmath  $\eta^\prime\ra\gamma e^+e^-$~\cite{Ablikim:2015wnx}}
\label{sub-etap-ggpio}

Dalitz decays of light pseudoscalar mesons,
$P\to \gamma e^+e^-$ where $P=\pi^0$, $\eta$, $\eta'$, play an
important role in revealing structure of the hadrons and the interaction
mechanism between photons and the hadrons~\cite{Landsberg:1986fd}.  The
decay rates can be calculated
in Quantum Electrodynamics (QED) where the inner structure of
the mesons is encoded by the transition form factor (TFF), $F(q^2)$,
where  $q^2$ is the invariant
mass of the lepton pair squared. A recent summary and discussion of this subject
can be found in Ref.~\cite{Czerwinski:2012ry}.

The knowledge of the TFF is also important in studies of the muon
anomalous magnetic moment, $a_\mu = (g_\mu -2)/2$, which is one of the most
precise low-energy test of the Standard Model and an important
probe for new physics. The theoretical uncertainty on the SM
calculation of $a_\mu$ is dominated by hadronic corrections and
therefore limited by the accuracy of their
determination~\cite{Blum:2013xva}.  In particular, the hadronic
light-by-light (HLbL) scattering contribution to $a_\mu$ includes two
meson-photon-photon vertices that can be related to the TFF.
Thus, models describing these transitions
should be tested to reduce the uncertainty in
the SM prediction for $a_\mu$.

 The conversion
decay $\eta'\to \gamma e^+e^-$ is closely related to
$\eta'\to\gamma\pi^+\pi^-$, and in particular the transition form
factor could be predicted from the invariant mass distribution of the
two pions and the branching ratio of the $\eta'\to\gamma\pi^+\pi^-$
decay in a model independent way using a dispersive integral~\cite{Hanhart:2013vba}.

The differential decay width~\cite{Landsberg:1986fd} is calculated with,
\begin{eqnarray}\label{eq:decaywidth}
\frac{d\Gamma(\eta'\to\gamma l^+l^-)}{dq^2\Gamma(\eta'\to \gamma\gamma)}&=&
[\mbox{QED}(q^2)] \times |F(q^2)|^2,
\end{eqnarray}
\noindent
where $[\mbox{QED}(q^2)]$ represents the QED part for a point-like
meson which includes $1/q^2$ term due to photon propagator.
Therefore for conversion decays involving electron-positron
pair the distribution peaks at the lowest invariant masses $q=2 m_e$.
 The TFF  can be experimentally determined
from the ratio of the measured di-lepton invariant mass spectrum
and the  $[\mbox{QED}(q^2)]$ term. In the VMD model,
it is assumed that interactions between virtual
photon and hadrons are described by a superposition of neutral vector meson states~\cite{Sakurai:1960ju,Budnev:1979fz}.
The dominant contribution  is expected to come from $\rho^0$ meson
and the form factor could be described by:
\begin{eqnarray}\label{eq:FFVMD}
F(q^2)=N\frac{m^2_V}{m^2_V-q^2-i\Gamma_V m_V},
\end{eqnarray}
where $N$ is a factor ensuring that $F(0)=1$ and
$m_V\approx m_\rho$, $\Gamma_V\approx\Gamma_\rho$ where
$m_\rho$, $\Gamma_\rho$ is the mass and width of the $\rho^0$ meson respectively.
In the case of the $\eta'$, the mass of the pole lies within the
kinematic boundaries of the decay an therefore the $-i\Gamma_V m_V$
term cannot be neglected.
A parameter often extracted experimentally is the slope of the
form factor, $b$, defined as
\begin{eqnarray}\label{eq:slope}
 b=\left.\frac{d|F|}{dq^2}\right|_{q^2=0}=\frac{1}{m_V^2+\Gamma_V^2}.
\end{eqnarray}

Before the BESIII result, only the $\etap \to \gamma
\mu^+\mu^-$ process has been observed with the TFF slope measured to be
$b_{\eta'} = (1.7\pm0.4)$~GeV$^{-2}$~\cite{Landsberg:1986fd, Dzhelyadin:1979za}.  In
the VMD model, $\Gamma(\eta'\to\gamma e^+e^-)/\Gamma(\eta'\to\gamma\gamma)=(2.06\pm0.02)\%$
~ \cite{Petri:2010ea} to be compared to 1.8\% if the TFF is set to one.
The TFF slope is expected to be $b_{\eta'} =
1.45$~GeV$^{-2}$~\cite{Bramon:1981sw,Ametller:1983ec} in the VMD model, while in ChPT it is $b_{\eta'}=1.60$~GeV$^{-2}$~\cite{Ametller:1991jv}. A recent
calculation based on a dispersion integral gives $b_{\eta'} =
1.53^{+0.15}_{-0.08}$~GeV$^{-2}$~\cite{Hanhart:2013vba}.

\begin{figure*}[hbtp]
  \includegraphics[width=0.32\linewidth]{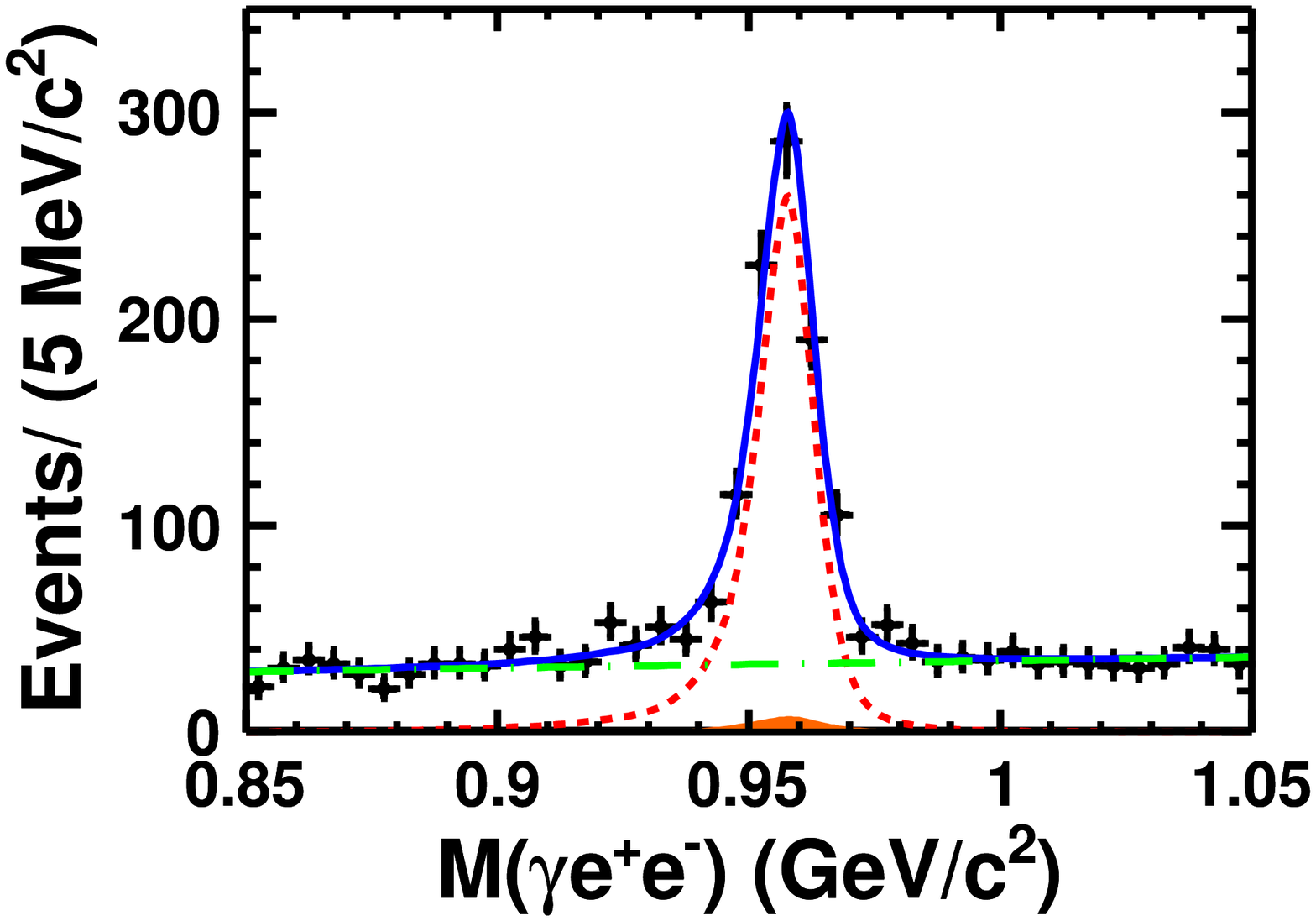}
  \put(-110,90){\bf(a)}
 \includegraphics[width=0.32\linewidth]{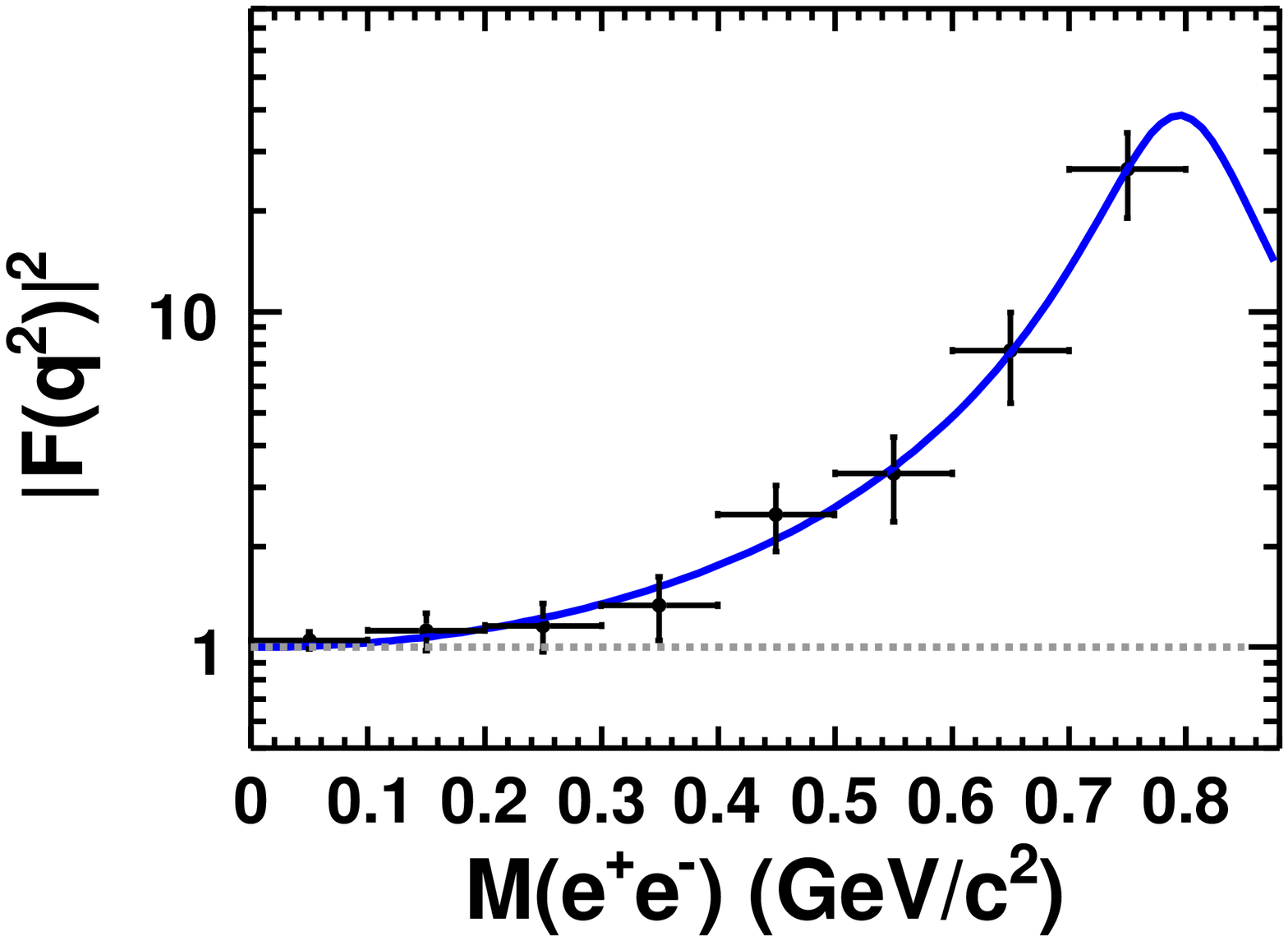}
    \put(-110,90){\bf(b)}
     \includegraphics[width=0.32\linewidth]{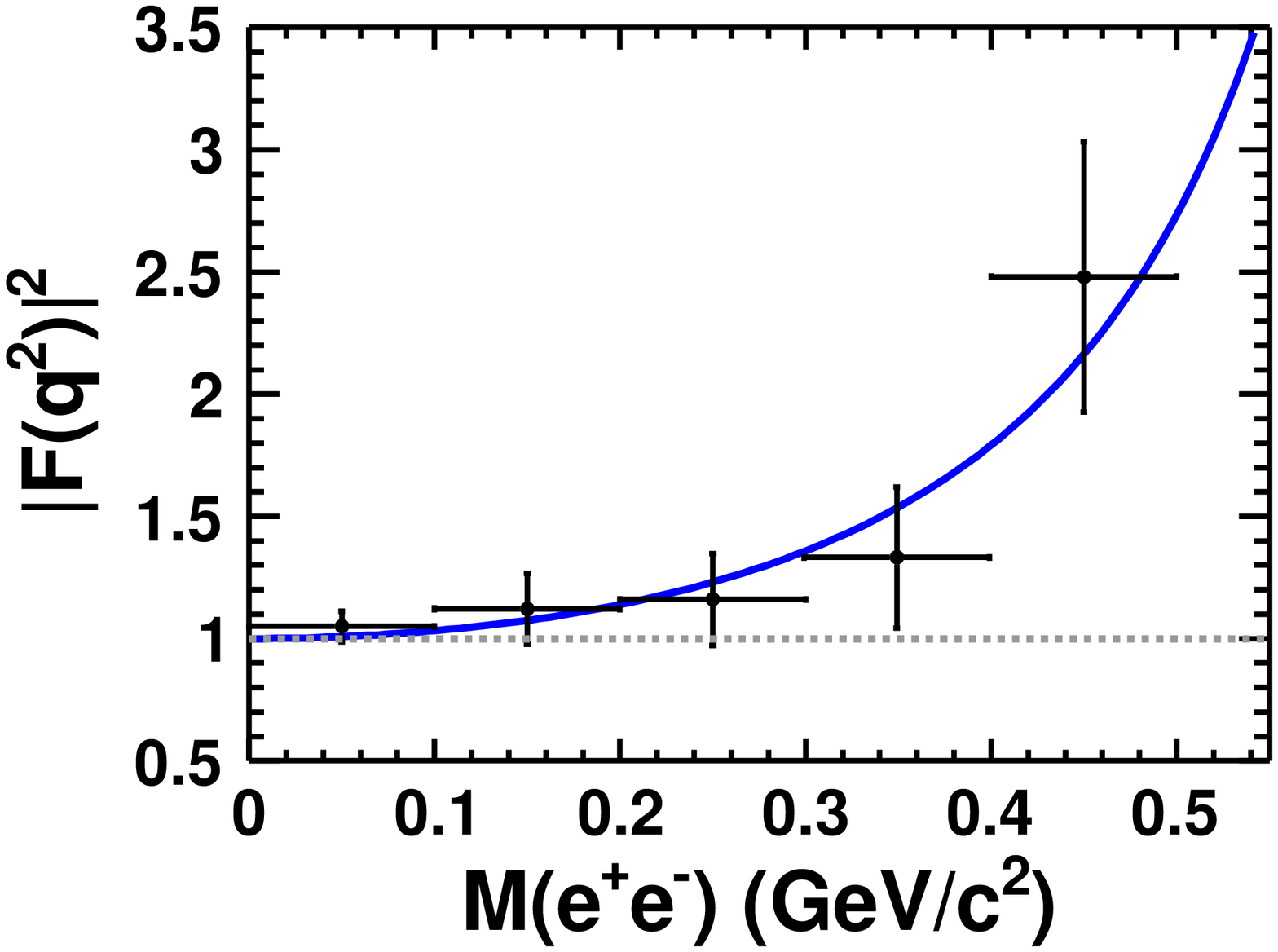}
       \put(-110,90){\bf(c)}%
  \caption{(a) Invariant $\gamma e^+e^-$ mass distribution for the
    selected signal events. The (black) crosses
           are the data, the (red) dashed line represents the signal, the
(green) dot-dashed curve shows
           the non-peaking background shapes, the (orange) shaded component
is the shape of the
           peaking
           background events.  (b) Fit to the single pole form factor  $|F|^2$. (c) Determination of the form factor slope by fitting to $|F|^2$.\label{etap:fit}}
\end{figure*}

In the BESIII experiment the largest background comes from QED processes
and $J/\psi\to e^+e^-\gamma\gamma$ decays.  For these channels, the
combination of the $e^+e^-$ with any final-state photon produces a
smooth $M(\gamma e^+e^-$) distribution.  The primary peaking
background comes from the decay $\eta' \to
\gamma\gamma$ followed by a $\gamma$ conversion in the material in
front of the main drift chamber.  The distance from the
reconstructed vertex point of the electron-positron pair to the $z$
axis is used to reduce the background down to $42.7\pm8.0$ events.
  The resulting
$M(\gamma e^+e^-)$ distribution after the selection criteria is shown
in Fig.~\ref{etap:fit}(a) and exhibits a clear peak at the $\eta'$ mass.
A fit is performed to
determine the signal yield with  the signal shape represented by the MC.
The
non-peaking background is described by a first-order Chebychev
polynomial.  The fraction of the peaking background
is fixed from the simulation.  The signal yield and the detection
efficiency is summarized in Table~\ref{tab:BIIIres}.  The decay
$\eta' \to \gamma \gamma$ from
the
same data set  is used for normalization and the result is quoted
in terms of the ratio
$\Gamma(\eta'\to\gamma e^+e^-)/\Gamma(\eta'\to \gamma\gamma)=
(2.13\pm0.09\pm0.07)\times10^{-2}$.
Using the branching fraction of $\eta'\to \gamma \gamma$ in PDG~\cite{PDG}, we obtain
the first measurement of the $\eta' \to \gamma e^+e^-$ branching fraction reported in Table~\ref{tab:BIIIres}.

The TFF is extracted from the bin-by-bin efficiency corrected signal
yields for eight $M(e^+e^-)$ bins for $M(e^+e^-)<0.80$ GeV/c$^2$.  The bin widths of 0.1~GeV\ are
used and are much wider than the $M(e^+e^-)$ resolution
(5$\sim$6~MeV~depending on $M(e^+e^-)$).  The signal yield in each
$M(e^+e^-)$ bin is obtained by repeating the fits to the $M(\gamma
e^+e^-)$ mass distributions.


The result for $|F|^2$ is obtained by dividing the acceptance
corrected yields by the integrated QED prediction in each $M(e^+e^-)$
bin and it is shown in Figs.~\ref{etap:fit} (b) and (c).
 The parameters from the fit of the TFF to the
parametrization of Eq.~\ref{eq:FFVMD} are $m_V =
(0.79\pm0.04\pm0.02)$~GeV and $\Gamma_V = (0.13\pm0.06\pm0.03)$~GeV.
The single pole parameterization provides a good description of data
as shown in Fig.~\ref{etap:fit} (b).
The corresponding value of the slope parameter is
$b_{\eta'}=(1.56\pm0.19)$~GeV$^{-2}$, in agreement with the result
from $\eta'\to \gamma \mu^+\mu^-$~\cite{Landsberg:1986fd}.
The slope agrees also within errors with the VMD model predictions and
the uncertainty matches the best determination in the space-like
region from the CELLO collaboration $b_{\eta'} =
(1.60\pm0.16)$~GeV$^{-2}$~\cite{Behrend:1990sr}.


\subsection[]{\boldmath $\eta^\prime\rightarrow e^+e^-\omega$~\cite{Ablikim:2015eos}}
\label{sub-etap-eeomega}

The decay $\eta^\prime\to
\pi^+\pi^-e^{+}e^{-}$~\cite{Ablikim:2013wfg} is dominated by
${\eta}^{\prime}$ ${\to}\rho^0 e^{+}e^{-}$, in agreement with
theoretical predictions~\cite{Faessler:1999de, Borasoy:2007dw}. The
corresponding
decay ${\eta}^{\prime}$ ${\to}$ ${\omega}e^{+}e^{-}$ was not observed
before BESIII measurements.
Theoretical models~\cite{Faessler:1999de,Terschlusen:2012xw} predict the
branching fraction to be around $2.0{\times}10^{-4}$.

A parallel analysis of $\eta^\prime\rightarrow e^+e^-\omega$ and $\eta^\prime\rightarrow\gamma\omega$ decays
allows to reduce the impact of systematic errors for the ratio of the branching fractions. For
$\eta^\prime\rightarrow e^+e^-\omega$ decay, candidate events with four well-reconstructed charged tracks and
at least three photons are selected.
The external conversion background from  $\eta^\prime\rightarrow\gamma\omega$ is removed by requiring the
distance of the vertex from the $z$ axis
to be less than 2 cm  (according to simulation only $2.6\pm0.3$ background events will survive the cut).
In the selected data sample both the $\omega$ peak in $M(\pi^{0}\pi^{+}\pi^{-})$ and
the $\eta^\prime$  peak in  $M(\pi^{0}\pi^{+}\pi^{-}e^{+}e^{-})$ is clearly seen in the scatter plot
shown in Fig.~\ref{fit3}(a). The best identification of the
process is achieved in  the $M(\pi^{0}\pi^{+}\pi^{-}e^{+}e^{-})- M(\pi^{0}\pi^{+}\pi^{-})$ distribution.
This distribution is used in a fit to extract the signal yields as indicated in Fig.~\ref{fit3}(b).
The decay of
$\eta^{\prime}\to\omega e^{+} e^{-}$ is observed with a statistical significance
of 8$\sigma$, and its branching fraction is measured to be
$\mathcal{B}(\eta^{\prime}\to\omega
e^{+}e^{-})=(1.97\pm0.34\pm0.17)\times10^{-4}$,
consistent with theoretical predictions~\cite{Faessler:1999de,Terschlusen:2012xw}.

\begin{figurehere}
   \includegraphics[width=6.6cm]{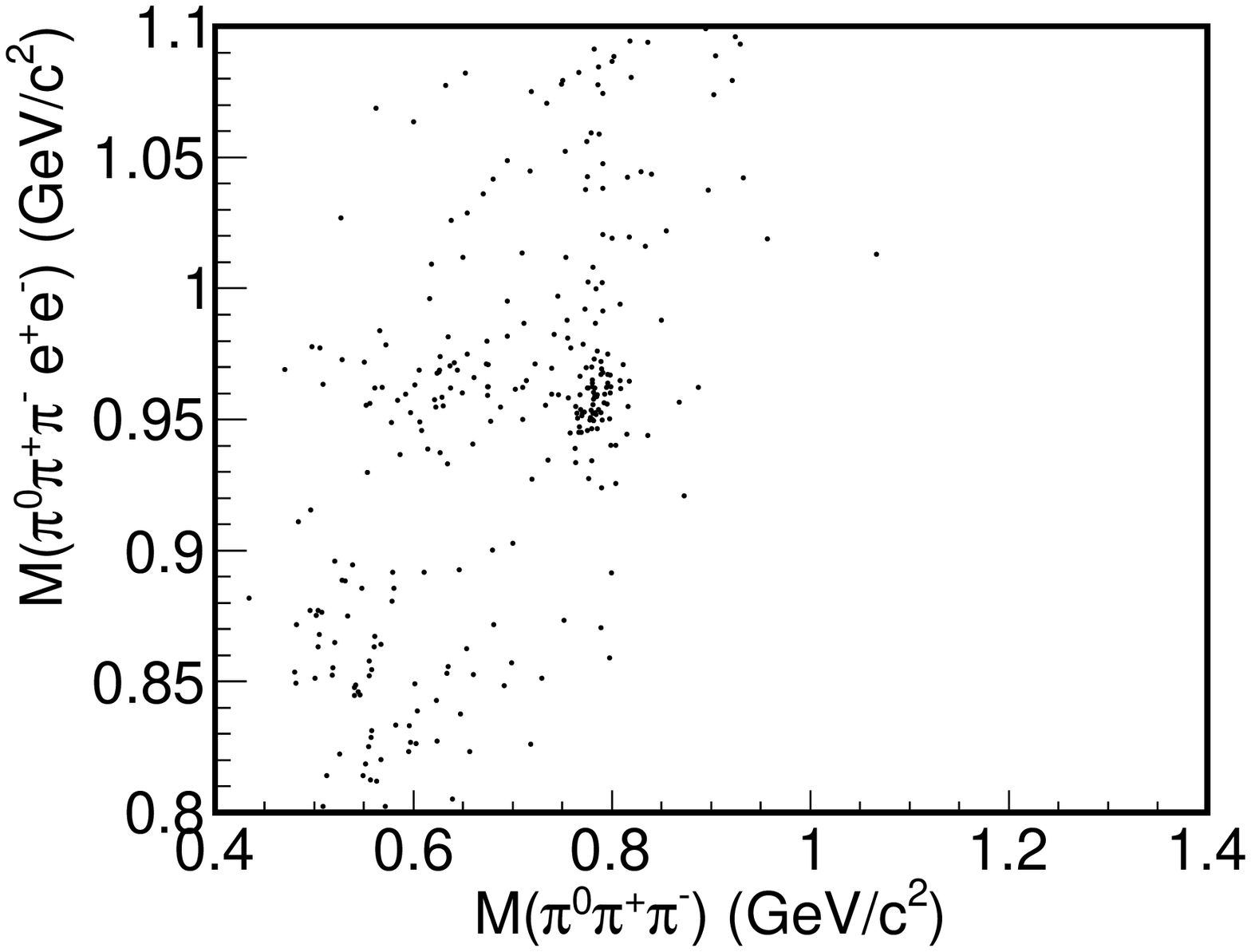}
   \put(-50,114){\bf (a)}\\
 \hspace*{0.5cm}\includegraphics[width=6.6cm]{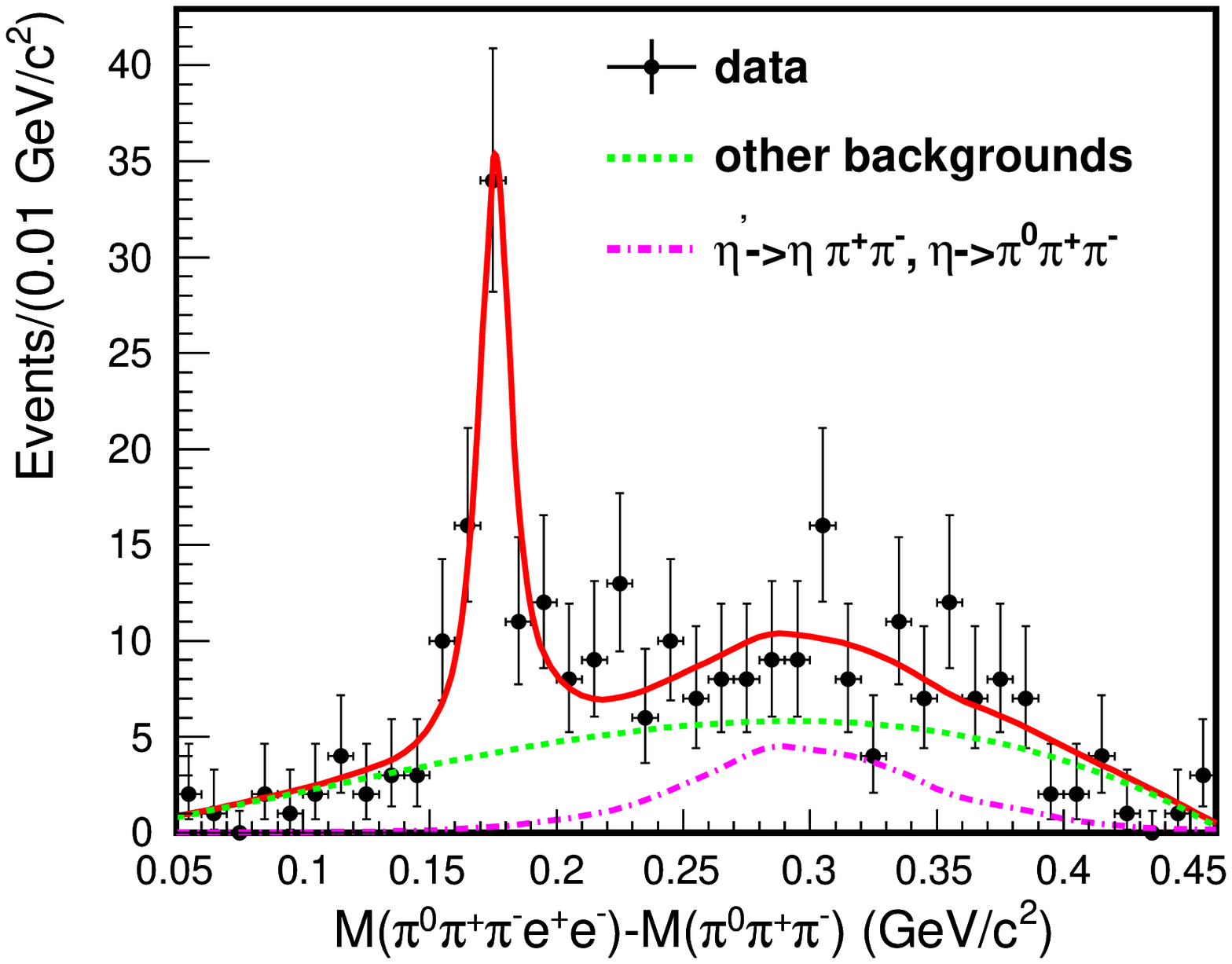}
   \put(-50,114){\bf (b)}
 \caption{ (a)~Distribution of $M(\pi^0\pi^+\pi^-e^+e^-)$ versus $M(\pi^0\pi^+\pi^-)$. (b)
   Distribution of $M(\pi^0\pi^+\pi^- e^{+}
  e^{-})-M(\pi^0\pi^+\pi^-)$ with the fit results.
The dash-dotted line is the $\eta^\prime\to\pi^+\pi^-\eta$ background
contributions and the dotted line
   is the remaining background. \label{fit3}}
\end{figurehere}

\subsection[]{\boldmath $\etap\to\gamma\gamma\pio$~\cite{Ablikim:2016tuo}}

\begin{figure*}
  \centering
  \includegraphics[width=5.5cm]{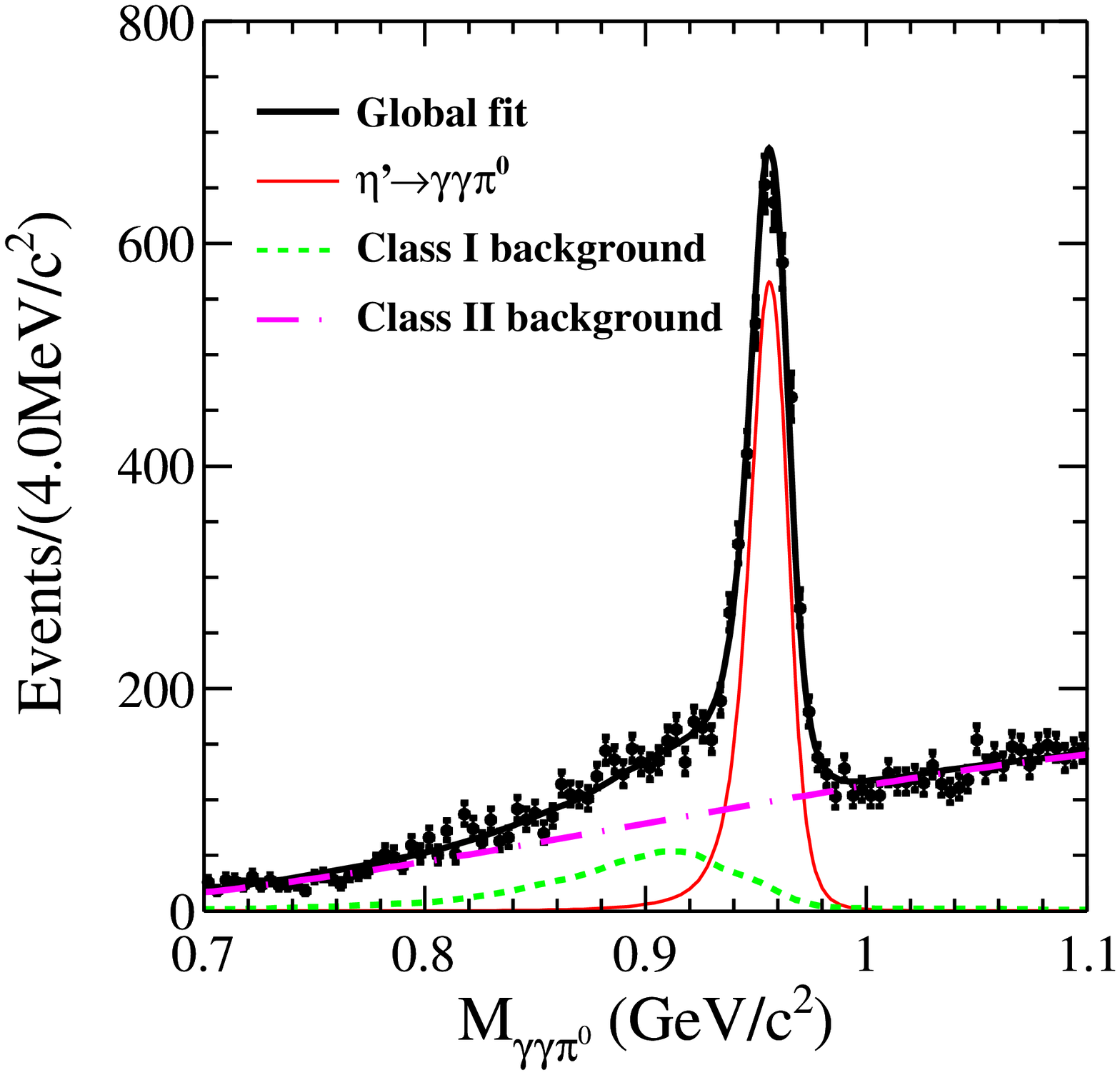}\put(-35,130){\bf (a)}
 \includegraphics[width=5.5cm]{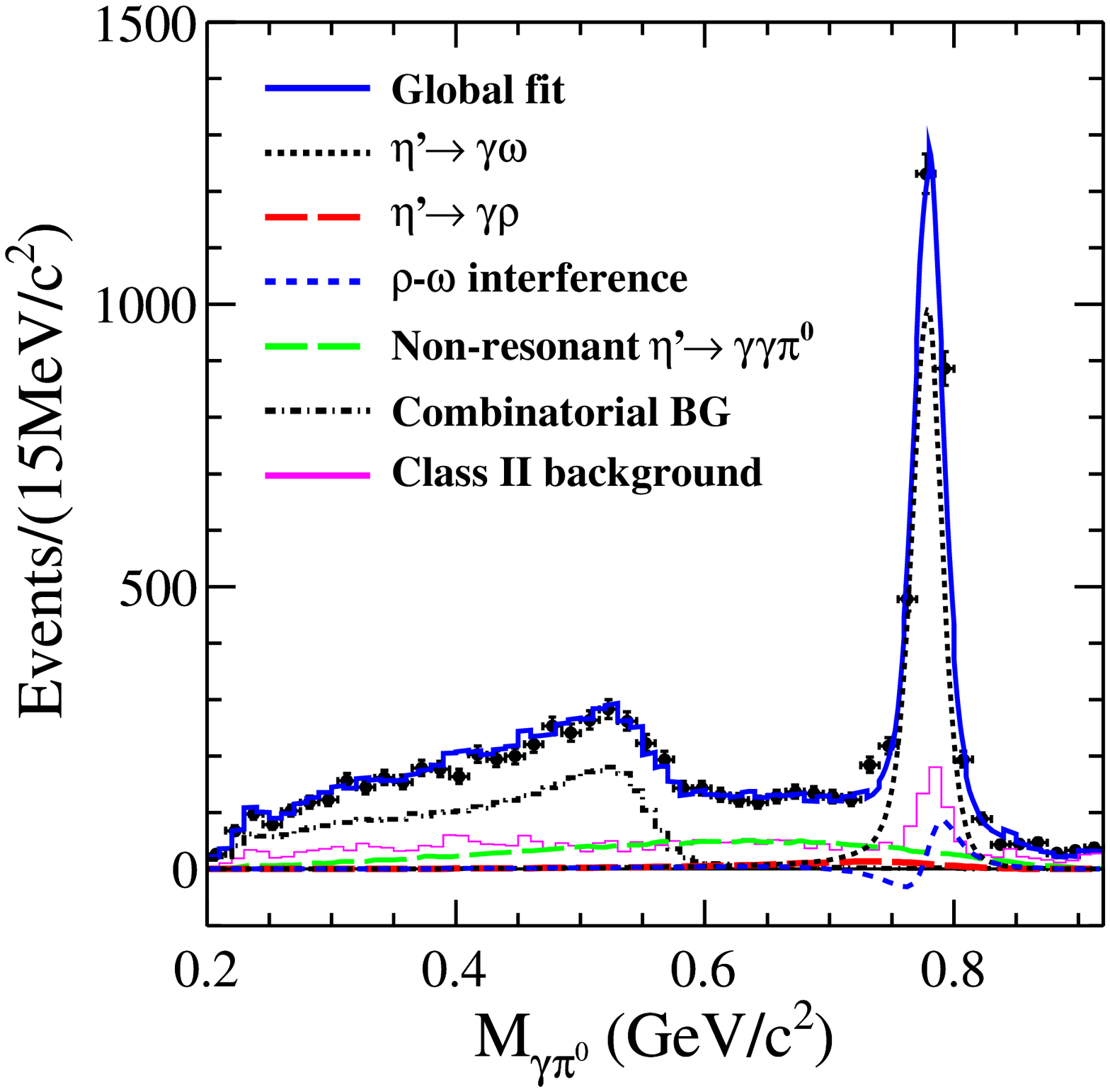}\put(-35,130){\bf (b)}
  \includegraphics[width=5.5cm]{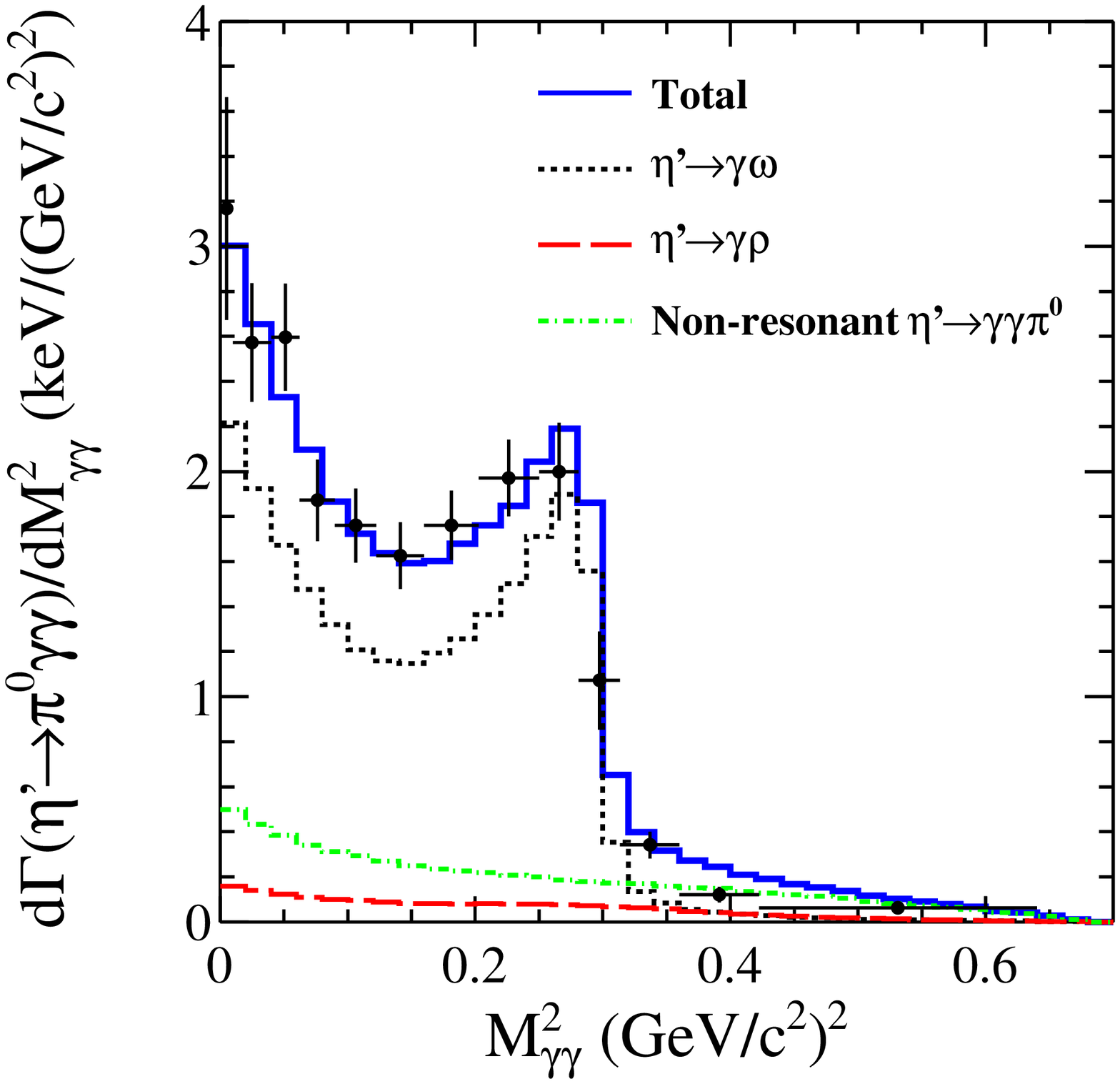}\put(-35,130){\bf (c)}
    \caption{ (a) Distribution of $M({\gamma\gamma\pi^0})$ for the
      selected inclusive $\eta^\prime\to \gamma\gamma\pi^0$ signal
      events. The dotted and dot-dashed curve are the of the
      background contributions. The total fit result is shown as the
      solid line. (b) Distribution of $M(\gamma\pi^0)$ and fit result
      (solid line). The dotted-curve is the $\omega$-contribution; the
      long dashed-curve $\rho^0$-contribution; the short dashed-curve
      is the $\rho^0$-$\omega$ interference. The other curves are
      described in the picture.
(c)     Acceptance corrected and normalized to the  partial width (in keV)
$M^2({\gamma\gamma})$
distribution for the inclusive $\eta^\prime\to \gamma\gamma\pi^0$ decay.
The error includes
    the statistic and systematic uncertainties. The (blue) histogram is the
incoherent sum of $\rho^0$ and $\omega$ and the
    non-resonant components from MC simulations; the (back) dotted-curve is $\omega$-contribution; the (red) dot-dashed-curve is
    the $\rho^0$-contribution; and the (green) dashed-curve is the non-resonant contribution. }
  \label{etafit_R}
  \vspace{0.0cm}
\end{figure*}

The
 $\etap\to\gamma\gamma\pio$ decay should be dominated
by the sequential process $\etap\to\gamma\omega\to\gamma\gamma\pio$.
The interesting question is to determine a non-resonant
contribution to the decay.
At present there is only a preliminary theoretical analysis
which uses
combination of the linear sigma model and VMD. The
prediction for branching fraction of
$\eta^\prime\rightarrow\gamma\gamma\pi^0$ is
$\sim6\times 10^{-3}$~\cite{Jora:2010zz,Escribano:2012dk}.
This is quite puzzling result giving two times larger
value than the $\gamma\omega$ sequence.
The first observation of
$\eta^\prime\rightarrow\gamma\gamma\pi^0$ is reported by the BESIII
experiment.  Fig.~\ref{etafit_R}(a) shows the $\gamma\gamma\pi^0$
invariant mass spectrum, where the clear $\eta^\prime$ peak is
observed.  By assuming that the inclusive decay $\eta^\prime\to
\gamma\gamma\pi^0$ can be attributed to the vector mesons $\rho^0$ and
$\omega$ and the non-resonant contribution, a unbinned maximum
likelihood fit to the $\gamma\pi^0$ invariant mass
[Fig.~\ref{etafit_R} (b)] is performed to determine the signal yields
for the non-resonant $\eta^\prime\to \gamma\gamma\pi^0$ decay using
the $\eta^\prime$ signal events with $|M(\gamma\gamma\pi^0) -
m_{\eta^\prime}|<25$~MeV/c$^{2}$.  In the fit, the $\rho^0$-$\omega$
interference is considered, but possible interference between the
$\omega$ ($\rho^0$) and the non-resonant process is neglected.

The branching fraction of the inclusive decay is measured
to be ${\cal B}(\eta^{\prime}\to \gamma\gamma\pi^{0})_{\text{Incl.}} = (3.20\pm0.07\pm0.23)\times 10^{-3}$,
which is much lower than the theoretical predictions \cite{Jora:2010zz,Escribano:2012dk}.  In addition, the branching fraction for the non-resonant decay is determined to be
${\cal B}(\eta^{\prime}\to \gamma\gamma\pi^{0})_{\text{NR}}$ = $(6.16\pm0.64\pm0.67)\times 10^{-4}$, which
agrees with the upper limit measured by the GAMS-2000 experiment~\cite{Alde:1987jt}. As a validation of the fit, the product branching
fraction with the omega intermediate state involved is obtained to be ${\cal B}(\eta^{\prime}\to \gamma\omega)\cdot{\cal B}(\omega\to \gamma\pi^{0})$
= $(2.37\pm0.07 \pm0.18)\times 10^{-3}$, which is consistent with the PDG value~\cite{PDG}. Hopefully this first result will trigger new theory analyses
of the decay. In particular a combined analysis of this decay with
$\etap\to\pip\pim\pio\gamma$ might provide more details about
role of isoscalar mesons and isospin violating processes to the $\etap$
transition form factor.

\section{Rare decays}

\subsection[]{Invisible decays~\cite{Ablikim:2012gf}}
Studies of $\eta$, $\eta^\prime$ decays
where one or more
products escape detection is a sensitive probe for
new light particles beyond the SM.
A two-body hadronic decay
$J/\psi\rightarrow\phi\eta(\etap)$ is
well suited to tag production of $\eta/\etap$ mesons since
the presence  of undetected particles could be
 established by missing four-momentum.
The $\phi$ meson is reconstructed
efficiently and with good resolution from the $K^+K^-$ decay.
  The method was first applied for
searches of the invisible decays of $\eta/\eta^\prime$
({\it i.e.} where none of the decay products
is observed) at BESII
experiment in 2005 \cite{Ablikim:2006eg}.

At BESIII the search for the invisible decays of $\eta$ and
$\etap$ was repeated using 2009 data set {\it i.e.} with the statistics of
four times larger than
BESII. The event selection requires exactly two tracks with
opposite charges, identified as kaons.
Figure~\ref{eta-etap-invidata}(a) shows
invariant mass of the kaons, $M(K^+K^-)$, with a
clear $\phi$ peak, while no evident $\eta$ or $\eta^\prime$ signal is
observed in the mass spectrum for the $\phi$ recoil system as shown in
Fig.~\ref{eta-etap-invidata}(b). To reduce the systematic
uncertainty, the $\eta(\eta^\prime) \to \gamma
\gamma$ decay is also identified in $J/\psi \to \phi \eta (\eta^\prime)$, and
the ratios of $\mathcal{B}(\eta(\eta^\prime)\to
invisible)$ to $\mathcal{B}(\eta(\eta^\prime)\to \gamma
\gamma)$ are determined. Using the world averages \cite{PDG}
for the two photon branching fractions of $\eta$ and
$\etap$,
the following 90\% C.L. upper limits  are obtained $\BR(\eta \to
invisible) < 1.32\times 10^{-4}$ and $\BR(\etap
\to invisible) < 5.31\times 10^{-4}$.

\begin{figurehere}
\begin{center}
  \includegraphics[width=5.5cm,height=4.6cm]{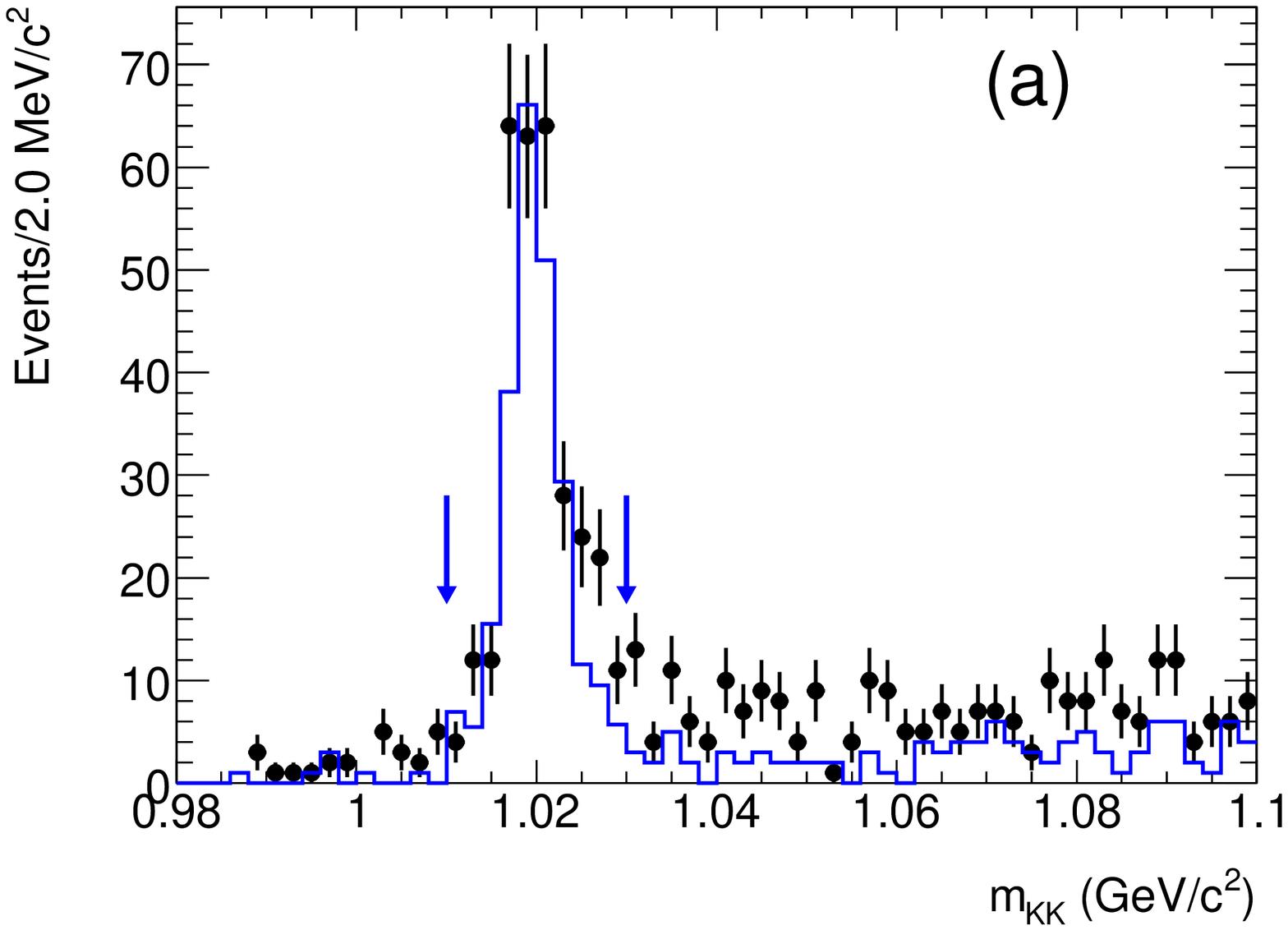}

  \includegraphics[width=5.5cm,height=4.6cm]{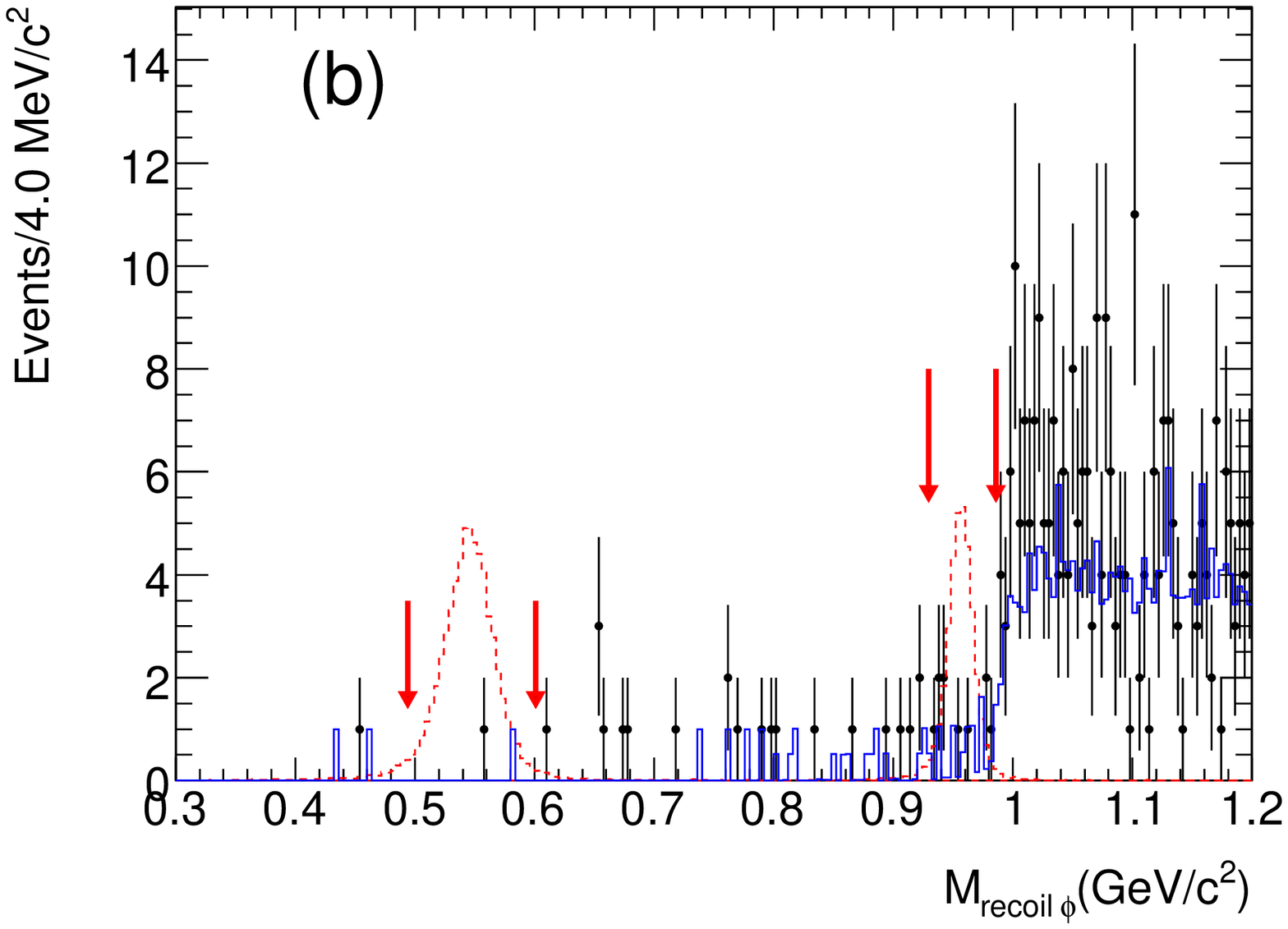}
  \caption{(a) $M(K^+K^-)$ distribution; (b) Recoil mass distribution against $\phi$
candidates,
  $M^{recoil}_{\phi}$, for events with $1.01$ GeV/$c^2$ $< M(K^+K^-) <1.03$ GeV/$c^2$ in (a).
Points with error bars are
  data;
  the (blue) solid histogram is the sum of the expected backgrounds;
  the dashed histograms (with arbitrary scale) are signals of $\eta$ and
  $\eta^\prime$ invisible decays from MC simulations; the arrows on the plot
indicate the signal regions of the $\eta$ and $\eta^\prime \rightarrow
invisible$. }
 \label{eta-etap-invidata}
\end{center}
\end{figurehere}

\subsection[]{\boldmath $\eta/\eta^\prime\to \pi^+ e^- \bar{\nu}_e +c.c.$~\cite{Ablikim:2012vn}}

Within the framework of the chiral perturbation theory, the upper bound of the branching fraction
$\eta \to \pi^+ l^-\bar{\nu}_l$  is predicted to be $2.6 \times 10^{-13}$.  After considering scalar or vector type interaction, the branching
fraction of $\eta \to \pi^+ l^-
 \bar{\nu}_l$ was estimated to be $ 10^{-8}-10^{-9}$~\cite{Fayet:2006sp,Fayet:2007ua}, which is a few orders of magnitudes higher than that in the SM.
 Therefore, searches for the $\eta \to \pi^+ l^-
 \bar{\nu}_l$ and  $\eta^\prime \to \pi^+ l^-
 \bar{\nu}_l$  at the branching fractions level of $ 10^{-8}-10^{-9}$ and below will provide information on the new physics beyond the SM.

At BESIII the searches for the decays of $\eta$ and $\eta^\prime\to\pi^+ e^- \bar{\nu}_e
+c.c.$ were performed  using  $\jpsi \to \phi \eta$ and $\phi\eta^\prime$ with the $\phi$ meson reconstructed using $K^+K^-$ decay. No signals are observed in the $\pi^+ e^- \bar{\nu}_e$ mass spectrum shown in Fig.~\ref{eta}
for either $\eta$ or $\eta^\prime$,
and upper limits at the 90\% C.L. are determined to be
$7.3\times 10^{-4}$ and $5.0\times 10^{-4}$ for the ratios
${\BR(\eta\to \pi^+ e^- \bar{\nu}_e +c.c.)}/{\BR(\eta \to \pi^+\pi^-\pi^0)}$ and
${\BR(\eta^\prime\to\pi^+ e^-\bar{\nu}_e +c.c.)}/{\BR(\eta^\prime \to
\pip\pim\eta)}$, respectively.
 Using the known values of $\BR(\eta \to \pi^+\pi^-\pi^0)$ and $\BR(\eta^\prime\to
\pi^+\pi^-\eta)$, the 90\% C.L. upper limits for the
semileptonic decay rates are $\BR(\eta\to \pi^+ e^-
\bar{\nu}_e +c.c.)<1.7\times 10^{-4}$ and $\BR(\eta^\prime\to \pi^+ e^- \bar{\nu}_e +c.c.)<2.2\times 10^{-4}$.

\begin{figurehere}
\begin{center}
  {\label{fig:subfig:a}\includegraphics[width=0.68\columnwidth]{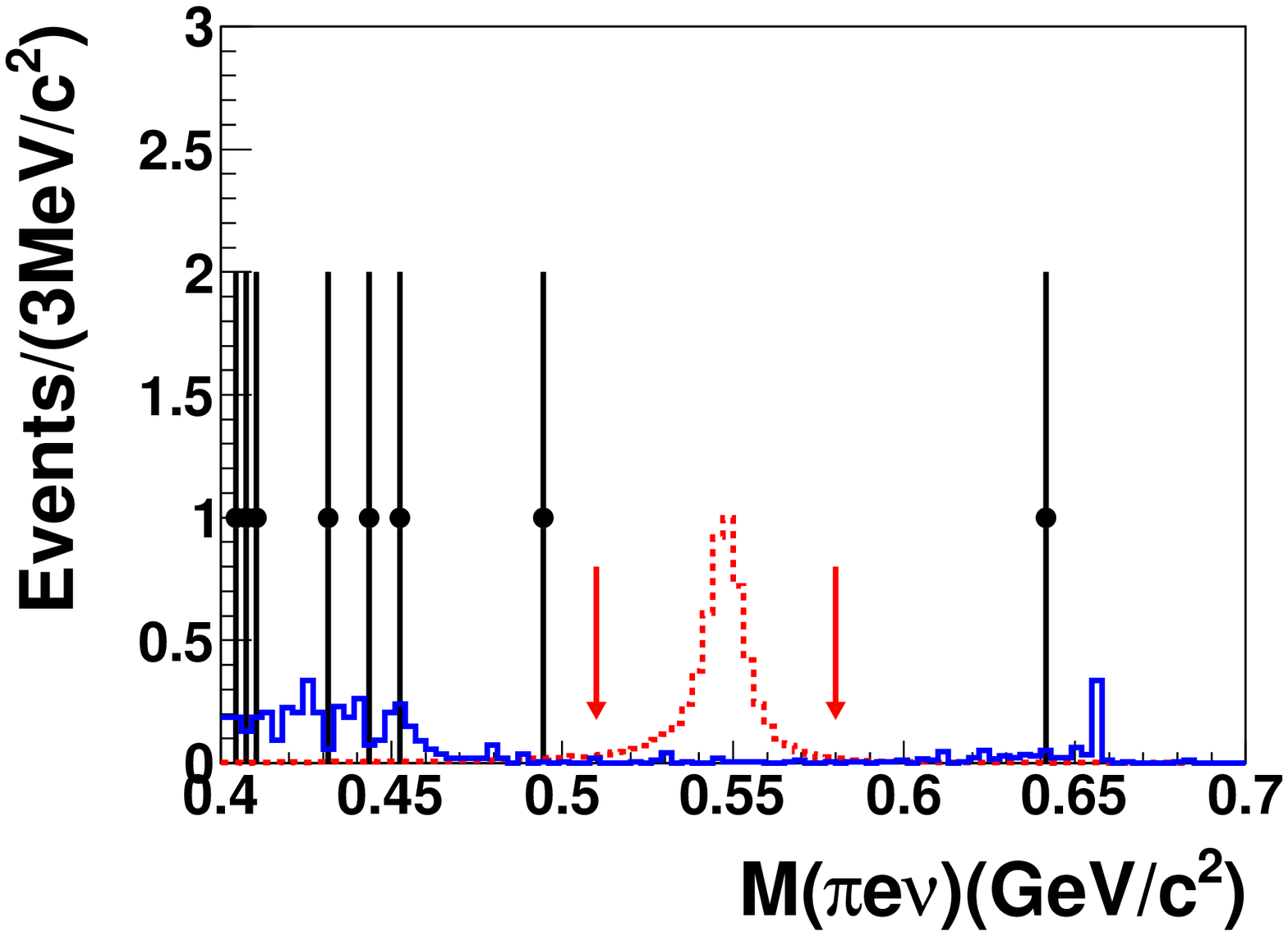}}
  \put(-35,100) {\bf \scriptsize (a)}

  {\label{fig:subfig:b}\includegraphics[width=0.68\columnwidth]{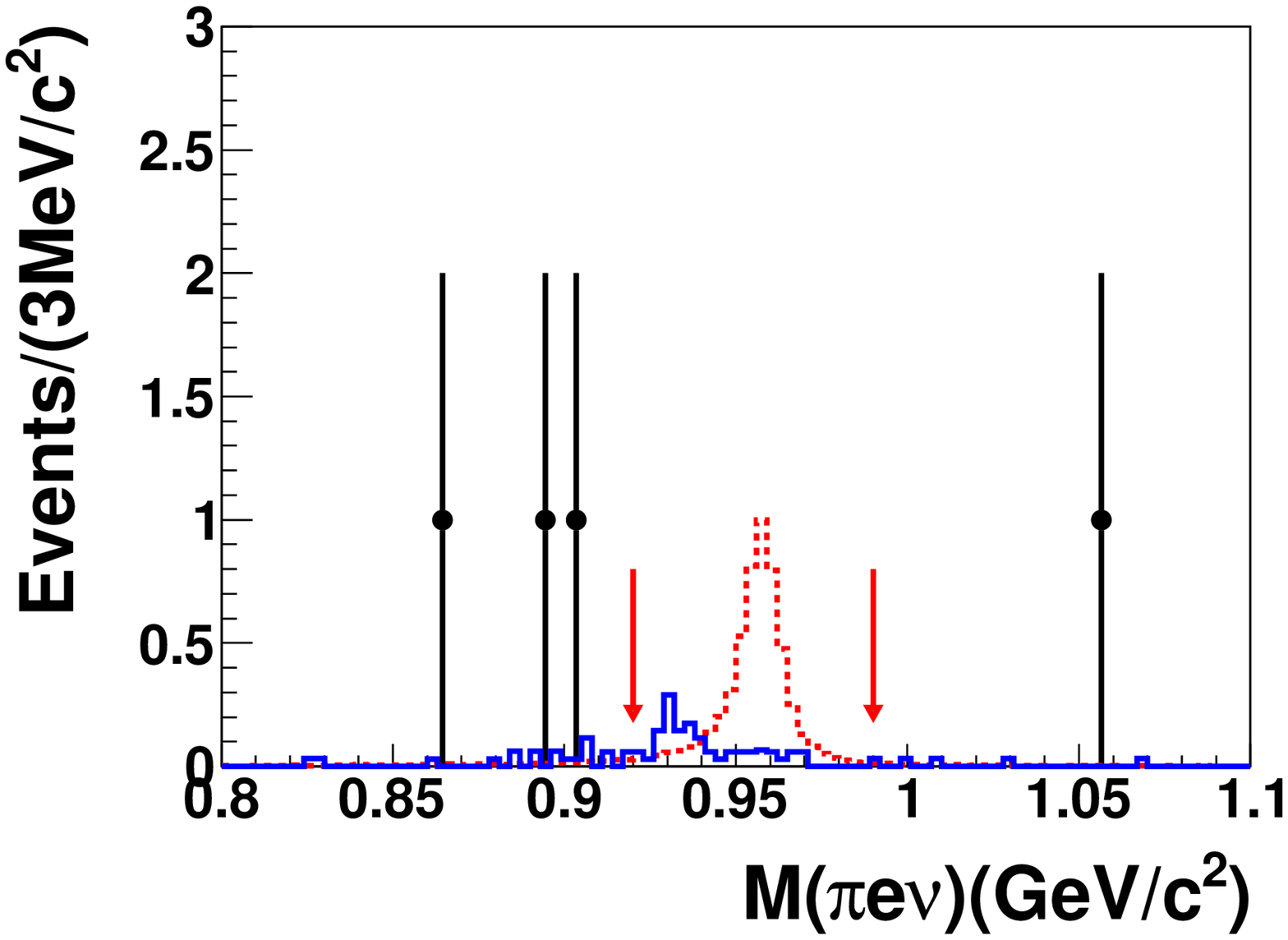}}
  \put(-35,100) {\bf \scriptsize (b)}
\caption{The $M({\pi^+ e^- \bar{\nu}_e})$ distributions of candidate
events:
 (a) for $J/\psi \to \phi \eta$ ($\eta \to \pi^+ e^- \bar{\nu}_e$); (b) for $J/\psi \to \phi
 \eta^\prime$ ($\eta^\prime\to \pi^+ e^- \bar{\nu}_e$).  For both (a) and (b): the data (dots with error bars) are
compared to the signal MC samples (red dashed histogram) and the
expected backgrounds (solid blue histogram). The arrows on the plots
indicate the signal regions of $\eta$ and $\eta^\prime$ candidates.}
\label{eta}
\end{center}
\end{figurehere}

\subsection[]{\boldmath $\eta/\etap\to \pi\pi $~\cite{Ablikim:2011vg} }

In the SM, these processes can proceed via the weak interaction with
a branching fraction of order $10^{-27}$ according to
Ref.~\cite{Jarlskog:2002zz}. Higher branching fractions are possible either by
introducing a $CP$ violating term in the QCD lagrangian (a branching
fraction up to $10^{-17}$ can be obtained in this scheme) or allowing
$CP$ violation in the extended Higgs sector (with  $\BR$ up to $10^{-15}$),
 as described in Ref.~\cite{Jarlskog:2002zz}. The detection of
these decays at any level accessible today would signal $P$ and $CP$
violations from new sources, beyond any considered extension of the
SM.

In BESIII analysis of the 2009 data set, the $CP$ and $P$
violating decays of $\eta/\eta^\prime \rightarrow$ $\pi^-\pi^-$ and $\pi^0 \pi^0$ were searched in $J/\psi$ radiative decays.
The mass spectra of $\pi^+\pi^-$ and $\pi^0\pi^0$ are shown in
Fig.~\ref{up-data-pippim} and Fig.~\ref{up-data-2pi0}, respectively.
 No significant $\eta$ or $\eta^\prime$ signal is observed. Using the Bayesian method,
the 90\% C.L. upper limits are determined to be  $\BR(\eta \rightarrow
  \pp)<3.9\times 10^{-4}$, $\BR(\etap \rightarrow \pp)<5.5\times 10^{-5}$, $\BR(\eta \rightarrow \pi^0
  \pi^0)<6.9\times 10^{-4}$ and $\BR(\etap \rightarrow \pi^0 \pi^0)<4.5\times
  10^{-4}$.

\begin{figurehere}
\begin{center}
\includegraphics[height=6.2cm, angle=-90]{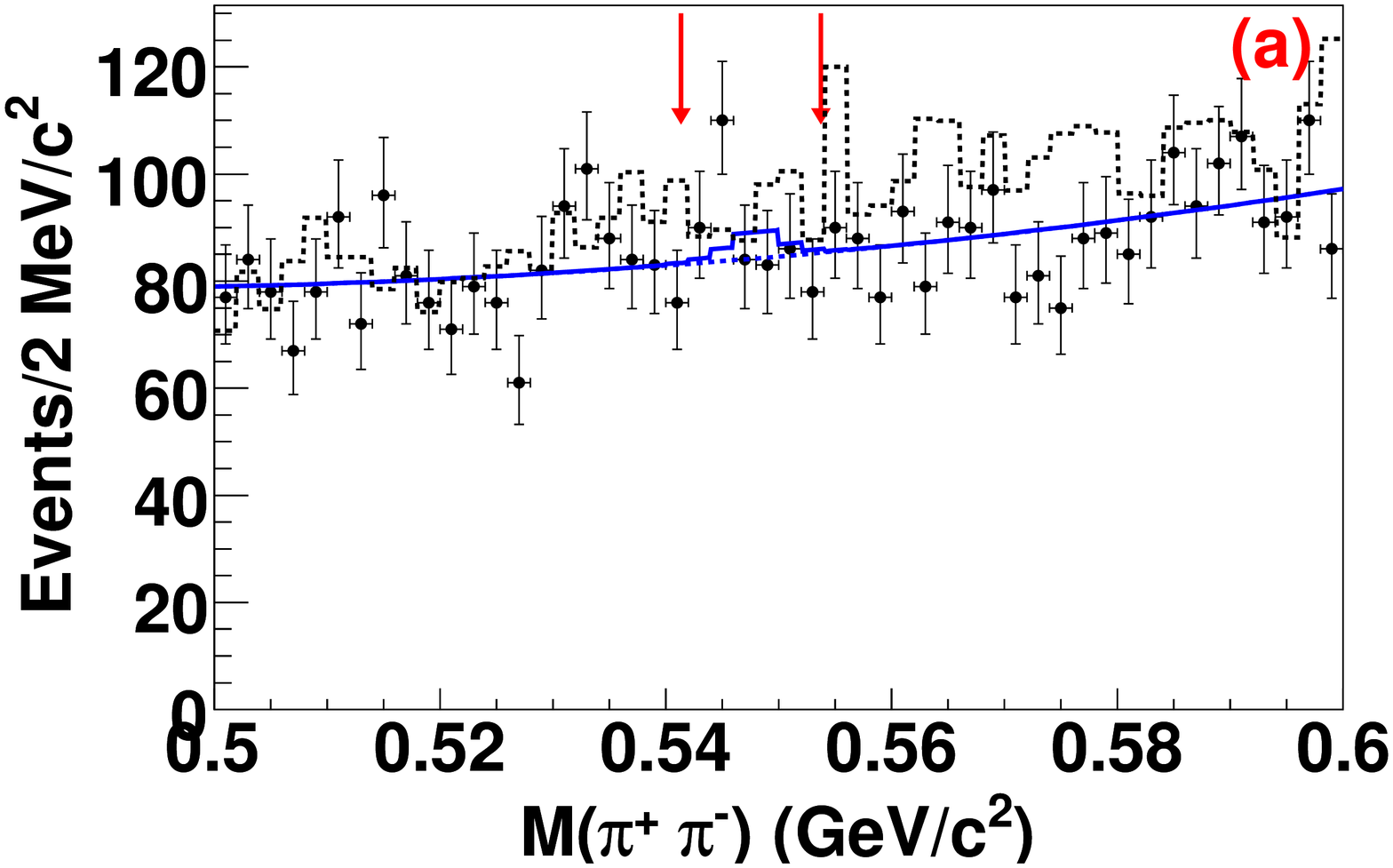}\hspace{0.2cm}

\includegraphics[height=6.2cm, angle=-90]{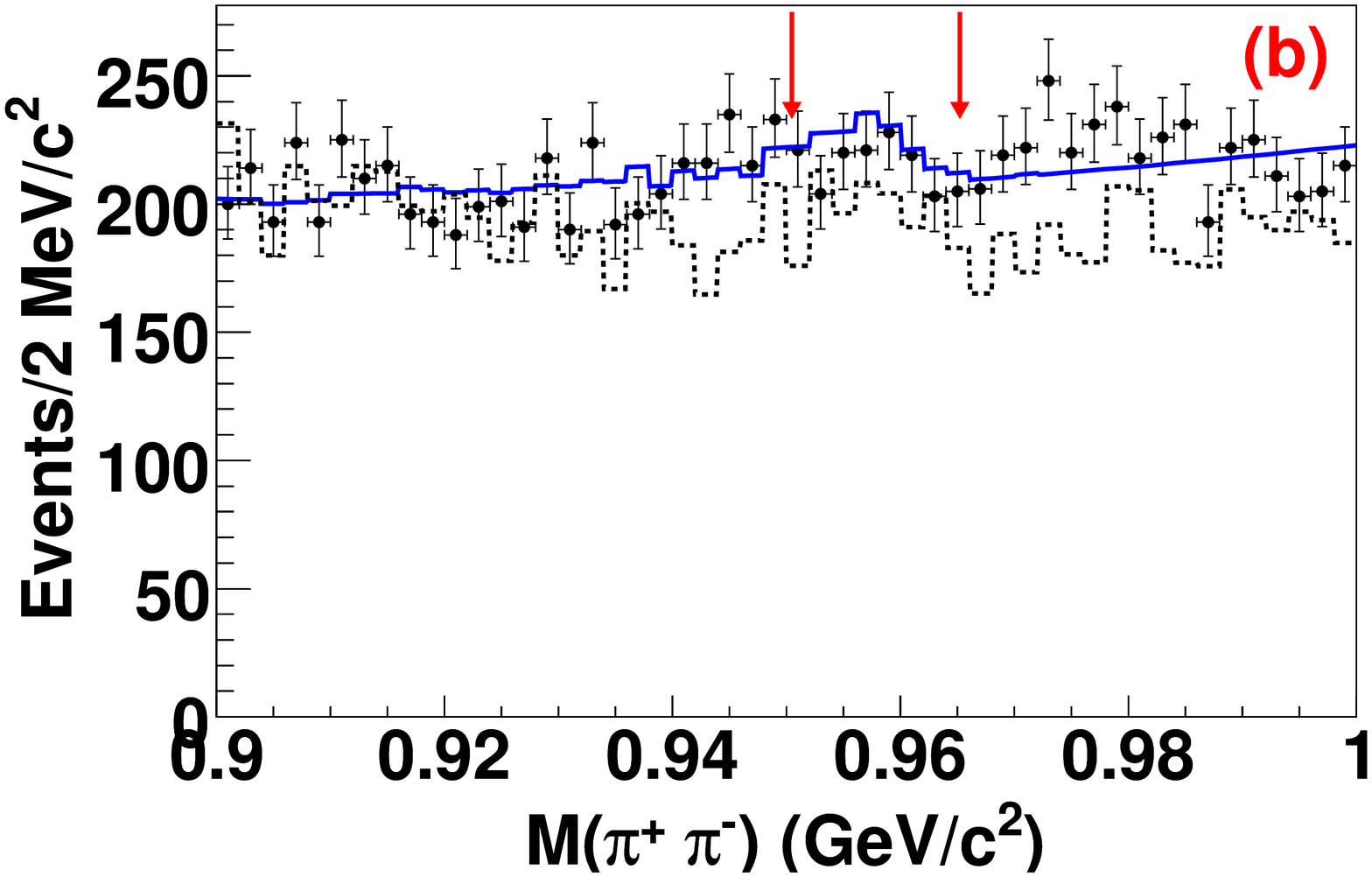}\hspace{0.2cm}
\end{center}
\caption{\label{up-data-pippim} The $\pi^+\pi^-$ invariant
mass distributions of the final candidate events, (a) $\eta$ mass region, (b)
$\eta^\prime$
mass regions. The dots with error bars are data,
the solid lines are the fit described in the text, and the
dashed histograms are the sum of all the simulated normalized backgrounds.
The arrows show mass regions
which should contain 95\% of the signal according to MC simulations.
}
\end{figurehere}

\begin{figurehere}
\begin{center}
\includegraphics[height=6.2cm, angle=-90]{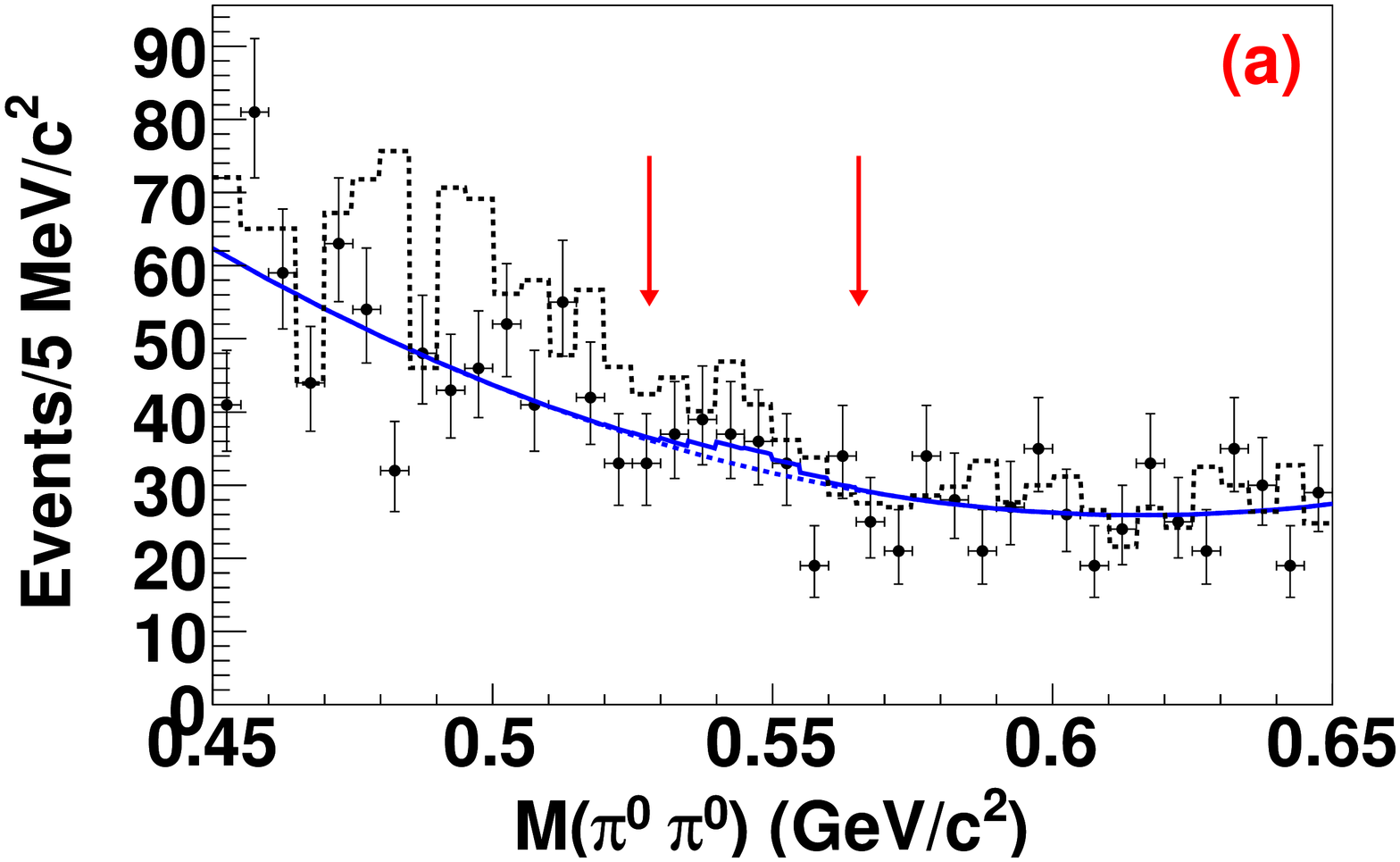}\hspace{0.2cm}

\includegraphics[height=6.2cm, angle=-90]{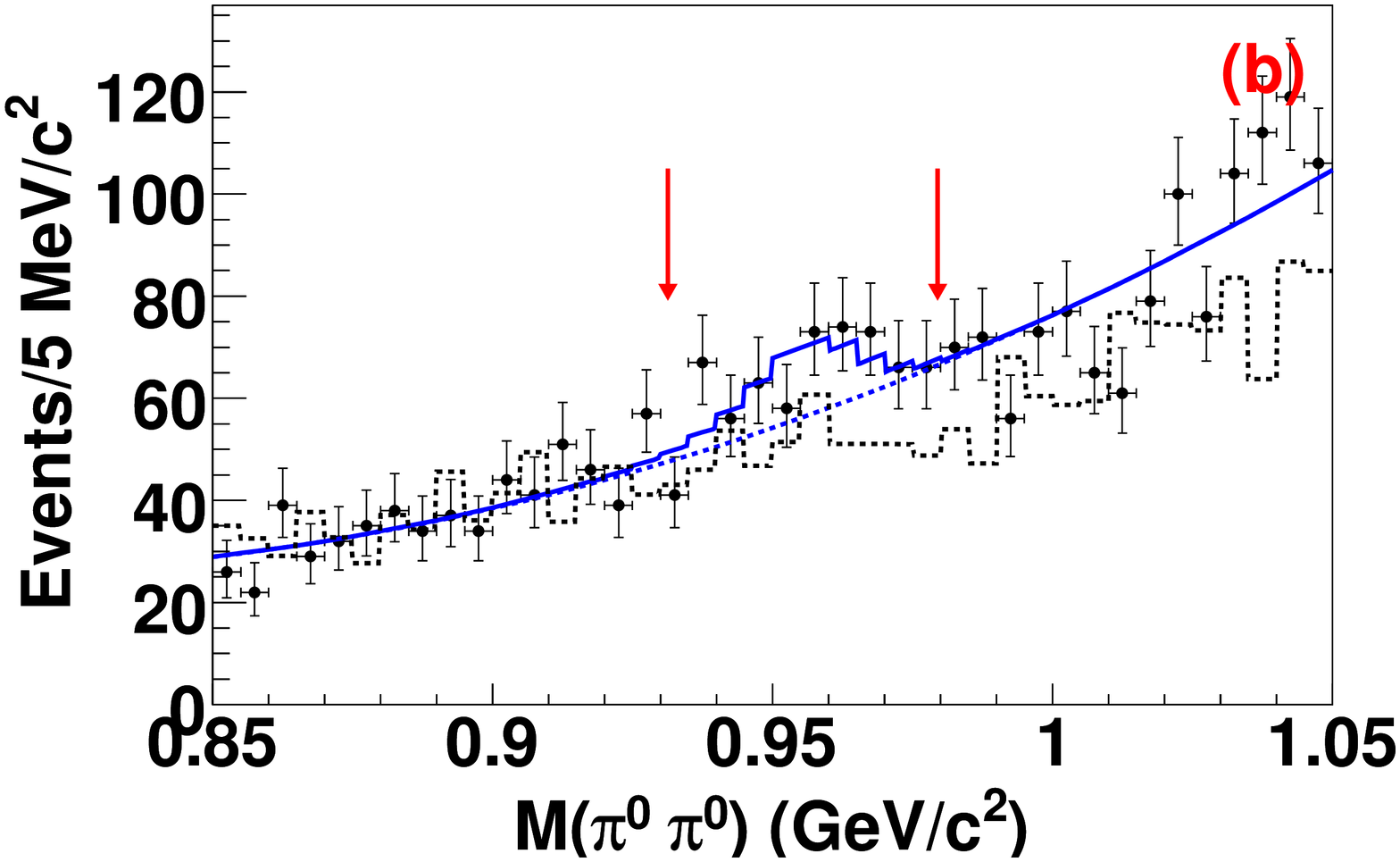}\hspace{0.2cm}
\end{center}
\caption{\label{up-data-2pi0} The $\pi^0 \pi^0$  invariant
mass distributions of the final candidate events, (a) $\eta$ mass region, (b) $\eta^\prime$
 mass regions. The dots with error bars are the data,
the solid lines are the fit described in the text, and the
dashed histograms are the sum of all the simulated normalized backgrounds.
The arrows show mass regions
which should contain 95\% of the signal according to MC simulations.
}
\end{figurehere}
\subsection[]{\boldmath $\etap\to K^\pm\pi^\mp$~\cite{Ablikim:2016bjc}}
Non-leptonic weak decays are valuable tools for exploring physics
beyond the SM. Among the non-leptonic decays, the decay
 $\eta'\to K^{\pm}\pi^{\mp}$ is of special interest due to
the long-standing problem of the
$\Delta I=1/2$ rule in weak non-leptonic interactions. The branching fraction
of $\etap\to K^{\pm}\pi^{\mp}$ decay is predicted to be of the order
of $10^{-10}$ or higher~\cite{Bergstrom:1987py}, with a large
long-range hadronic contribution expected.

 A search for the non-leptonic weak decay
$\eta'\to K^{\pm}\pi^{\mp}$ is performed for the first time through
the $\jpsi\to\phi\eta'$ decay, while no evidence for
$\etap\to K^{\pm}\pi^{\mp}$ is seen in the $K\pi$ mass spectrum (Fig.~\ref{PhiKPi_plot}). Thus the 90\% C.L. upper limit on
$\mathcal{B}(\eta^\prime\to K^\pm\pi^\mp)$ of $3.8\times10^{-5}$ is reported.

\begin{figurehere}
\begin{center}

  \includegraphics[width=0.45\textwidth]{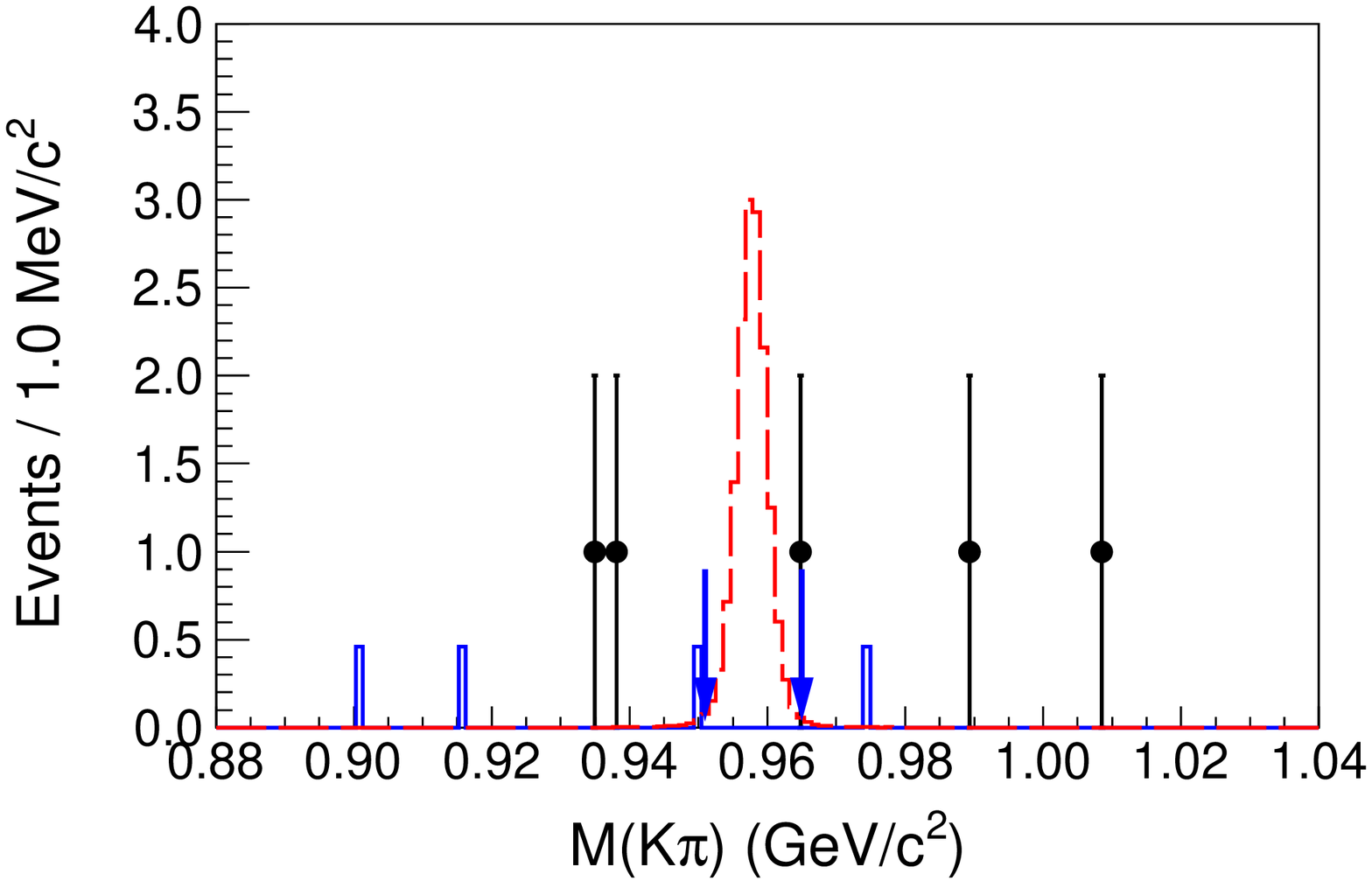}
  \caption{\label{PhiKPi_plot}
   The $K^ \pm
    \pi^\mp$ invariant mass distribution, where the arrows show the
    signal region. The dots with error bars are the data, the dashed
    histogram is for the signal MC with arbitrary normalization, and
    the solid histogram is the background contamination from a MC
    simulation of $J/\psi \to \phi\pi^+\pi^-$.}
\end{center}
\end{figurehere}

\section{Summary}
$J/\psi$ decays provide a clean source of $\etap$ for the
decay studies.  Based on the world largest sample of $J/\psi$ events,
the recent results on $\eta/\eta^\prime$ decays achieved at the BESIII
experiment are presented.  In addition to the improved accuracy on the
branching fractions of $\eta^\prime$,  observation of
$\eta^\prime$ new decay modes, including $\eta^\prime\ra\pip\pin\pip\pin$,
$\eta^\prime\ra\pip\pin\pio\pio$,
$\eta^\prime\rightarrow\rho^\mp\pi^\pm$ and $\eta^\prime\ra \gamma
e^+e^-$, were reported for the first time.
Precision of  $\eta'\to \pi^+\pi^-\gamma$  $M(\pip\pim)$ distribution
from BESIII with clear $\rho^0-\omega$ interference is comparable
to the $\EE\to\pip\pim$ data and allows to compare these two reactions
in both model dependent and model independent way. In particular
a competitive extraction of $\omega\to\pip\pim$ branching fraction is possible.
It is found
that an extra contribution is necessary to describe data besides the
contributions from $\rho^0(770)$ and $\omega$.


Despite the impressive progress, many $\eta/\eta^\prime$
decays are still to be observed and explored.
 The
  BESIII detector will collect a sample of $10^{10}$ $J/\psi$
  events in the near future, which offers great prospects for research
  in $\eta/\eta^\prime$ physics with unprecedented precision.

A list of the specific decay channels
where important impact of the new data is expected includes:
\begin{itemize}
\item Larger data samples of $\eta \ra \pi^{+}\pi^{-}\pi^0$ and
  $\eta\ra\pi^0\pi^0\pi^0$ decays from BESIII experiment are needed to
  provide independent check of the analyses carried out at other
  experiments. The Dalitz plot distributions will be made available
  for the direct fits to theory.  In particular data from the planned
  run will allow to collect approximately $0.5\times10^6$ events of $\eta \ra
  \pi^{+}\pi^{-}\pi^0$ with negligible background. In addition one could
probably also  use other $J/\psi$ decay modes as sources of
$\eta$ mesons. Some possible examples are  $J/\psi\to\omega\eta$
($\BR=(1.74\pm0.20)\times10^{-3}$) and $J/\psi\to p\bar{p}\eta$
($\BR=(2.00\pm0.12)\times10^{-3}$).
\item For $\etap\rightarrow \pi^+\pi^-\eta$ and
  $\etap\ra\pi^0\pi^0\eta$ decays the analysis could be extended not
  only by using the new data but also by reconstructing the final
  states with three pion decay modes of $\eta$ in addition to
  $\eta\to\gamma\gamma$.
The increased statistics would allow {\it e.g.} to search for the cusp
in  $\etap\ra\pi^0\pi^0\eta$ and provide further information
about $\eta\pi$ scattering \cite{Isken:2017dkw}.
In addition the use of the same final state topology
$\eta^\prime\rightarrow \pip\pim(\eta\to\pio\pio\pio)$
and  $\eta^\prime\rightarrow \pio\pio(\eta\to\pip\pim\pio)$ will enable to determine
ratio of the two decay modes with low systematic uncertainty.
\item For $\etap\ra\pi^0\pi^0\pi^0$ and $\etap \ra \pi^{+}\pi^{-}\pi^0$ larger statistics is crucial
to carry out amplitude analysis of the processes. At present it is impossible to differentiate
between $S$ and $D$ waves. A detailed understanding of this process dynamics is a prerequisite
for a program of light quark masses determination from comparison to
the $\etap\to \pi\pi\eta$ processes, a method which do not rely on the full decay width value.
Several theory groups has expressed interest in description of the decay within
dispersive approach.  The overall goal is understanding of all parity even processes of
$\eta$ and $\etap$.
\item Hadronic parity odd processes $\eta^\prime\rightarrow
  \pi^+\pi^-\pi^+\pi^-$, $\pi^+\pi^-\pi^0\pi^0$ offers a window to study
  double off shell transition form factor of $\eta'$. The ultimate
  goal would be to carry out amplitude analysis of the reactions.  The
  new data with expected about 1200 events would allow for a
   first stage of such analysis. However, the collected data on
these decays
  together with  $\etap\to\pi^+\pi^-e^+e^-$ decay will already provide
  unique constraints and checks for the models of the $\eta'$
  double off shell transition form factor. In addition,
since the predicted branching fraction for the
related $\etap\to \pi^+\pi^- \mu^+\mu^-$ process is about $2\times 10^{-5}$,
the decay likely will be observed with the new data. The branching
fraction value of this decay is sensitive to the $\eta'$ transition form factor.
\item For $\eta^\prime\rightarrow \pi^+\pi^-e^+e^-$ decay much
  progress is expected.  A combined analysis of 2012 and the new run
  data will allow for a sample close to $2\times 10^4$ events. In
  particular a $CP$ symmetry test by measurement of asymmetry between
  the lepton and the pion decay planes \cite{Geng:2002ua,Gao:2002gq}
  as well as studies of the $M(\EE)$ and $M(\pip\pim)$ distributions would
  be possible.
\item Among the discussed in this review very rare decays of $\eta/\etap$
the largest impact of the new data is expected  for
invisible decays, $\eta/\etap\to \pi^+ e^- \bar{\nu}_e$ and
 $\etap\to K^\pm\pi^\mp$ where no background was observed.
Therefore the sensitivity should scale with luminosity. In
addition the
results for the first two decay modes
are  based  on the  2009 data set only.
\end{itemize}

Both $\eta$ and $\eta^\prime$ decays are important tools for studies
of strong interactions in non-perturbative region and for
determination of some SM parameters.  In addition they provide an
indirect way to probe physics beyond the standard model.  In
particular the pursued at BESIII $\eta$ and $\eta^\prime$ decay
program, where the data collected at $J/\psi$ are used for wealth
of other studies, represents smart and resource efficient
research strategy.

\bibliographystyle{lisstyle}
\bibliography{lit}
%
%
\end{multicols}{}
\end{document}